\documentclass{article}

\usepackage{arxiv}
\usepackage[utf8]{inputenc}
\usepackage[T1]{fontenc}
\usepackage[svgnames]{xcolor}
\usepackage[colorlinks]{hyperref}
\usepackage{url}
\usepackage{booktabs}
\usepackage{amsfonts}
\usepackage{nicefrac}
\usepackage{microtype} 
\usepackage{amsmath}
\usepackage{dsfont}
\usepackage{amsthm}
\usepackage{tikz}
\usetikzlibrary{decorations.pathreplacing}
\usepackage{hhline}
\theoremstyle{definition}

\newtheorem{remark}{Remark}[section]
\newtheorem{definition}{Definition}[section]
\newtheorem{assumption}{Assumption}[section]

\usepackage{graphicx}
\usepackage{enumitem}
\usepackage{natbib}
\usepackage{appendix}
\usepackage{doi}
\usepackage{lscape}
\usepackage{multirow}
\usepackage[justification=centering, hypcap=false]{caption} 
\usepackage{colortbl}

\makeatletter
\newcommand{\subalign}[1]{
  \vcenter{
    \Let@ \restore@math@cr \default@tag
    \baselineskip\fontdimen10 \scriptfont\tw@
    \advance\baselineskip\fontdimen12 \scriptfont\tw@
    \lineskip\thr@@\fontdimen8 \scriptfont\thr@@
    \lineskiplimit\lineskip
    \ialign{\hfil$\m@th\scriptstyle##$&$\m@th\scriptstyle{}##$\hfil\crcr
      #1\crcr
    }
  }
}
\makeatother

\title{The Short-Term Predictability of Returns in Order Book Markets: a Deep Learning Perspective}

\author{Lorenzo Lucchese\thanks{This research has been supported by the EPSRC Centre for Doctoral Training in Mathematics of Random Systems: Analysis, Modelling and Simulation (EP/S023925/1). We would like to thank Thomas Oliver (InferStat) for helpful discussions on the subject.} \\
	Department of Mathematics \\
	Imperial College London \\
	\texttt{lorenzo.lucchese17@imperial.ac.uk} \\
	\And
	Mikko S. Pakkanen\\
	Department of Statistics and Actuarial Science\\
	University of Waterloo\\
	and\\
	Department of Mathematics \\
	Imperial College London \\
	\texttt{mikko.pakkanen@uwaterloo.ca} \\
        \And
        Almut E. D. Veraart\\
	Department of Mathematics \\
	Imperial College London \\
	\texttt{a.veraart@imperial.ac.uk}\\
}

\renewcommand{\headeright}{}
\renewcommand{\undertitle}{}
\renewcommand{\shorttitle}{The Short-Term Predictability of Returns in Order Book Markets}

\hypersetup{
pdftitle={The Short-Term Predictability of Returns in Order Book Markets: a Deep Learning Perspective},
pdfauthor={Lorenzo Lucchese}
}

\begin{document}
\hypersetup{
  linkcolor=LightGreen,
  urlcolor=LightGreen,
  citecolor=LightGreen
}
\maketitle

\begin{abstract}
    In this paper, we conduct a systematic large-scale analysis of order book-driven predictability in high-frequency returns by leveraging deep learning techniques. First, we introduce a new and robust representation of the order book, the volume representation. Next, we carry out an extensive empirical experiment to address various questions regarding predictability. We investigate if and how far ahead there is predictability, the importance of a robust data representation, the advantages of multi-horizon modeling, and the presence of universal trading patterns. We use model confidence sets, which provide a formalized statistical inference framework particularly well suited to answer these questions. Our findings show that at high frequencies predictability in mid-price returns is not just present, but ubiquitous. The performance of the deep learning models is strongly dependent on the choice of order book representation, and in this respect, the volume representation appears to have multiple practical advantages.
\end{abstract}

\section{Introduction} \label{sec:intro}
\subsection{Financial markets and exchanges: the rise of High-Frequency traders} \label{sec:market_context}
A financial market is an ensemble of market agents willing to buy or sell a certain financial security, such as a stock, bond, or derivative. Today most trades take place on electronic exchanges, virtual places that bring together buyers and sellers, facilitating the occurrence of transactions. At the time of writing, the two largest U.S.\ equity exchanges by market capitalization are the NYSE, a hybrid (floor and electronic) auction market accounting for 20\% of the U.S.\ equities market transactions, and the Nasdaq, a fully electronic dealer market executing about 16\% of such trades (source: Cboe Exchange, Inc.). Both of these markets allow traders to access live order book information, i.e.\ the collection of all standing orders for a given security. In theory, this allows for symmetric information across traders, which should all have access to the same data regarding market depth, liquidity, and price discovery dynamics.

In practice, traders have access to different technology, receiving market data and submitting orders at different latencies. In the quest to exploit the advantages gained by faster access to markets, almost two decades ago, a new market participant emerged. These market players are today known as High-Frequency Traders (HFTs) and, over the years, have rapidly grown to represent a significant share of the market \citep{HFT2014}. Their role has since been the object of a fierce debate: their critics claim HFTs engage in predatory -- and sometimes illegal -- behavior, while their supporters believe HFTs to be overall beneficial to the market by providing liquidity, reducing spreads, and helping price discovery dynamics. In this paper, we will not delve into questions regarding the legitimacy of HFT practices, but we will instead aim to independently analyze the value of immediate access to order book information. Over the past few decades, HFT companies have engaged in a fierce race to zero latency, making vast economic efforts to reduce their latency by just a few microseconds. What we aim to explore in this research is one of the possible reasons why such a race happened in the first place. Specifically, we will be analyzing the predictive value of order book data, i.e.\ to what extent can a trader with immediate access to the order book predict the future direction of the market?

It is also important to note that, over the past couple of decades, the trading process has become increasingly complex due to market fragmentation, availability of new technologies such as smart order routers (SORs), and -- sometimes controversial -- regulation, e.g. RegNMS \citep{regNMS} and MiFID \citep{MiFID}. While practices that exploit arbitrage between competing trading venues exist, in this research, we will assume the trader has access to a single electronic order book-based exchange, namely the Nasdaq.

\subsection{Order book predictability: asking the right questions} \label{sec:questions}
As discussed in the opening paragraph, we would like to explore whether, contrary to low-frequency returns, ultra-high-frequency returns tend to display predictability. Empirical studies \citep{universalLOB} have shown that price formation dynamics, i.e.\ next mid-price moves, are predictable. In this paper, we will try to understand whether such predictability persists at longer horizons. Intuitively, predictability in high-frequency returns may be understood to arise simply from the way an order book market is structured, i.e.\ the side of the order book with less liquidity is more likely to erode faster, resulting in a price increase/decrease, or as the product of recurring trading patterns in response to liquidity information. The approaches considered in the literature for forecasting high-frequency returns from order book data can be roughly divided into two categories: relatively simple models built on carefully handcrafted features \citep{AitSahalia2022} and more sophisticated architectures applied directly to raw order book data \citep{deepLOB, deepLOB_multihorizon, deepOF}. In this research, we will focus on the latter class of models, leveraging the ability of deep learning techniques to learn complex dependence structures. We will consider a specific class of deep learning models, introduced by \citet{deepLOB}, designed to extract features from order book data. There is empirical evidence \citep{representation_learning_Bengio} which suggests that, although deep learning models can extract complex features, the way data is represented may have a significant impact on model performance. We will hence explore how model performance varies when changing the way the order book data is arranged. Equipped with this class of models, we will aim to investigate the following questions, which naturally arise from our preceding discussion:
\begin{enumerate}
\item Do high-frequency returns display order book-driven predictability? If so, how far ahead can we predict?
\item Which order book representations perform best? 
\item Can we use a single model across multiple horizons?
\item Can we use a single model across multiple stocks?
\end{enumerate}
We aim to answer all these questions in a formalized statistical inference framework based on model confidence sets \citep{mcs}.

\subsection{Related work and contributions} \label{sec:literature}
It is important to note that mathematical modeling of order book dynamics is a very broad and active area of research: ranging from Hawkes process models for order and trade events in continuous time \citep{Hawkes_LOB_Bacry, Hawkes_LOB_Large} to discrete-time synthetic data-driven order book generation \citep{ABIDES, LOB_simulation_GANs}. The work presented in this paper can be seen as contributing to two parallel research streams in the literature for order book-driven mid-price predictions. On one hand, we expand on the deep learning ideas discussed in \citet{deepLOB, deepLOB_multihorizon, deepOF} by introducing new data representations and carrying out a disciplined comparison between the specifications. On the other, we explore a set of questions related to short-term price predictability in a similar spirit to \citet{AitSahalia2022} but under a different class of models.

In the broader context of the questions addressed in this paper, related works are those of \citet{universalLOB}, exploring the universality of order book dynamics, and \citet{robust_representation}, advocating for robust representations of order books. We base our experimental procedure on model confidence sets, introduced by \citet{mcs}. We believe this formalized statistical inference framework perfectly suits our aim of addressing questions that require comparisons between multiple models and benchmarks.

The two main contributions of this paper are summarized as follows. First, we introduce a deep learning model for mid-price forecasting based on a more robust representation of the order book, which we will refer to as deepVOL. This representation allows us to easily adapt the model to the setting where more granular L3 data\footnotemark{} is available\footnotetext{See Section \ref{sec:order_book} for the definitions of L1, L2 and L3 order book data.}. Second, we provide new empirical results addressing essential questions regarding short-term price predictability in a disciplined experimental framework.

\section{The space of models under consideration} \label{sec:models}
\subsection{The order book} \label{sec:order_book}
\begin{figure}[!htb]
    \centering
    \includegraphics[width = 12cm]{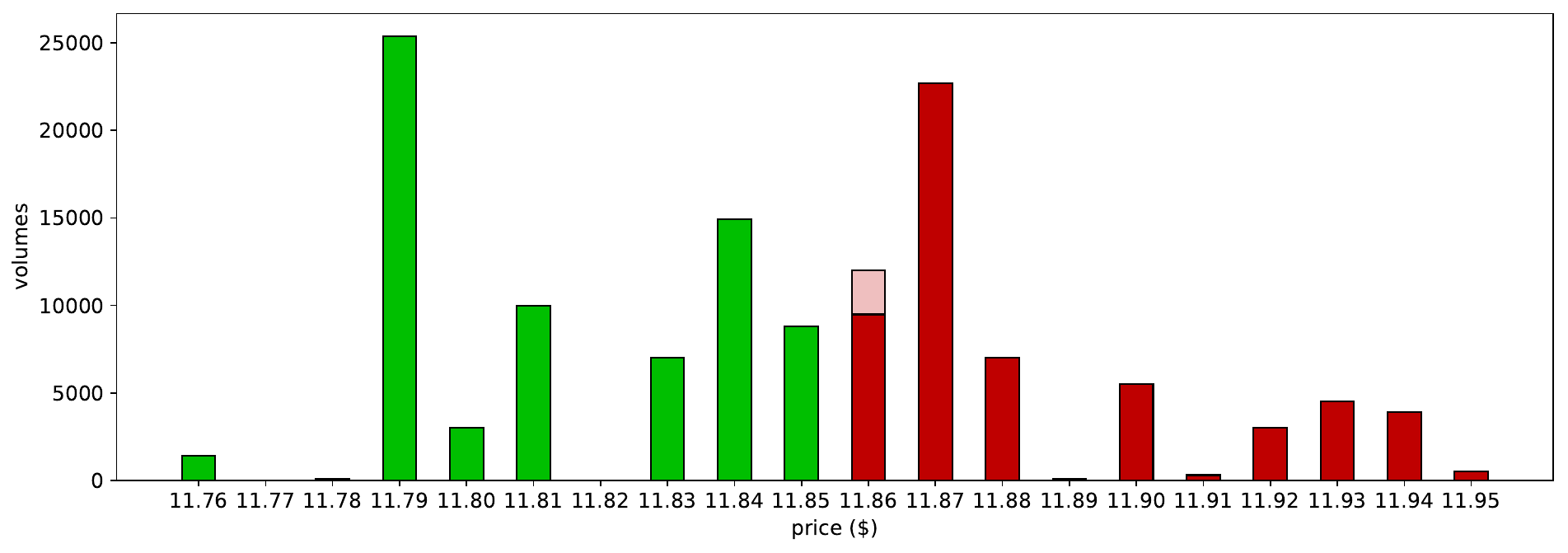}
    \caption{A sample snapshot of an order book. Ask (resp.\ bid) volumes are denoted by red (resp.\ green) bars. The lighter green shaded area represents the change in the order book shape when some of the liquidity at the best bid price is removed.} \label{fig:orderbook}
\end{figure}

At a given point in time, an order book contains all the (visible) buy and sell orders placed for a given security on a specific exchange. The lowest ask price, resp.\ the highest bid price, is known as the first ask order book level, resp.\ the first bid order book level. Subsequent levels are defined accordingly. An example of a 10-level order book snapshot is displayed in Figure \ref{fig:orderbook}. We note that in electronic exchanges, orders can be submitted on an evenly spaced discrete set of prices, known as ticks. The smallest price increment is known as the tick size, for most U.S.\ traded stocks, it corresponds to \$0.01. It is important to note that not all tick prices may have standing orders; therefore, the first 10-level ask/bid prices may not coincide with the first 10 ask/bid tick prices.

There are three main types of actions that traders can request on an exchange:
\begin{itemize}
\item a limit order: an order to buy a given quantity of the security at a given price;
\item a market order: an order to buy a given quantity of the security at the best available price;
\item a cancellation or partial deletion: an order to fully or partially delete a standing limit order.
\end{itemize}
Note that while market orders are always immediately executed once posted, passive limit orders, i.e.\ orders which do not cross the spread, will sit in the order book until they are matched. A limit order entering the market at a tick price where other limit orders are already present will be added at the end of the standing queue. Trades occur each time a market order or aggressive limit order is posted, the requested volume is matched to the standing limit orders according to price-time priority. We will not delve into the detailed characteristics of all the different order types which exist, but it suffices to point out that the three actions described above represent the fundamental drivers of the evolution of order books. For example, in Figure \ref{fig:orderbook}, the change in order book shape may be due to either a (partial) deletion of a first-level buy limit order or the execution of a sell order (either a market order or an aggressive limit order).

During trading hours, electronic exchanges operate continuously, and for this reason, order books are sometimes referred to as continuous books. Exchanges such as the Nasdaq time stamp each event at nanosecond precision. Between different stocks, the level of trading activity may vary significantly, and the time elapsed between consecutive events may differ by orders of magnitude. For this reason, we define an alternative time clock: a discrete order book clock, which increments by one each time an action is executed on the order book. Throughout our discussion, we will explore questions of predictability with respect to this order book clock, which is the same as the one considered in \citet{FI2010}. Some authors, for example, \citet{AitSahalia2022} consider alternative order book-driven time clocks, such as transaction clocks and volume clocks. 

We understand that models based on order book-specific clocks might be challenging to use in practical trading applications. However, we believe order book-based clocks provide a more natural measure of time for exploring predictability and are more suitable for comparing results across stocks than physical time: a 100ms time horizon has a significantly different meaning for stocks with different levels of trading activity.

\begin{figure}[!htb]
    \centering
    \includegraphics[width = 12cm]{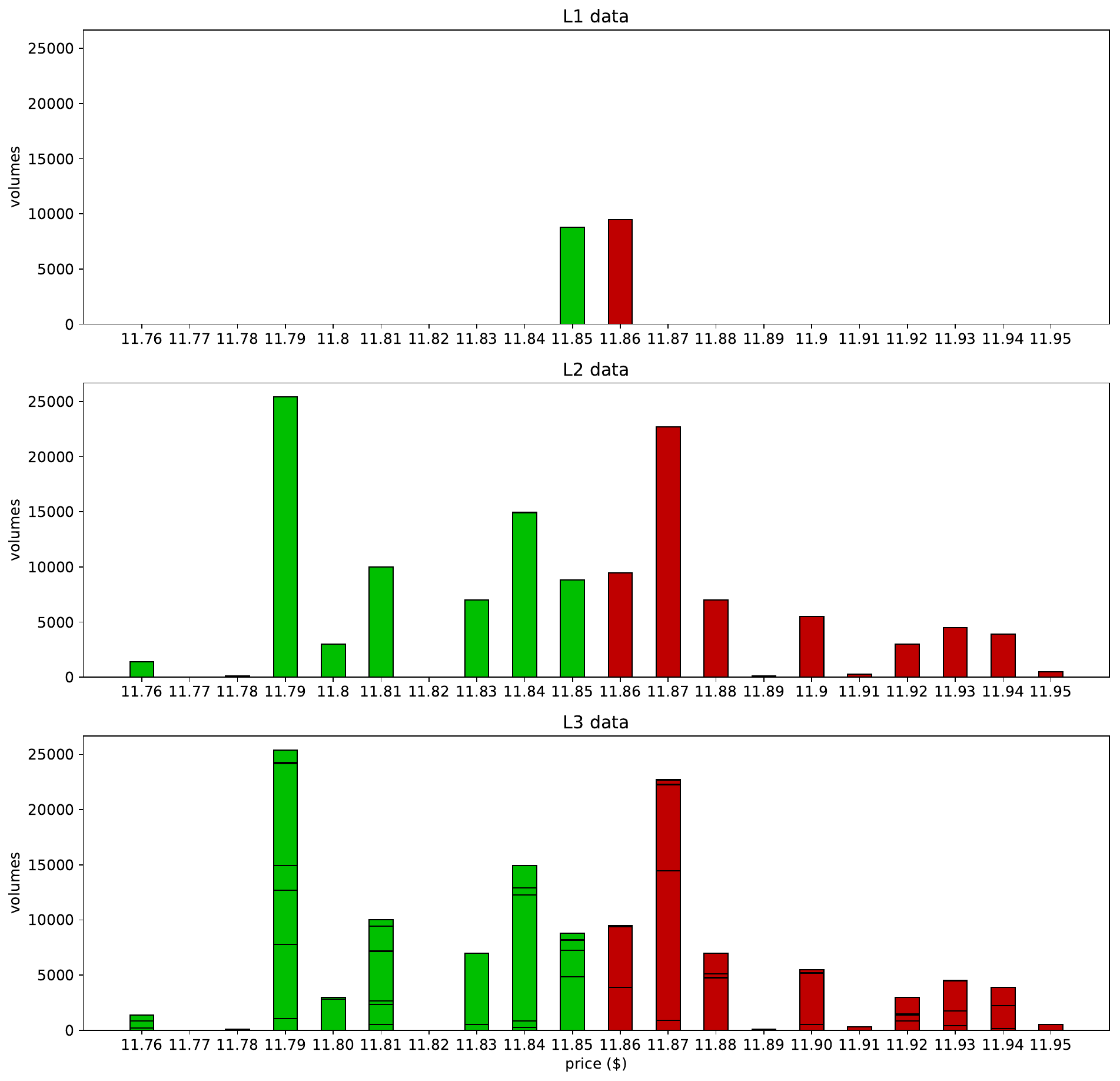}
    \caption{L1, L2 and L3 representations of the order book in Figure \ref{fig:orderbook}.}
    \label{fig:orderbook_L1L2L3}
\end{figure}

Another important observation is that order book data might not be accessed by all market participants at the same level of granularity. The Nasdaq Quotation Dissemination Service makes the following distinctions:
\begin{itemize}
    \item L1 data: the best bid and ask prices and corresponding volumes;
    \item L2 data: all available bid and ask prices and corresponding volumes;
    \item L3 data: all available bid and ask prices and corresponding volumes split among the orders in the queue.
\end{itemize}
We will be comparing the performance of models when different levels of data granularity are available. Figure \ref{fig:orderbook_L1L2L3} provides a visual comparison of L1, L2, and L3 data.

We introduce the following notation: at time $t\in\mathbb{Z}$ we denote by
\begin{itemize}
    \item $p^{(l)}_{x, t}, v^{(l)}_{x, t}$ for $l=1, 2, \ldots$ the $l$-th level ask/bid price and volume, for $x\in\{a,b\}$;
    \item $m_{t} = \dfrac{p^{(1)}_{b, t} + p^{(1)}_{a, t}}{2}$ the mid-price;
    \item $\pi^{(j)}_{x, t}, s^{(j)}_{x, t}$ for $j=1, 2, \ldots$ the $j$-th ask/bid tick price from the mid and corresponding volume for $x\in\{a,b\}$, i.e.\ this is defined recursively as
    \begin{align*}
        \pi^{(1)}_{x, t} &= 
            \begin{cases}
                m_t &\text{if } m_{t} \in \mathcal{T}, \\
                m_t \pm \frac{\vartheta}{2} &\text{if } m_{t} \notin \mathcal{T},
            \end{cases} \\
        \pi^{(j)}_{x, t} &= \pi^{(j-1)}_{x, t} \pm \vartheta \ \text{ for } j =2, 3, \ldots,
    \end{align*}
    where $\vartheta$ is the tick size, $\mathcal{T}=\{k\vartheta:k\in\mathbb{N}\}$ is the set of possible tick prices and $\pm$ depends on $x\in\{a,b\}$.
    \item $q^{(i, k)}_{x, t}$ for $k=1, 2, \ldots$ the queue corresponding to volume 
    $s^{(i)}_{x, t}$ and ordered by time priority.
\end{itemize}

\subsection{Predictability of returns} \label{sec:predictability_returns}
We will explore models that aim to identify short-term predictability in returns. We first introduce the familiar regression framework for return predictions. We then rephrase the task in terms of a classification problem, extending the definition of predictability to this setting. 

We will be exploring predictability arising from past order book data and thus we define the information $\sigma$-algebra $\mathcal{F}_t$ to be
\begin{equation*}
\mathcal{F}_t = \sigma(\mathbf{x}_t, \ldots, \mathbf{x}_{t-T+1}),
\end{equation*}
where $\mathbf{x}_t, \ldots, \mathbf{x}_{t-T+1}$ are order book derived features at times $t,\ldots, t-T+1$ for some look-back window of length $T$.

\paragraph{Predictions in the regression framework}
Let us first consider the regression setting. At time $t$ we denote by $r_{t,t+h}\in\mathbb{R}$ the $h$-step ahead mid-price return, as defined in \ref{sec:responses}. Given order book information at time $t$, $\mathcal{F}_t$, there exists a measurable function $g$ such that
\begin{equation*}
\mathbb{E}[r_{t,t+h}|\mathcal{F}_t] = g(\mathbf{x}_t, \ldots, \mathbf{x}_{t-T+1}),
\end{equation*}
or, equivalently, 
\begin{equation*}
r_{t,t+h} = g(\mathbf{x}_t, \ldots, \mathbf{x}_{t-T+1}) + \epsilon_t,
\end{equation*}
for some mean-zero noise variable $\epsilon_t$ orthogonal to the space of $\mathcal{F}_t$-measurable random variables. A prediction is defined to be any $\mathcal{F}_t$-measurable random variable and the best prediction is the $\mathcal{F}_t$-measurable random variable $r^{(t)}_{t,t+h}$ which minimizes the expected cost 
\[\mathbb{E}[C(r^{(t)}_{t,t+h}, r_{t,t+h})],\] 
for some appropriate cost function $C:\mathbb{R}\times\mathbb{R}\rightarrow \mathbb{R}.$ In the case of quadratic cost $C(r_1, r_2) = (r_1 - r_2)^2$, we have 
\[r^{(t)}_{t,t+h} = \mathbb{E}[r_{t,t+h}|\mathcal{F}_t] = g(\mathbf{x}_t, \ldots, \mathbf{x}_{t-T+1}).\]
Different choices of cost function are possible, for example, when $C$ is absolute cost, i.e. $C(r_1, r_2) = |r_1 - r_2|$, the best prediction is given by the conditional median of $r_{t,t+h}$ given $\mathcal{F}_t$. Given a parametric family of models $\{g_\theta(\cdot)\}_{\theta\in\Theta}$ and an observed training data set $\mathcal{D}_{\text{train}} = \{(\mathbf{x}_t, \ldots, \mathbf{x}_{t-T+1}, r_{t,t+h})\}_{t \in \mathcal{I}_{\text{train}}}$ one can first learn a function ${g}_{\hat\theta}$ approximating $g$ and then produce the return predictions 
\[\hat{r}^{(t)}_{t,t+h} = {g}_{\hat\theta}(\mathbf{x}_t, \ldots, \mathbf{x}_{t-T+1}), \]
for test data points $\mathcal{D}_{\text{test}} = \{(\mathbf{x}_t, \ldots, \mathbf{x}_{t-T+1}, r_{t,t+h})\}_{t \in \mathcal{I}_{\text{test}}}$. Assuming returns to be stationary, we say that there is order book-driven predictability if the learned predictions outperform an \textit{unpredictive benchmark} prediction on the testing set with respect to the chosen cost function $C(\cdot, \cdot)$.

\paragraph{Predictions in the classification framework}
In this paper, we will discretize the space of returns by grouping mid-price movements as downward, no-change, and upward\footnote{We believe that, for high-frequency returns, a classification framework is more natural: mid-prices move in (half)-tick increments, and hence the corresponding returns live on a discrete lattice (see Section \ref{sec:responses}). Moreover, grouping returns in three classes (downward, no-change, and upward) significantly reduces the effects of idiosyncratic noise while preserving the ability to investigate questions regarding predictability.}. We hence introduce the discretized return random variable 
\begin{equation*}
c_{t,t+h} = \begin{cases}
\downarrow &\text{if } r_{t,t+h} \in (-\infty, -\gamma), \\
= &\text{if } r_{t,t+h} \in [-\gamma, +\gamma], \\
\uparrow &\text{if } r_{t,t+h} \in (+\gamma, +\infty), \\
\end{cases}
\end{equation*}
for an appropriate choice of $\gamma>0$, cf.\ Section \ref{sec:responses}.
Instead of modeling only the expected conditional return, in the classification setting, one aims to approximate the whole conditional distribution, i.e.\ find measurable functions $p_{\downarrow}, p_{=}, p_{\uparrow}$ such that
\begin{equation*}
\mathbb{P}(c_{t,t+h} = * |\mathcal{F}_t) = p_{*}(\mathbf{x}_t, \ldots, \mathbf{x}_{t-T+1}),
\end{equation*}
for $* \in \{\downarrow, =, \uparrow\}$. The discretized return prediction for an unobserved sample is then given by the minimizer of the expected misclassification cost:
\begin{equation*}
    \hat{c}_{t,t+h} = \underset{* \in \{\downarrow, =, \uparrow\}} {\text{argmin}} \sum_{\star \in \{\downarrow, =, \uparrow\}} c_{*|\star} {p}_{\star, \hat\theta}(\mathbf{x}_t, \ldots, \mathbf{x}_{t-T+1}),
\end{equation*}
where the parametric conditional probabilities ${p}_{\downarrow, \hat\theta}, {p}_{=, \hat\theta}, {p}_{\uparrow, \hat\theta}$ are learnt from a training data set $\mathcal{D}_{\text{train}} = \{(\mathbf{x}_t, \ldots, \mathbf{x}_{t-T+1}, c_{t,t+h})\}_{t \in \mathcal{I}_{\text{train}}}$ and $c_{*|\star}$ is the misclassification cost of a $*$ observation classified as a $\star$.
Assuming the return process is stationary, we say that there is order book-driven predictability if the learned predictions outperform an \textit{unpredictive benchmark} prediction on a test set in terms of total misclassification cost.

Specifying an appropriate cost function is task-dependent. For example, a trader using the predictions as trading signals might be more interested in the correctness of up and down predictions than no-change ones. On the other hand, a market maker might prioritize the correctness of no-change predictions when using these to decide whether to tighten quoted spreads. In both cases, the consequences of different types of errors are asymmetric and heavily impact the choice of the cost matrix $C = \{c_{*|\star}\}_{*,\star\in\{\downarrow, =, \uparrow\}}$. In the classification framework an alternative approach is to compare the predicted conditional distributions, $\{p_{*}(\mathbf{x}_t, \ldots, \mathbf{x}_{t-T+1})\}_{*\in\{\downarrow, =, \uparrow\}}$ to the realized outcomes directly. This approach does not require specifying a cost matrix $C$ nor a return prediction. The predicted conditional distribution (the output of our model) is directly compared with the observed data via a suitable ``distance'' on the space of probability measures $\mathcal{P}(\{\downarrow, =, \uparrow\})$, where realized returns are encoded as Dirac measures. A natural choice of such ``distance'' is given by the categorical cross-entropy, as this can be interpreted (under appropriate assumptions) as the (log)-likelihood of the test set:
\[ - \log \mathbb{P}(\mathcal{D}_{\text{test}}|\theta) \propto - \frac{1}{|\mathcal{I}_{\text{test}}|}\sum_{t\in\mathcal{I}_{\text{test}}} \sum_{*\in\{\downarrow, =, \uparrow\}} \mathds{1}_{\{c_{t,t+h} = *\}} \log p_{*,\hat{\theta}}(\mathbf{x}_t, \ldots, \mathbf{x}_{t-T+1}), \]
where the model parameters $\hat{\theta}$ are learnt from the training set $\mathcal{D}_{\text{train}}$. As this paper does not target a specific trading strategy but is more concerned with general predictability questions, we will be evaluating our models based on the categorical cross-entropy loss. In the following, we will thus say that there is order book-driven predictability if the learned conditional distributions outperform an \textit{unpredictive benchmark} distribution on a test set relative to categorical cross-entropy loss. As discussed in Appendix \ref{app:deep_learning}, the natural choice for the loss used in training will also be categorical cross-entropy.

Note that in both the regression and classification framework, we compare the learned predictions with those obtained from an \textit{unpredictive benchmark} model to determine whether there is predictability by using a score/cost function. Simply comparing point estimates of the scores does not provide a sound statistical argument answering the question of whether there is predictability: the difference in score may simply be due to statistical variability. This is where the model confidence set \citep{mcs} procedure comes in. As we will discuss in Section \ref{sec:experiments} this provides a statistical testing framework to determine whether the learned models statistically outperform the \textit{unpredictive benchmark}, i.e.\ whether there is predictability.

\begin{remark} \label{rem:benchmark}
It is clear that the definition of predictability is intrinsically tied to that of \textit{unpredictive benchmark}. Different choices for \textit{unpredictive benchmark} models are possible, for example, in the regression framework, a natural choice is given by (a version of) the efficient market hypothesis (EMH) \citep{fama_emh}. In this case, the unpredictive hypothesis assumes the conditional expected return to be 0. Under the classification framework, the simple EMH does not translate into a clear-cut model for the conditional distribution of returns. In this setting, we consider as a natural unpredictive hypothesis a slightly stronger version of the EMH: under the \textit{unpredictive benchmark}, returns are assumed to be IID and independent of any information up to time $t$. The \textit{unpredictive benchmark} conditional distribution will therefore be given by the empirical distribution of the training set. As a side remark, we believe it is important to note the EMH was originally proposed in a very different market environment and at significantly higher latencies than the ones considered in this paper. Nevertheless, we believe the EMH provides a natural unpredictive benchmark hypothesis when testing for predictability.
\end{remark}

As mentioned in Section \ref{sec:questions}, there are two main approaches considered in the literature for exploring the short-term predictability of returns in order book markets: the first considers carefully handcrafted features $\mathbf{x}_t, \ldots, \mathbf{x}_{t-T}$ and relatively simple specifications for the prediction functions $g$ or $(p_{\downarrow}, p_{=}, p_{\uparrow})$, e.g.\ linear specifications or decision trees, the second -- which we will explore in this paper -- uses raw order book features $\mathbf{x}_t, \ldots, \mathbf{x}_{t-T}$ as inputs to more complex prediction functions, e.g.\ deep neural network architectures. We will assume the reader is familiar with deep learning techniques and thus give a brief overview of the relevant concepts in Appendix \ref{app:deep_learning}, more detailed expositions can be found in \citet{deep_learning_Goodfellow}.

\subsection{Deep learning models for short-term return predictions in order book markets} \label{sec:DL_models}
In the previous section, we set up the learning framework for the return prediction task. We now discuss how the neural networks covered in Appendix \ref{app:deep_learning} can be combined to model price formation mechanisms and predict $h$-step ahead high-frequency returns. We will consider a specific class of deep learning models based on the deepLOB architecture \citep{deepLOB}.

\subsubsection{deepLOB, \texorpdfstring{\citep{deepLOB}}{deepLOB}} \label{sec:deepLOB}
This network acts on raw order book input. A CNN module and an inception module feed into an LSTM layer which produces the final classification output. The CNN and inception module aim to extract short-term spatio-temporal features in the data, while the LSTM module works on longer-term dependencies.

\begin{figure}[!htb]
    \centering
    \includegraphics[width = 14cm]{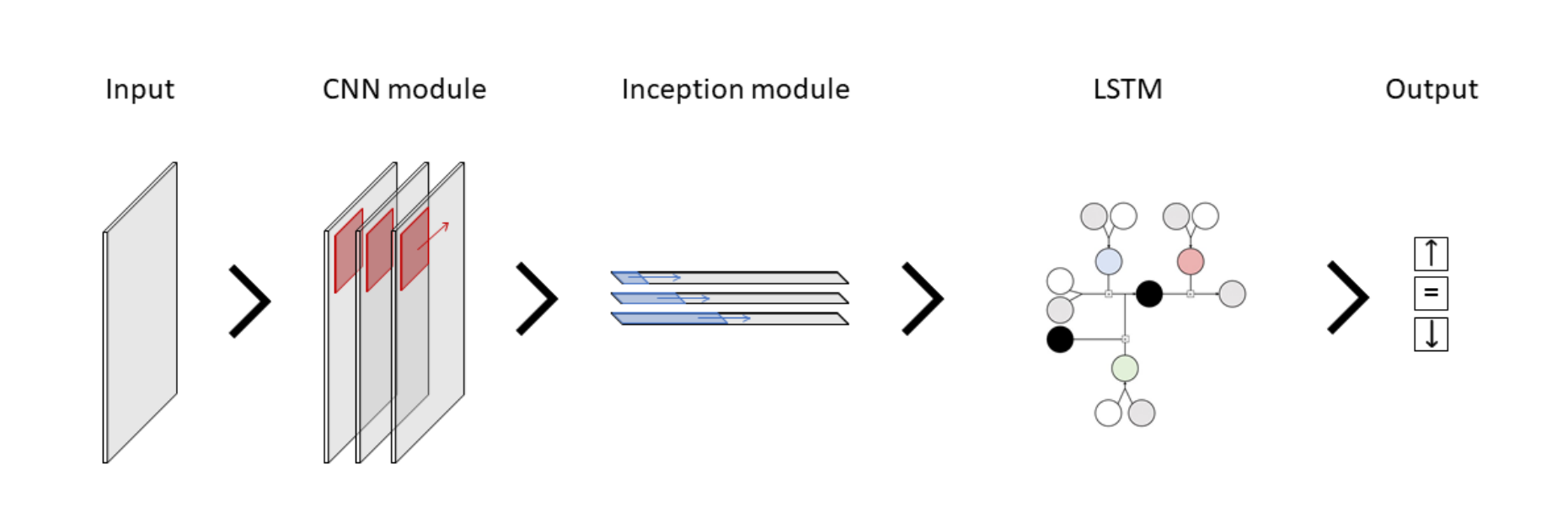}
    \caption{Core modules of network architectures.} \label{fig:core}
\end{figure}

The deepLOB architecture is summarized in Figure \ref{fig:core} and is made up of the following components:
\begin{itemize}
    \item \textbf{Input}. The first $L$ level raw order book information - price and volume - with a look-back window of length $T$ is used as input. The input at time $t$ is thus a $(T \times 4L)$ array given by 
        \[
        \{\mathbf{x}_{t-\tau}\}_{\tau=0,\ldots,T-1} = \left\{ \left(p^{(l)}_{a,t-\tau},v^{(l)}_{a,t-\tau}, p^{(l)}_{b,t-\tau}, v^{(l)}_{b,t-\tau}\right)  \right\}_{\tau=0,\ldots,T-1,\,l=1,\ldots,L}\in\mathbb{R}^{T\times 4L}.
        \]
    A feature-wise rolling window z-score standardization is applied to the input, i.e.\ $v^{(1)}_{a,t - \tau}$ is standardized using the mean and standard deviation of the first level ask volumes over the previous five days.
    \item \textbf{CNN module}. Convolutions are applied to the data in both the spatial and temporal dimensions. The spatial convolutions aim to aggregate information across order book levels, and the temporal convolutions can be understood as smoothing operations. The CNN module is summarized in Figure \ref{fig:CNN_module}.
    \begin{figure}[!htb]
        \centering
        \centerline{
        \includegraphics[height = 5.75cm]{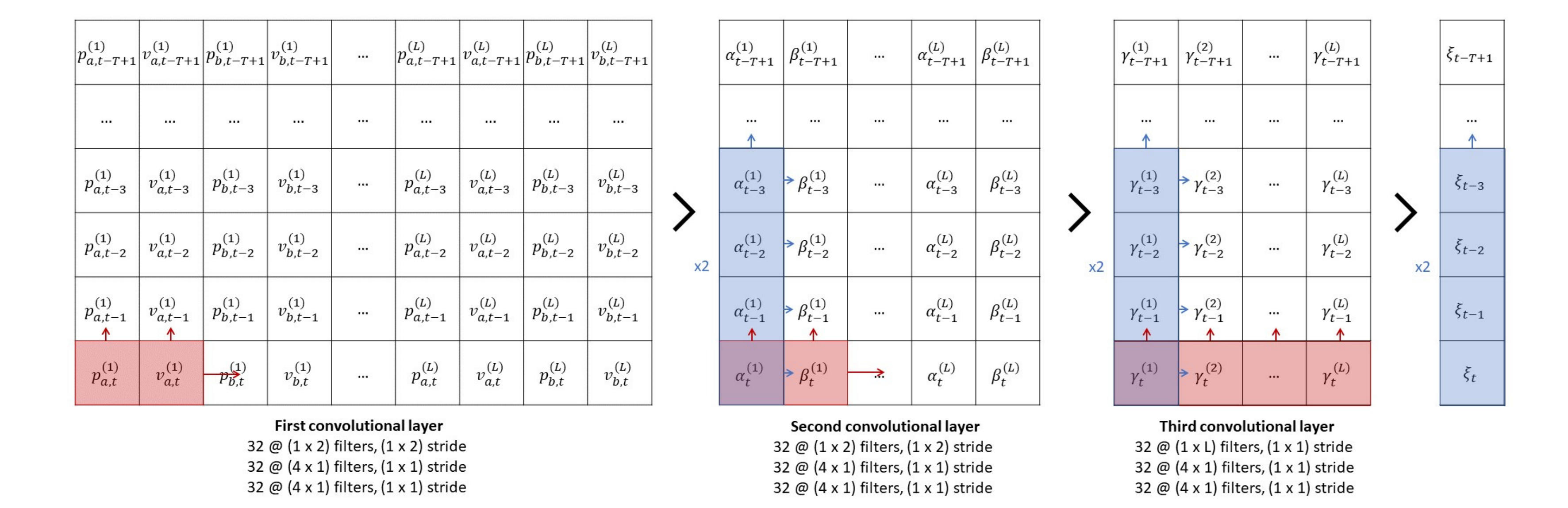}}
        \caption{CNN module deepLOB.} \label{fig:CNN_module}
    \end{figure}
    \item \textbf{Inception module}. This module up-samples the convoluted data by applying various temporal convolutions with different filter lengths (time-window). Each temporal convolution can be interpreted as a (weighted) moving average. It is similar in spirit to the computation of technical indicators, but the frequencies at which this is applied are substantially different.
    \item \textbf{LSTM}. The Long Short-Term Memory layer takes the multidimensional time series produced by the inception module and feeds it through a recurrent network structure aimed at extracting longer-term dependencies among the data. The last hidden state of the LSTM is passed through a dense layer with a softmax activation function to produce the $h$-step ahead return prediction $\downarrow, =$ or $\uparrow$.
    \begin{figure}[!htb]
        \centering
        \includegraphics[height = 11cm]{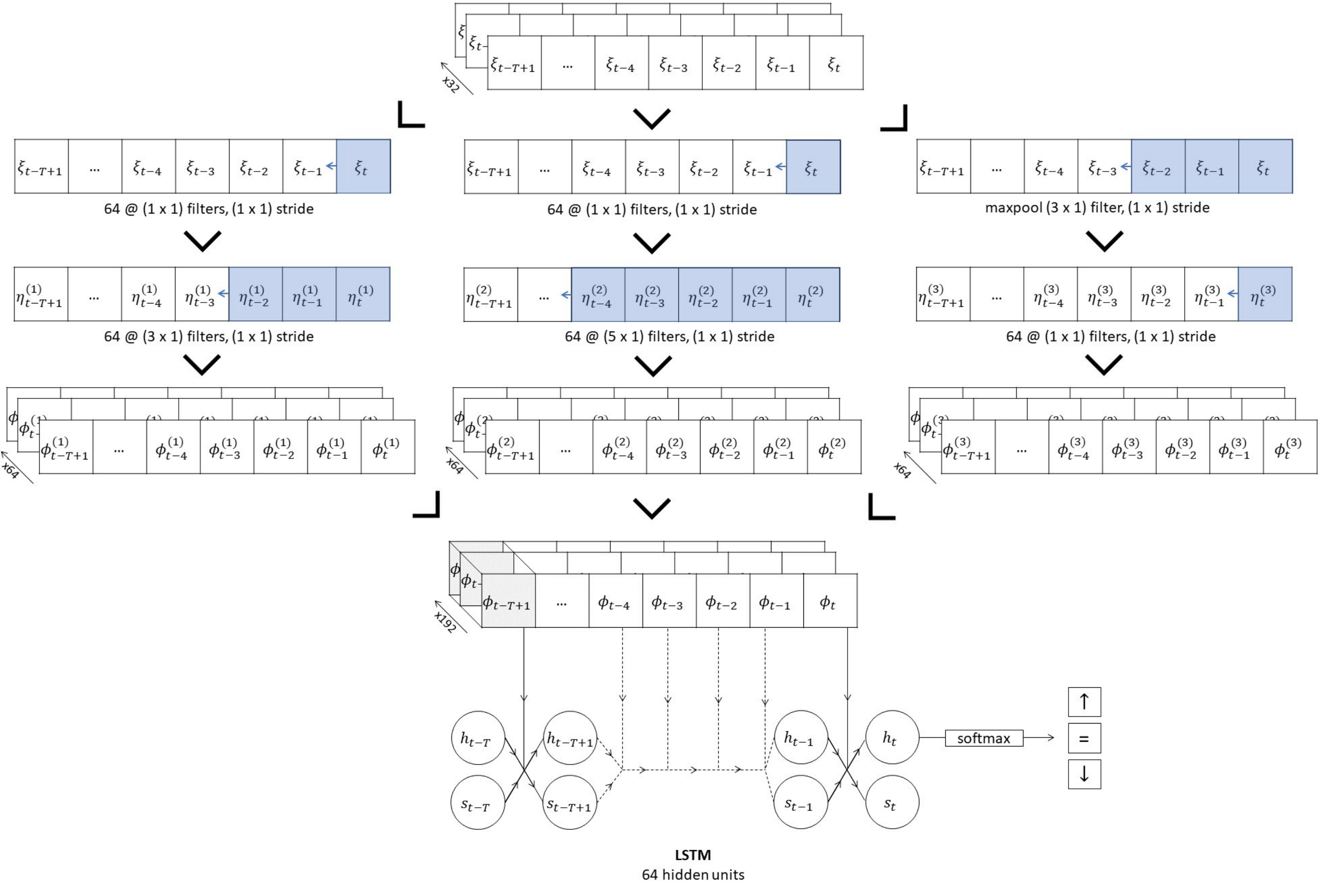}
        \caption{Inception module and LSTM layer.} \label{fig:Inception_LSTM}
    \end{figure}
\end{itemize}
The exact details of the deepLOB architecture can be found in Table \ref{table:deepLOBdeepOF}.

\subsubsection{deepOF, \texorpdfstring{\citep{deepOF}}{deepOF}} \label{sec:deepOF}
In the deepLOB model order book states, which are non-stationary, are mapped to a stationary quantity, returns. While in theory, this shouldn't be a problem (due to universality property of deep neural networks), \citet{deepOF} argue that using some form of stationary input might improve model performance by facilitating the training procedure. 

A stationary order book quantity is order flow. First-level order flow describes the net flow of orders at the best bid and ask. This was introduced in \citet{order_flow} to explore the price impact of order book events. This single quantity parsimoniously models the instantaneous effect of order book events on prices. 
The first-level bid and ask order flows corresponding to the order book event occurring at time $t$ (in order book time) are defined by:
\begin{equation*}
    bOF^{(1)}_t = \begin{cases}
    v^{(1)}_{b,t} &\text{if } p^{(1)}_{b, t} > p^{(1)}_{b, t-1},\\
    v^{(1)}_{b,t} - v^{(1)}_{b,t-1}  &\text{if } p^{(1)}_{b, t} = p^{(1)}_{b, t-1},\\
    - v^{(1)}_{b,t-1}  &\text{if } p^{(1)}_{b, t} < p^{(1)}_{b, t-1},\\
    \end{cases}
    \quad \text{and} \quad
    aOF^{(1)}_t = \begin{cases}
    -v^{(1)}_{a,t-1} &\text{if } p^{(1)}_{a, t} > p^{(1)}_{a, t-1},\\
    v^{(1)}_{a,t} - v^{(1)}_{a,t-1}  &\text{if } p^{(1)}_{a, t} = p^{(1)}_{a, t-1},\\
    v^{(1)}_{b,t}  &\text{if } p^{(1)}_{a, t} < p^{(1)}_{a, t-1}.\\
    \end{cases}
\end{equation*}
The difference between the two is known as order flow imbalance:
\[ OFI^{(1)}_{t} = bOF^{(1)}_t - aOF^{(1)}_t.\]
Bid order flow corresponds to the net change in volume at the best bid level; the three cases in the definition can be understood as:
\begin{itemize}
    \item a new bid order being placed at a higher price than the current best bid;
    \item the order volume at the best bid price increasing or decreasing;
    \item the entire volume at the best bid price being consumed, thus decreasing the best bid price.
\end{itemize}
Note that, as discussed in Section \ref{sec:order_book}, volumes can increase or decrease due to orders being placed, executed, or canceled. A similar interpretation holds for ask order flow.

In \citet{multilevel_order_flow} the authors explore flows of volumes at deeper levels in the order book, by introducing multi-level order flow. The definitions are exactly as above with superscript $(1)$ replaced by general $(l)$. We note that in both \citet{order_flow} and \citet{multilevel_order_flow} the authors investigate the explanatory power of (multi-level) order flow imbalance for price changes, i.e.\ the relationship between contemporaneous order flow imbalance and price changes. 

In our setting, as in \citet{deepOF}, we are instead interested in exploring the predictive power of order flow, i.e.\ the relationship between past order flow and future price changes. We will hence refer to deepOF as the deepLOB architecture with stationary order flow input, i.e.\ a $(T \times 2L)$ array given by 
\[
\{\mathbf{x}_{t-\tau}\}_{\tau=0,\ldots,T-1} = \left\{ \left(aOF^{(l)}_{t-\tau},bOF^{(l)}_{t-\tau}\right)  \right\}_{\tau=0,\ldots,T-1,\,l=1,\ldots,L} \in \mathbb{R}^{T\times2L}.
\]
The order flow input enters the CNN module in Figure \ref{fig:CNN_module} at the second convolutional layer. The rest of the architecture is exactly the same as deepLOB, as detailed in Table \ref{table:deepLOBdeepOF}. We note that the first convolutions applied to the order flow input, i.e.\ the second convolutional layer in Figure \ref{fig:CNN_module}, aggregate information across bid and ask order flows, essentially computing a weighted order flow imbalance.

\begin{remark}
The original deepOF specification in \citet{deepOF} was structured as a (multi-horizon) regression task. In their setting, the last layer of the deep neural network maps each prediction to $\mathbb{R}$ instead of $\mathcal{P}(\{\downarrow, =, \uparrow \})$. Moreover, as discussed in Section \ref{sec:multihorizon_models}, the way multi-horizon predictions are produced in the original work does not rely on the encoder-decoder structure we use for our multi-horizon models. Another slight difference with the experiments in \citet{deepOF} is in the standardization procedure: instead of standardizing each feature by its training mean and standard deviation, we use a rolling window approach.
\end{remark} 

\subsection{deepVOL} \label{sec:deepVOL}
\subsubsection{The need for a robust representation} \label{sec:robust_VOL}
As discussed in the previous Section the main difference between deepLOB and deepOF is in the way the data is fed into the model. While deepLOB uses raw order book data as input, deepOF uses a derived quantity, order flow. In general, the success of deep learning tasks is highly dependent on the way the data is represented. The task of predicting returns in order book markets is no exception. In order to achieve the best possible results, one should adopt a robust representation of the data. \citet{robust_representation} identify five main desiderata for a robust representation of order book data:
\begin{itemize}
    \item Region of interest: the entire order book may contain a wide range of prices, the data representation should select a region of interest based on a complexity-performance trade-off.
    \item Efficiency: the data representation should avoid excessive dimensionality.
    \item Validity: the data representation should have a simple definition of valid manipulations.
    \item Smoothness: the data representation should be robust to small perturbations.
    \item Compatibility: the data representation should be compatible with the deep learning architecture.
\end{itemize}
We note the order book representations used in deepLOB and deepOF do not conform to these desiderata. Order book states organized by `level' do not have a simple validity (price and volume information are intrinsically tangled and would loose their significance if treated separately in a black box algorithm), are not robust to small perturbation (small orders added at empty ticks completely change the order book feature vector), and are incompatible with deepLOB's CNN module (the spatial structure is not homogeneous as there is no fixed interval between levels). It turns out that while the `level' representation of the order book may be easily understandable by humans it is less so for statistical models. In addition to not satisfying the desiderata, this representation does not respect the following fundamental but implicit assumption of deep learning models: signals at the same entry of the input should come from the same source. In the `level' representation, as new order book events happen, the same signal (i.e.\ a posted order) may move between levels.

\begin{figure}[!htb]
    \centering
    \includegraphics[width = 14cm]{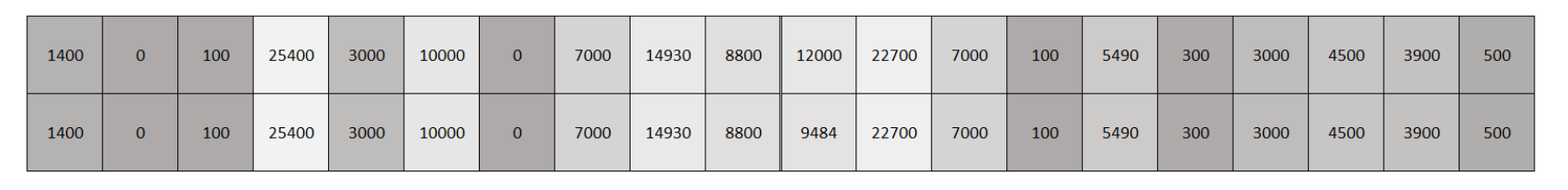}
    \caption{Gray-scale visualization of the volume representation of the order book in Figure \ref{fig:orderbook}.} \label{fig:orderbook_grayscale}
\end{figure}

We, therefore, introduce volume features, which provide a robust representation of order book data. Fixing a window of size $W>0$ we define:
\[
\{\mathbf{x}_{t-\tau}\}_{\tau=0,\ldots,T-1} =
\left\{
\left(s^{(W)}_{b, t-\tau}, \ldots, s^{(1)}_{b, t-\tau}, s^{(1)}_{a, t-\tau}, \ldots, s^{(W)}_{a, t-\tau}\right) 
\right\}_{\substack{\tau=0,\ldots,T-1}},
\]
where $s^{(j)}_{x, t-\tau}$ for $x\in\{a,b\}$ are the bid/ask volumes corresponding to the $j$-th price from the mid $\pi^{(j)}_{x, t-\tau}$, as defined in Section \ref{sec:order_book}. We note that the volume representation indeed satisfies the five desiderata: a region of interest is identified (via the window $W>0$), it is efficient (for the same dimension of input it may convey more or less information than the `level' representation, depending on how sparse the orders are placed in the order book), it has a simple validity (all entries are in the same units), it is robust to small perturbations (new orders at empty levels minimally affect the feature) and is compatible with the CNN architecture (the spatial structure of the volumes is homogeneous). An intuitive visualization of this representation as a one-dimensional gray-scale strip is given in Figure \ref{fig:orderbook_grayscale}, when including a time dimension this naturally becomes a two-dimensional gray-scale image. The main drawback of the volume representation is that when orders are placed far apart in the order book it is sparse and a larger window $W>0$ may be required. Our definition of volume features is similar to the mid-price-centered moving window representation of \citet{robust_representation}, with the latter living in $\mathbb{R}^{2W+1}$ instead of $\mathbb{R}^{2W}$ and using $\pm$ signs to distinguish between bid and ask volumes. The need for a new, more robust representation of the order book was reached independently.

\begin{figure}[!htb]
    \centering
    \includegraphics[height = 5.75cm]{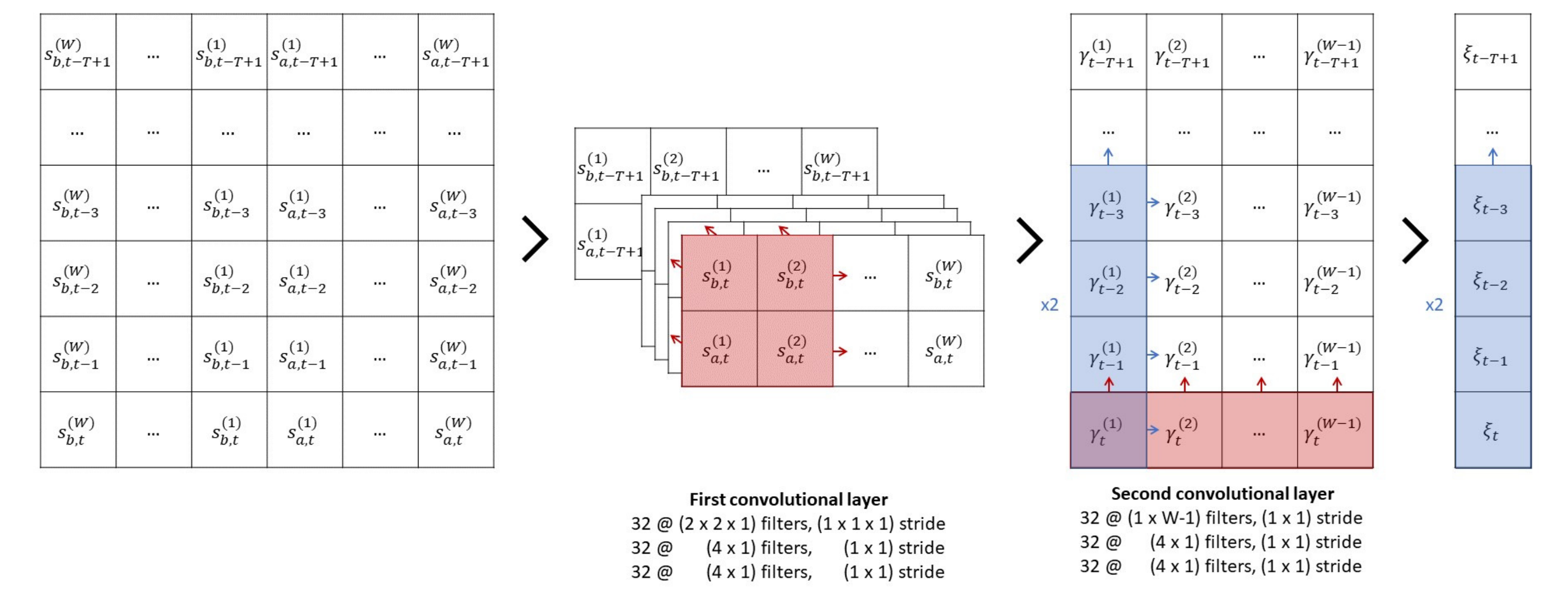}
    \caption{CNN module deepVOL.} \label{fig:CNN_module_deepVOL}
\end{figure}

In order to use the deepLOB architecture with this new data representation, we must adapt the CNN module. In particular, we fold the volume input into a three-dimensional array
\[ 
\left\{ s^{(j)}_{x, t-\tau} \right\}_{\tau=0,\ldots,T-1,\, j= 1,\ldots, W,\, x\in\{ a, b\}} \in \mathbb{R}^{T\times W \times 2},
\]
and feed this into a three-dimensional convolutional layer with a  $(2 \times 2 \times 1)$ filter and $(1 \times 1 \times 1)$ stride. This layer aims to extract imbalances in the order book by comparing volumes on the two sides of the mid-price. The CNN module with the appropriate changes is depicted in Figure \ref{fig:CNN_module_deepVOL}. The rest of the deepVOL architecture is exactly the same as deepLOB with one slight difference in the way the data is normalized. Thinking of the volume representation as a gray-scale image a natural choice of normalization is
\[ \left\{ s^{(j)}_{x, t-\tau} \right\}_{\tau,j, x} \mapsto \frac{1}{\underset{\tau, j, x}{\max}\ s^{(j)}_{x, t-\tau} } \left\{s^{(j)}_{x, t-\tau}\right\}_{\tau, j, x}, \] 
instead of the rolling window standardization we apply to deepLOB and deepOF features.

\begin{remark}
    One could define a volume flow quantity based on the distance from the mid, in the same way order flow describes the flow of orders based on the level. Following the same motivation for considering order flow in deepOF presented in Section \ref{sec:deepOF}, one could consider using a volume flow quantity as input to the deep learning architectures with the desired robust representation properties.
\end{remark}

\subsubsection{L3 volume features} \label{sec:deepVOL_L3}
All models considered so far, deepLOB, deepOF, and deepVOL use L2 data as input. If one has access to more granular data, i.e.\ L3 data breaking down each volume queue into single orders, one might be interested in trying to leverage this information to obtain higher predictive performance. We thus define a natural extension of the volume representation considered in the previous section.

Let us denote by $(q_{x,t}^{(j,1)}, q_{x,t}^{(j,2)}, \ldots) \in \mathbb{R}^\mathbb{N}$ for $x\in\{a,b\}$ the queue at the $j$-th bid/ask price from the mid $\pi^{(j)}_{x,t}$ as introduced in Section \ref{sec:order_book}. Here $q_{x,t}^{(j,k)}$ denotes the volume of the $k$-th order in the queue sorted by time priority and is set to zero if there is no such order. The aggregated volume at $\pi^{(j)}_{x,t}$ is given by:
\[ s^{(j)}_{x,t} = \sum_{k\geq 1} q_{x,t}^{(j,k)}.\]
A natural extension of the volume representation would therefore be to consider 
\[
\{ (q_{x,t-\tau}^{(j,1)}, q_{x,t-\tau}^{(j,2)}, \ldots) \}_{ \tau=0,\ldots,T-1,\, j=1,\ldots,W,\, x\in\{ a, b\}} \in \mathbb{R}^{T \times W \times 2 \times \mathbb{N}}.
\] 
Unfortunately, this is an infinite-dimensional array that cannot be directly fed into machine learning models. We therefore cut off the queue at a given depth level. In order to avoid discarding precious information, we aggregate all orders sitting past the maximum depth level at the end of the queue. For a given depth level $D>0$, we thus consider the L3 volume feature
\[
\left\{ \left(q_{x,t-\tau}^{(j,1)}, \ldots, q_{x,t-\tau}^{(j, D-1)}, \sum_{k\geq D} q_{x,t-\tau}^{(j, k)} \right) \right\}_{\tau=0,\ldots,T-1,\, j=1, \ldots, W,\, x\in\{ a, b\}} \in \mathbb{R}^{T \times W \times 2 \times D}.
\] 

\begin{remark}
    A similar approach to the one used for cutting off the queue might be applied to aggregate volumes sitting deeper in the order book when deriving L2 representations. This would give a better idea of the total liquidity in the order book.
\end{remark}

\begin{figure}[!htb]
    \centering
    \includegraphics[height = 4.5cm]{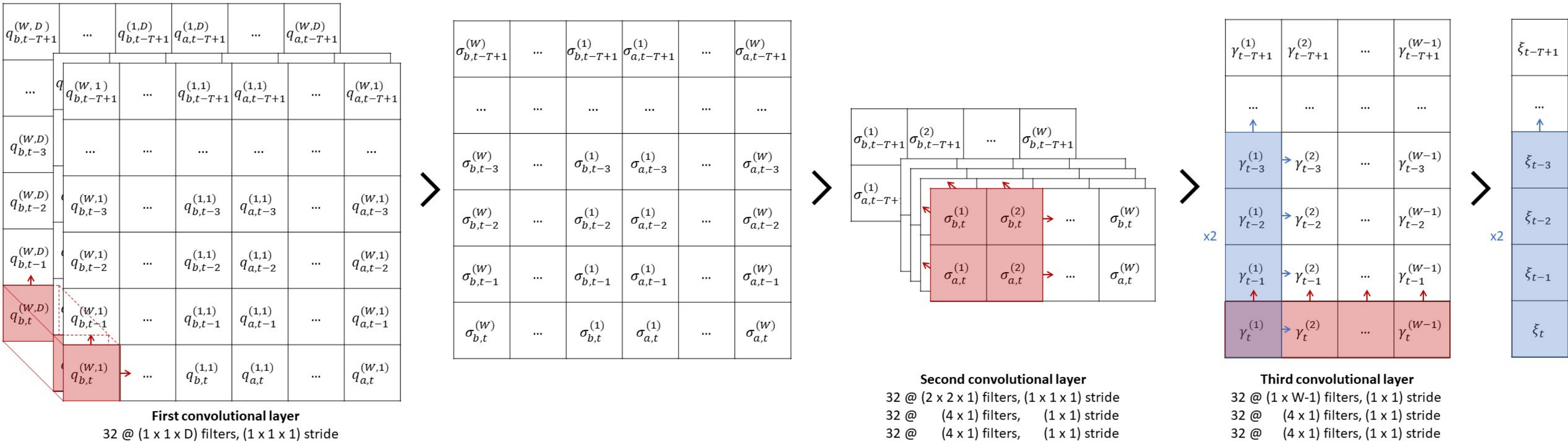}
    \caption{CNN module deepVOL L3. Abusing notation slightly we denote $\sum_{k\geq D} q_{x,t-\tau}^{(j, k)}$ by $q_{x,t-\tau}^{(j, D)}$.} \label{fig:CNN_module_deepVOL_L3}
\end{figure}

In order to extract relevant information from the queue, we add an initial convolutional layer, which maps each queue to a weighted sum of the order sizes. The weighted aggregated volumes are then fed into the deepVOL architecture as above. The resulting CNN module is summarized in Figure \ref{fig:CNN_module_deepVOL_L3}.

The full architectures for deepVOL and deepVOL L3 are detailed in Table \ref{table:deepVOL}.

\subsection{Multi-horizon models} \label{sec:multihorizon_models}
So far we considered single horizon models, i.e.\ order book input ${x}_{t}$ is mapped to a three class distribution $(p_{\downarrow}, p_{=}, p_{\uparrow})$ corresponding to the discretized return $c_{t,t+h}$ at fixed horizon $h$. In the following, we consider a generalization of the architectures considered thus far to the setting where the forecasting horizon is a vector $\mathbf{h} = (h_1, \ldots, h_K)$. In this case the modelling task consists in predicting the distributions of the discretized returns $c_{t, t+\mathbf{h}} = (c_{t,t+h_1}, \ldots, c_{t,t+h_K})$.

The simplest way of adjusting the current models to the multi-horizon framework would be to replace the last softmax layer in Figure \ref{fig:Inception_LSTM} with $K$ parallel dense softmax layers. The last hidden state of the LSTM module would hence be mapped to an array of size $3 \times K$, corresponding to $K$ distributions over three classes. A similar architecture -- though in the regression framework -- is considered in \citet{deepOF}. While this approach is perfectly valid it does not make use of the sequential nature of the task, potentially neglecting an important structural feature of the data.

\begin{figure}[!htb]
    \centering
    \includegraphics[height = 6cm]{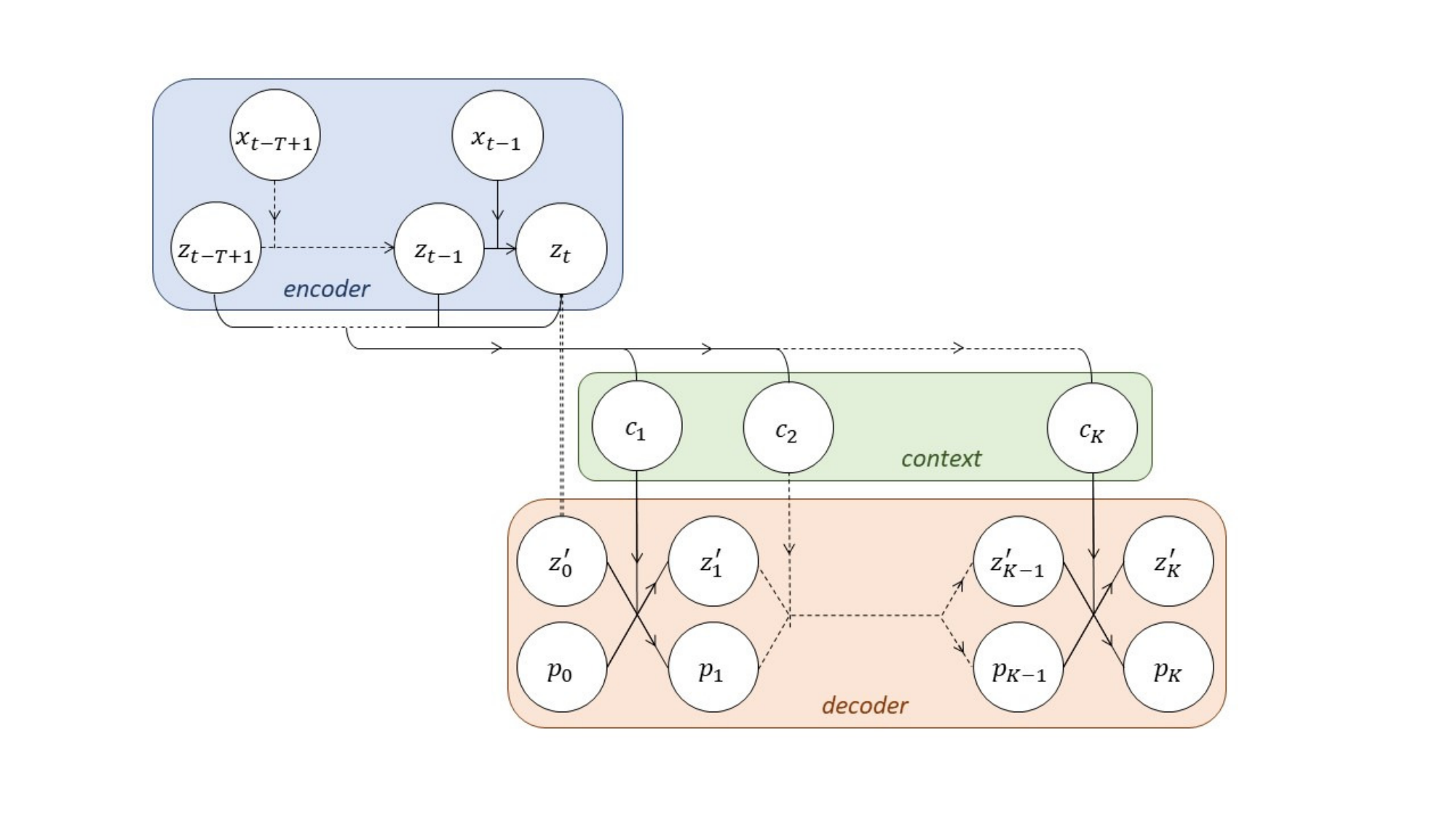}
    \caption{A general encoder-decoder architecture.} \label{fig:encoder_decoder}
\end{figure}

In this paper, we will be leveraging architectures inspired by machine translation which are naturally suited for sequential forecasting. Specifically, we will be considering encoder-decoder models: an encoder maps the input data to a latent summary state (also known as context vector) and then a decoder rolls forward predictions sequentially. In this context let ${z}_{t-T+1}, \ldots, {z}_t$ denote the last $T$ hidden states of an encoder at time $t$, then a decoder rolls forward predictions by:
\begin{align*} {z}'_0 &= {z}_t, {p}_{0} = {\hat{p}}_0, \\
	{c}_k & = h({z}_{t-T+1}, \ldots, {z}_t),\ {z}'_{k} = f([{z}'_{k-1},{c}_k], {p}_{k-1}),\ {p}_{k} = g({z}'_{k-1},{c}_k),
\end{align*}
for $k = 1,\ldots, K$. Here $h(\cdot)$ is a function acting on the hidden states of the encoder to extract the context vector ${c}_k$ (this may possibly depend on other inputs as well, such as previous hidden states of the decoder), $f({r}, {p})$ is a recurrent layer with recurrent input ${r}$ and exogenous input ${p}$, $g(\cdot)$ is an output layer depending on both the decoder hidden state and the context. The general mechanism of such encoder-decoder architecture is visualized in Figure \ref{fig:encoder_decoder}.

In our experiments, we will consider a sequence-to-sequence decoder \citep{seq2seq_decoder}, the simplest example of such architecture. In this setting we set $ {c}_k \equiv {z}_t$, implicitly assuming that at every forecasting horizon, the last hidden state of the encoder summarizes all the relevant information required to make the prediction. More complex architectures exist, for example, \citet{attention_decoder} introduce an attention-based decoder that uses a weighted combination of all the hidden states of the encoder as context vector. In this setting different weights are used at different forecasting horizons, selectively accessing hidden states of the encoder during decoding. While attention-based networks have been successfully applied for high-frequency mid-price predictions, cf.\ \cite{deepLOB_multihorizon} and \cite{attention_financial_timeseries}, in this work, we wish to exemplify the potential of multi-horizon models and thus restrict ourselves to simple seq2seq decoders only.

In the experiments, we set $f(\cdot)$ to be an LSTM and $g(\cdot)$ to be a dense layer with softmax activation. Moreover ${p}_0$ is initialized at ${\hat{p}}_0 = (0, 1, 0)$.

\begin{figure}[!htb]
    \centering
    \includegraphics[height = 11cm]{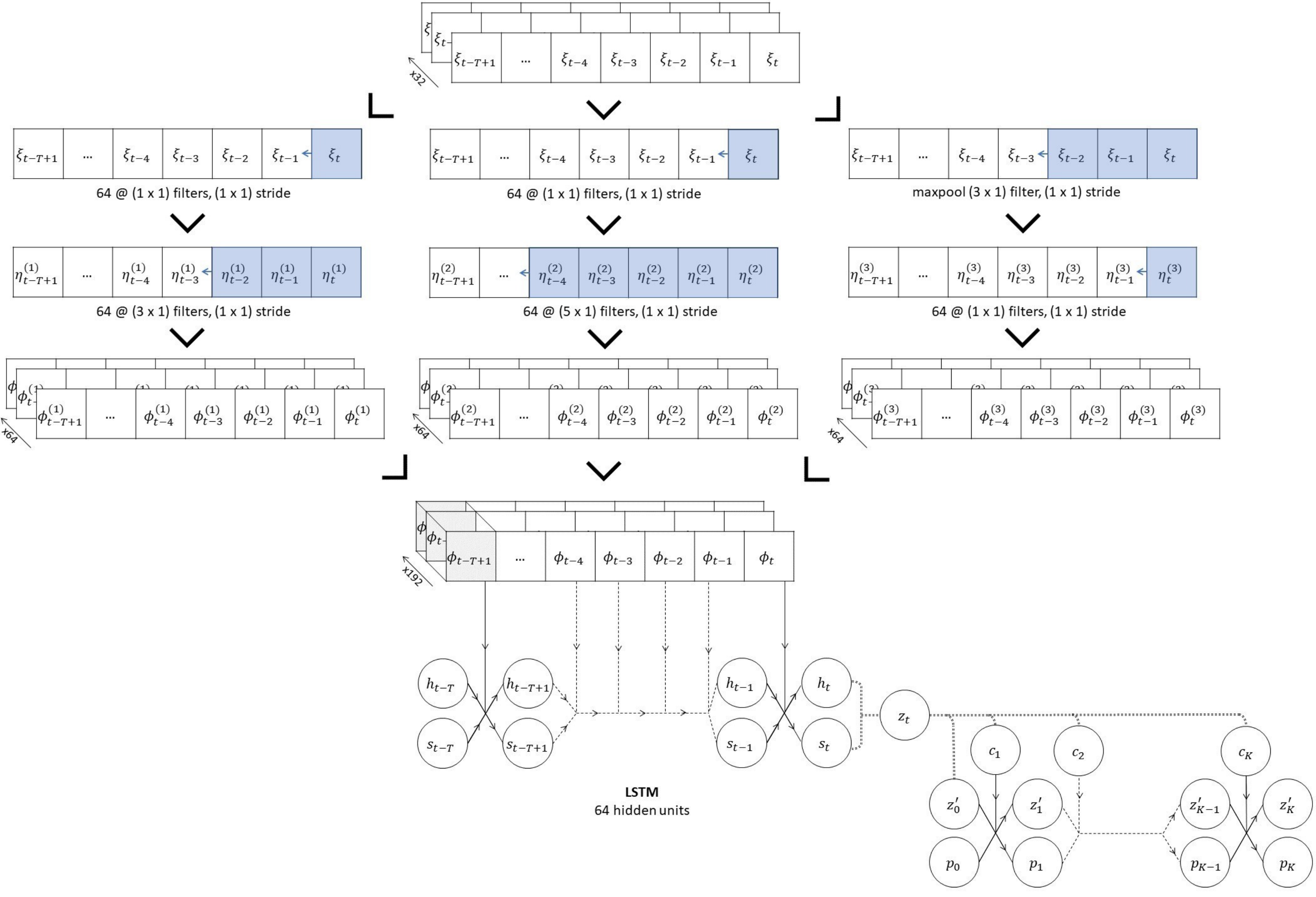}
    \caption{Inception module and LSTM layer with seq2seq decoder for multi-horizon forecasting.} \label{fig:inception_LSTM_seq2seq}
\end{figure}

Using multi-horizon forecasting was first proposed for deepLOB in \citet{deepLOB_multihorizon}: the LSTM module of the deepLOB architecture discussed in Section \ref{sec:deepLOB} acts as an encoder, mapping order book states to the LSTM final latent vector ${z}_t = [h_t, s_t]$. A seq2seq decoder then rolls forward the prediction producing distributional forecasts at horizons $\mathbf{h} = (h_1, \ldots, h_K)$. The LSTM module with seq2seq decoder is illustrated in Figure \ref{fig:inception_LSTM_seq2seq}. Clearly, there is nothing stopping us from applying the same multi-horizon structure to the output of deepOF and deepVOL convolutional modules.

Full details of the multi-horizon architectures can be found in Table \ref{table:deepLOBdeepOF} and Table \ref{table:deepVOL}.

\begin{remark}
    We adopt the same multi-horizon framework as in \citet{deepLOB_multihorizon}, where the response $c_{t,t+h_k}$ at horizon $h_k \in \{h_1, \ldots, h_K\}$ corresponds to the return from time $t$ to time $t+h_k$. Alternatively, one might wish to consider as multi-horizon responses the subsequent incremental returns $c_{t+h_{k-1}, t+h_k}$, i.e.\ the returns between time $t+h_{k-1}$ and time $t+h_k$. By choosing evenly spaced $h_k$’s one would obtain more consistent responses across prediction time steps $\{1,\ldots, K\}$ and possibly improve the performance of the multi-horizon models.
\end{remark}

\section{Data set} \label{sec:dataset}
In this section, we introduce the data set used in the experiments presented in Section \ref{sec:experiments}. First, we briefly describe how LOBSTER order book data is related to the Nasdaq ITCH feed. Next, we give details of how we process LOBSTER data to obtain features and responses for the experiments. We also provide some descriptive statistics of the data.

We consider the same universe of stocks and trading period as in \citet{deepOF}: through LOBSTER \citep{lobster} we access one year of open (9:30 EST) to close (16:00 EST) trading data for 115 Nasdaq tickers from January 2, 2019, to January 31, 2020. To produce results in feasible computational time, we select a subset of 10 stocks, trying to preserve a sufficiently varied set of liquidity characteristics. The ten tickers and their liquidity characteristics are summarized in Table \ref{table:stock_characteristics}. For full details on how the 10 stocks were selected, see Appendix \ref{app:stock_selection}.

\vspace{1cm}
\begin{table}[htb]
\small
\centerline{
    \begin{tabular}{lrrrrrr}
    \toprule
    \textbf{Ticker} & \multicolumn{1}{l}{\textbf{Updates (000)}} & \multicolumn{1}{l}{\textbf{Trades (000)}} & \multicolumn{1}{l}{\textbf{Price Changes (000)}} & \multicolumn{1}{l}{\textbf{Price (USD)}} & \multicolumn{1}{l}{\textbf{Spread (bps)}} & \multicolumn{1}{l}{\textbf{Volume (USD MM)}} \\
    \toprule
    LILAK & 49.89 & 1.81  & 3.77  & 18.09 & 15.92 & 2.63 \\
    \midrule
    QRTEA & 121.32 & 5.23  & 2.79  & 13.54 & 9.75  & 9.53 \\
    \midrule
    XRAY  & 83.07 & 5.41  & 7.11  & 52.32 & 4.35  & 21.67 \\
    \midrule
    CHTR  & 80.97 & 6.75  & 19.21 & 400.68 & 6.22  & 111.71 \\
    \midrule
    PCAR  & 131.38 & 6.85  & 13.03 & 70.63 & 3.82  & 34.43 \\
    \midrule
    EXC   & 298.32 & 7.58  & 7.21  & 47.35 & 2.45  & 37.55 \\
    \midrule
    AAL   & 398.13 & 13.15 & 10.99 & 30.65 & 3.88  & 50.29 \\
    \midrule
    WBA   & 328.5 & 13.2  & 15.18 & 57.81 & 2.54  & 75.19 \\
    \midrule
    ATVI  & 423.01 & 17.97 & 20.25 & 49.87 & 2.81  & 91.82 \\
    \midrule
    AAPL  & 1137.36 & 64.19 & 127.86 & 215.8 & 0.99  & 1156.3 \\
    \bottomrule
    \end{tabular}}
    \caption{Selected stocks' characteristics, daily averages.} \label{table:stock_characteristics}
\end{table}

\subsection{LOBSTER data, \texorpdfstring{\citet{lobster}}{lobster}} \label{sec:dataset_LOBSTER}
In Section \ref{sec:order_book} we described the basic mechanisms governing electronic exchanges such as the Nasdaq. Every day, trading activity on the Nasdaq alone results in hundreds of thousands of order book updates for each stock, cf.\ Table \ref{table:stock_characteristics}. To keep track of all events occurring on the exchange and communicate them efficiently to (subscribed) market participants the Nasdaq uses the TotalView-ITCH protocol. For efficiency, instead of streaming the entire state of the order book after each update, only information on the event changing the order book is sent out to market participants. An example of a (decoded) ITCH message is reported in Table \ref{table:message}, note these are timestamped at nanosecond precision. It is up to each market participant to store the current state of the order book and update it each time a new message arrives.

\begin{table}[htb]
\small
\centerline{
    \begin{tabular}{ccccccc}
    \toprule
    \textbf{Event ID} & \textbf{Timestamp (sec)} & \textbf{Event Type} & \textbf{Order ID} & \textbf{Size (shares)} & \textbf{Price} & \textbf{Direction} \\ 
    \midrule
    \vdots & \vdots & \vdots & \vdots & \vdots & \vdots & \vdots \\
    1312 & 34714.133632201 & deletion & 206833312 & 2516 & \$11.86 & sell \\
    \vdots & \vdots & \vdots & \vdots & \vdots & \vdots \\
    \midrule
    \end{tabular}}
  \caption{Example message from a LOBSTER message file. This example corresponds to the order book dynamics shown in Figure \ref{fig:orderbook}.} \label{table:message}
\end{table}

The service LOBSTER provides to academic researchers is to reconstruct the historic order book from Nasdaq's Historical TotalView-ITCH data. For each selected ticker and date, LOBSTER returns message and order book files subsampled at a given granularity. In practical terms, a $10$-level data granularity yields the set of messages corresponding to order book updates in the first 10 levels along with the corresponding reconstructed order book `chopped at 10 levels'. The evolution of the order book determined by the message in Table \ref{table:message} is reported in Table \ref{table:orderbook}. More information on the order book reconstruction algorithm used by LOBSTER can be found in \citet{lobster} and on the website \url{www.lobsterdata.com}.

\begin{table}[htb]
\small
\centerline{
    \begin{tabular}{ccccccccccc}
    \toprule
    \textbf{Event ID} & \textbf{Timestamp (sec)} & $p^{(1)}_a$ & $v^{(1)}_a$ & $p^{(1)}_b$ & $v^{(1)}_b$ & $\cdots$ & $p^{(10)}_a$ & $v^{(10)}_a$ & $p^{(10)}_b$ & $v^{(10)}_b$ \\ 
    \midrule
    \vdots & \vdots & \vdots & \vdots & \vdots & \vdots & $\cdots$ & \vdots & \vdots & \vdots & \vdots\\
    1311 & 34713.685155243 & \$11.86 & 12000 & \$11.85 & 8800 & $\cdots$ & \$11.95 & 500 & \$11.73 & 5500 \\
    1312 & 34714.133632201 & \$11.86 & 9484 & \$11.85 & 8800 & $\cdots$ & \$11.95 & 500 & \$11.73 & 5500 \\
    \vdots & \vdots & \vdots & \vdots & \vdots & \vdots & $\cdots$ & \vdots & \vdots & \vdots & \vdots\\
    \midrule
    \end{tabular}}
  \caption{Example order book update from a LOBSTER order book file. This example corresponds to the order book dynamics shown in Figure \ref{fig:orderbook}.} \label{table:orderbook}
\end{table}

\subsection{Processed data} \label{sec:dataset_processed}
From the historic LOBSTER order book data we build the features $\mathbf{x}_t,\ldots, \mathbf{x}_{t-T+1}$ and the target responses $c_{t,t+h}$. In the experiments in Section \ref{sec:experiments} we apply the learning framework of Section \ref{sec:predictability_returns} with the deep learning models introduced in Section \ref{sec:DL_models} to these feature-response pairs, using the model confidence set (MCS) procedure for model comparison. As previously discussed in Section \ref{sec:order_book} we will be measuring time $t$ (and prediction horizons $h$) using an order book-driven clock, ticking every time an event occurs on the order book. This clock corresponds to the Event ID in Table \ref{table:message} and Table \ref{table:orderbook}. Note we access LOBSTER data up to level 10, thus our order book clock is conditional on the updating event being in the first 10 levels. By construction, the data contains all information on new limit orders, market orders, and cancellations restricted to the first 10 levels. We apply some minor pre-processing steps to the LOBSTER data, summarized in Appendix \ref{app:data_proc}.

\subsubsection{Features: order book, order flow and volume} \label{sec:dataset_features}
We start by discussing how to derive the features $\mathbf{x}_t$ at each time point $t$ from the raw LOBSTER data, i.e.\ from Table \ref{table:message} and Table \ref{table:orderbook}. While it is quite simple to build L1/L2 order book, order flow and volume features, reconstructing L3 volume data requires a bit more work.

The raw order book input $\mathbf{x}_t$ used in deepLOB \citep{deepLOB} and described is Section \ref{sec:deepLOB} simply corresponds to the LOBSTER data in Table \ref{table:orderbook}, i.e.\ at $t=1312$ the 10-level order book feature is given by
\begin{align*}
\mathbf{x}_{t} &= \left(p^{(1)}_{a,t}, v^{(1)}_{a,t}, p^{(1)}_{b,t}, v^{(1)}_{b,t}, \ldots, p^{(10)}_{b,t}, v^{(10)}_{b,t}\right) = (11.86, 9484, 11.85, 8800, \ldots, 11.73, 5500).
\end{align*}
The features are standardized using a 5-day rolling window.

To compute the order flow input $\mathbf{x}_t$ used in deepOF \citep{deepOF} we apply the equations given in Section \ref{sec:deepOF} to the LOBSTER data in Table \ref{table:orderbook}, i.e.\ at $t=1312$ the 10-level order flow feature is given by
\begin{align*}
\mathbf{x}_{t} &= \left(aOF^{(1)}_{t}, bOF^{(1)}_{t}, \ldots, bOF^{(10)}_{t}\right) = (-2516, 0, \ldots, 0).
\end{align*}
Again, the features are standardized using a 5-day rolling window.

To construct the volume input $\mathbf{x}_t$ at L2 granularity for the deepVOL model described in Section \ref{sec:deepVOL}, we select only the volume information from Table \ref{table:orderbook} adding in zeros corresponding to empty price ticks, i.e.\ at $t=1312$ the volume feature with window size $W=10$ is given by
\begin{align*}
\mathbf{x}_{t} &= \left(s^{(10)}_{b,t}, \ldots, s^{(1)}_{b,t}, s^{(1)}_{a,t}, \ldots, s^{(10)}_{a,t}\right) = (1400, \ldots, 8800, 9484, \ldots, 500).
\end{align*}
Note that, in this example, $s^{(10)}_{b,t} = 1400 \neq 5500 = v^{(10)}_{b,t}$ since some bid price ticks are empty, cf.\ Figure \ref{fig:orderbook}. As discussed in Section \ref{sec:deepVOL}, volume features are normalized using max-scaling over the whole input array $(\mathbf{x}_t,\ldots, \mathbf{x}_{t-T+1})$.

Finally, to construct the L3 volume features, one needs to work with the message file in Table \ref{table:message} to keep track of the queues. For example, the volume queue at the first ask price, i.e.\ $\pi_{a, t}^{(1)} = \$11.86$, given at time $t=1311$ by
\[ \left(q_{a,t}^{(1, 1)}, q_{a,t}^{(1, 2)}, q_{a,t}^{(1, 3)}, q_{a,t}^{(1, 4)}, q_{a,t}^{(1, 5)}\right) = (2516, 2000, 1484, 4500, 1500), \]
is updated at time $t=1312$ to
\[ \left( q_{a,t}^{(1, 1)}, q_{a,t}^{(1, 2)}, q_{a,t}^{(1, 3)}, q_{a,t}^{(1, 4)} \right) = (2000, 1484, 4500, 1500), \]
by the deletion in Table \ref{table:message}. All other queues are left unchanged. For more details on the complexities of reconstructing volume features from LOBSTER data see Appendix \ref{app:data_proc_vol}. A deep dive into the distributions of the processed order book, order flow, and volume features is carried out in Appendix \ref{app:descriptive_stats_features}.

\subsubsection{Responses: categorical mid-price returns} \label{sec:responses}
In this paper, we are interested in answering questions regarding the predictability of market returns. Inevitably, the way returns, i.e.\ the target responses, are defined has a profound effect on this analysis. Here, in line with the related literature, we treat the mid-price as the ``true'' price and define returns relative to it, but it is important to note that, by definition, this is not a tradable price. We define the return at horizon $h$ as
\begin{align*}
    r_{t,t+h} = \frac{\overline{m}_{t+h}^{(k)} - m_t}{m_t},\
    \overline{m}_{t+h}^{(k)} = \frac{1}{2k+1}\sum_{i=-k}^{k} m_{t+h+i},
    \vspace{-0.4cm}
\end{align*}
where $m_t$ denotes the mid-price at time $t$ and $k$ is a fixed smoothing window. The mid-price $m_t$ is computed from Table \ref{table:orderbook} by $\dfrac{1}{2}\left(p^{(1)}_{b, t} + p^{(1)}_{a, t}\right)$, i.e.\ at times $t=1311$ and $t=1312$ we have $m_t=\$11.855$. This definition is subject to two possible interpretations. Treating the smoothed mid-price as a de-noised estimate of the true (latent) price $h$ steps ahead, we can understand the return as the percentage change of the true (latent) price relative to the current mid. Alternatively, the return can be understood as the average return one would experience by entering a position at the current mid and exiting it roughly $h$ steps ahead (assuming mid-mid trading). For all the horizons $h\in\{10, 20, 30, 50, 100, 200, 300, 500, 1000\}$ which are considered in Section \ref{sec:experiments} we fix $k=5$. In Appendix \ref{app:return_def} we discuss other methods for defining mid-price returns and their shortcomings.

\begin{remark}
    When defining the returns we assume immediate access to the order book. In practice though, hardware and software constraints lead to non-zero time lags when receiving messages and sending orders to the exchange. While, in our setting, such latencies have a negligible impact on the definition of returns, we discuss how their presence could be more precisely accounted for in Appendix \ref{app:latencies}.
\end{remark}

All our experiments are carried out in a classification framework where the discretized returns are defined as
\[ c_{t,t+h} = 
    \begin{cases}
    \downarrow &\text{if } r_{t,t+h} \in (- \infty, - \gamma), \\
    = &\text{if } r_{t,t+h} \in [-\gamma, +\gamma], \\
    \uparrow &\text{if } r_{t,t+h} \in (+\gamma, +\infty),
    \end{cases}
\]
for some $\gamma>0$. In order to make the three classes roughly symmetric and as balanced\footnote{Note that, especially for short horizons $h$, many of the returns $r_{t,t+h}$ can be exactly zero, leading to $\hat{Q}_{h}(0.33)=\hat{Q}_{h}(0.66)=0$. When this happens the three classes are unbalanced.} as possible, we empirically choose $\gamma$ from the training set $\mathcal{D}_{\text{train}}$ by
\[\hat{\gamma}_{h} = \frac{\lvert\hat{Q}_{h}(0.33)\rvert + \hat{Q}_{h}(0.66)}{2},\]
where $\hat{Q}_{h}$ is the empirical quantile function of the training set returns $\{r_{t,t+h}\}_{t \in \mathcal{I}_{\text{train}}}$. As discussed in Section \ref{sec:experiments}, we will be splitting our data $\mathcal{D}$ in disjoint windows $\mathcal{D}_w = \mathcal{D}_{w, \text{train}} \cup \mathcal{D}_{w, \text{val}} \cup \mathcal{D}_{w, \text{test}}$ for $w=1,\ldots,W$. The choice of $\gamma$ will, therefore, be window $w$ and horizon $h$ specific. Descriptive statistics of the target return labels for the first window $w=1$ are reported in Appendix \ref{app:descriptive_stats_responses}.


We note that, since stocks trade on a discrete grid of prices determined by the tick size\footnote{$\vartheta=\$0.01$ for many US listed stocks and all the stocks considered in this paper.} $\vartheta$, also the mid-price $m_t$ evolves on a discrete price (with steps of size $\vartheta/2$). One could thus define the dollar return from $t$ to $t+h$ by the number of (half) ticks the mid-price moves, i.e.\ 
\[R_{t,t+h} = m_{t+h} -m_t \in\{\ldots,-\vartheta,-\vartheta/2, 0, \vartheta/2, \vartheta, \ldots\} \equiv \{\ldots, -2, -1, 0, +1, +2, \ldots\}.\]
This return is by definition discretized, and thus one could directly apply classification models (grouping large negative and positive returns to obtain a finite number of classes). A similar approach is used in \citet{universalLOB} when predicting the next change in mid-price. In our work, we consider the estimate for the ``true'' mid-price to be 
\[\overline{m}_{t+h}^{(k)} = \frac{1}{2k+1}\sum_{i=-k}^{k} m_{t+h+i},\]
which lives on a much finer grid than $m_{t+h}$: over the, possibly quite long, time horizon $h$ multiple changes to the mid-price might occur. In this case the smallest change has little meaning and so we group the dollar returns \[R_{t,t+h} = \overline{m}_{t+h}^{(k)} - m_t \in\left\{\ldots,-\frac{\vartheta}{2k+1},-\frac{\vartheta}{4k+2}, 0, \frac{\vartheta}{4k+2}, \frac{\vartheta}{2k+1}, \ldots\right\}, \]
into larger classes given by 
\[(-\infty, -m_t\gamma), [-m_t\gamma, +m_t\gamma], (+m_t\gamma, +\infty).\]

\section{Experiments} \label{sec:experiments}
\begin{figure}[!htb]
    \centering
    \includegraphics[width = 14cm]{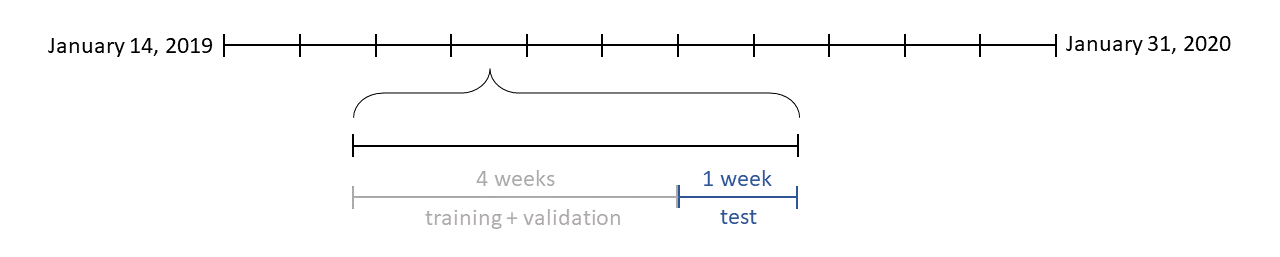}
    \caption{Data set windowing for experiments.} \label{fig:rolling_window}
\end{figure}

In this section, we explore the four questions introduced in Section \ref{sec:questions} and attempt to answer them via the statistical framework provided by model confidence sets (MCS). This inference procedure allows to compare a set of competing models $\mathcal{M}_0$ based on observed loss time series $\{L_{i,w}\}_{w=1}^W$ where $L_{i,w}$ denotes the loss of model $i\in\mathcal{M}_0$ at time $w\in W$. To apply the model confidence set procedure to the data described in Section \ref{sec:dataset}, we divide the 55 weeks from January 14, 2019 to January 31, 2020 into $W=11$ five-week periods\footnote{Data from the week starting on January 7, 2019, is used to start the rolling window standardization required for deepLOB and deepOF architectures}, as represented in Figure \ref{fig:rolling_window}. Each window of data $\mathcal{D}_w$ is divided into a training-validation set $\mathcal{D}_{w,\text{train}}\cup \mathcal{D}_{w,\text{val}}$, the first four weeks, and a test set $\mathcal{D}_{w,\text{test}}$, the fifth week. The joint training-validation set is then further split into training and validation sets by randomly selecting 5 days out of the four weeks for validation. First, the training dataset $\mathcal{D}_{w,\text{train}}$ is used to choose the $\gamma$ threshold for defining the return labels, as detailed in Section \ref{sec:responses} (note that the choice of $\gamma$ is specific to the choice of window, horizon, and ticker). Then, the joint training-validation dataset $\mathcal{D}_{w,\text{train}}\cup \mathcal{D}_{w,\text{val}}$ is used to train the models. For the \textit{unpredictive benchmark} model, this simply means determining the empirical distribution of returns in $\mathcal{D}_{w,\text{train}}\cup \mathcal{D}_{w,\text{val}}$, cf.\ Appendix \ref{app:descriptive_stats_responses}. For the deep neural network architectures, this amounts to finding the optimal parameters which minimize the training weighted cross-entropy loss. To do so, we use Adam optimization with validation-based early stopping as described in Appendix \ref{app:deep_learning}. Once the model has been trained we compute the out-of-sample losses on $\mathcal{D}_{w,\text{test}}$, i.e.\ for period $w\in\{1,\ldots,11\}$ and model $i\in\mathcal{M}_0$
\[ L_{i,w} = - \text{cce}(\hat{\mathbf{p}}^i_{w, \text{test}}, \mathbf{c}_{w, \text{test}})  = - \frac{1}{|\mathcal{I}_{w, \text{test}}|} \sum_{t\in \mathcal{I}_{w, \text{test}}} \sum_{*\in\{\downarrow, =, \uparrow\}} \mathds{1}_{\{c_{t, w, \text{test}} = * \}} \log \hat{p}^i_{*, t}, \]
is the categorical cross-entropy loss corresponding to the estimated probabilities $\hat{\mathbf{p}}^i_{w, \text{test}}$ when compared with the observations $\mathbf{c}_{w, \text{test}}$. $\hat{\mathbf{p}}^i_{w, \text{test}}$ are the class probabilities produced by model $i$ on the testing set $\mathcal{D}_{w, \text{test}}$ after being trained on the training-validation set $\mathcal{D}_{w,\text{train}}\cup\mathcal{D}_{w,\text{val}}$. The time series of test losses $\{L_{i,w}\}_{w=1}^{11}$ are then ``fed through'' the MCS procedure described in Appendix \ref{app:MCS} to obtain the set of MCS p-values $\{p^{\text{MCS}}_i\}_{i\in\mathcal{M}_0}$. The intuitive interpretation of these p-values is: if model $i\in\mathcal{M}_0$ has an MCS p-value lower than a prescribed confidence level then it is deemed statistically inferior to other models in $\mathcal{M}_0$ at that confidence level. This naturally justifies the following definition of order book-driven predictability.
\begin{definition} \label{def:predictability}
    For $\alpha\in(0,1)$ we say that there is order book-driven predictability at confidence level $1-\alpha$ if $p_{\text{benchmark}}^{\text{MCS}}<\alpha$. 
\end{definition}
In other words, if the \textit{unpredictive benchmark} is deemed to be statistically inferior to some of the other models in $\mathcal{M}_0$ at least one order book-driven model does better than the \textit{unpredictive benchmark}, i.e.\ such model is predictive.

One of the strengths of the way we apply the model confidence set procedure in this context is that it allows to fully account for parameter estimation error, while classic methods for comparing deep learning models do not. To make this statement precise let us make explicit the dependence of the test loss
\[ L_{i,w} = -\text{cce}(\mathbf{p}^i(\mathbf{X}_{w,\text{test}}; \hat\theta_i(\mathcal{D}_{w,\text{train}}\cup\mathcal{D}_{w,\text{val}}, \epsilon_{i,w})), \mathbf{c}_{w,\text{test}}),\]
on the training-validation data $\mathcal{D}_{w,\text{train}}\cup\mathcal{D}_{w,\text{val}}$, the fitting procedure initialization seed $\epsilon_{i,w}$ and the testing data $\mathcal{D}_{w,\text{test}} = \{\mathbf{X}_{w,\text{test}}, \mathbf{c}_{w,\text{test}}\}$. In the related literature, deep learning models are often compared by fitting the model to the same data $\mathcal{D}_{\text{train}} \cup \mathcal{D}_{\text{val}}$ starting with different seeds and reporting the out-of-sample average loss and standard error, This accounts only for the uncertainty in $\epsilon$ and not in the sampled training and testing data sets (cf.\ Remark \ref{rem:uncertainty_errors}). Our approach instead naturally considers overall parameter estimation uncertainty, comprised of both statistical and optimization errors, by comparing model losses as functions of the random processes $\mathcal{D}_{w,\text{train}}\cup\mathcal{D}_{w,\text{val}}$, $\epsilon_{w, i}$ and $\mathcal{D}_{w,\text{test}}$. This provides a more robust comparison of model performance, i.e.\ specification error. 

\begin{remark} \label{rem:hyperparameter_tuning}
Note that not all training procedures may converge to the optimal combination of parameters. This is a characteristic of any model learned via numerical optimization methods. We further note that in our experimental setup, no hyperparameter tuning is carried out for any of the models. When using these models in a production setting, one may obtain better results by selecting parameters using cross-validation on the training-validation set. Parameters which one may investigate tuning include: 
\begin{itemize}
\item Architecture hyperparameters:
\begin{itemize}
\item number of filters in each convolutional layer (we fix 32 channels);
\item number of weighted
averages and lengths of averaging windows;
\item number of LSTM hidden nodes (we fix 64 hidden nodes);
\item decoder type for multi-horizon models (we use seq2seq).
\end{itemize}
\item Feature hyper-parameters:
\begin{itemize}
\item number of levels in features (we fix $L=10$ levels);
\item queue depth in deepVOL(L3)  (we fix queue depth $D=10$);
\item look back window length (we fix $T=100$);
\end{itemize}
\item Training/Optimization hyper-parameters: 
\begin{itemize}
\item Adam learning rate (we fix $\eta=0.01$); 
\item Adam parameters (we fix $\epsilon=1, \beta_1=0.9, \beta_2=0.999$);
\item batch size (we fix batch size 256);
\item number of epochs (we fix 50 epochs);
\item early stopping patience (we fix 10 epochs validation patience);
\item training set downscale factor\footnote{It is worth noting that in all our experiments, we subset the training dataset by a factor of 10, i.e.\ subsequent feature-response observations in the dataset are obtained by sliding forward by 10 order book events. This serves multiple purposes: on one hand, it significantly reduces the computational effort required to train the models, on the other, it helps to prevent the risk of overfitting to specific patterns arising in the training dataset. Down-sampling the dataset might also have adverse effects, for example, reducing the available training data of an illiquid stock by a factor of 10 might result in insufficient information to robustly train these data-intensive models.} (we subset the training dataset by a factor of 10).
\end{itemize}
\end{itemize}
Here we are not interested in obtaining the best fit possible for a specific model but in comparing different models/benchmarks on a level playing field. We thus leave questions related to hyper-parameter tuning for future work.
\end{remark}



All code is developed in Python with the \texttt{tensorflow} library and the \texttt{keras} API, with some layers requiring custom \texttt{tensorflow} methods. Due to the computationally intensive nature of the experiments, specialized infrastructure was required to store the data (5TB) and train the models (GPUs). All computations were carried out on Imperial College's High-Performance Computing cluster \citep{imperial_hpc}, which provides access to several RTX6000 GPUs. 


The results discussed hereafter are specific to the experimental setup under consideration, i.e.\ they are specific to the selected stocks, time period, and models. Different experimental setups may lead to different results. 

\subsection{Do high-frequency returns display predictability? If so, how far ahead can we predict?} \label{sec:predictability_experiment}
In order to understand whether high-frequency returns display predictability we consider the following set of models:
\footnotesize
\begin{equation*}
    \mathcal{M}_0 = \big\{\text{benchmark},\ \text{deepLOB(L1)},\ \text{deepOF(L1)},\ \text{deepLOB(L2)},\ \text{deepOF(L2)},\ \text{deepVOL(L2)},\ \text{deepVOL(L3)}\big\},
\end{equation*}
\normalsize
where L1 models only use the first level of the order book, while L2 and L3 models use the first 10 levels. For each return horizon $h\in \{10, 20, 30, 50, 100, 200, 300, 500, 1000\}$ and \text{TICKER} $\in$ $\{$\text{LILAK}, \text{QRTEA}, \text{XRAY}, \text{CHTR}, \text{PCAR}, \text{EXC}, \text{AAL}, \text{WBA}, \text{ATVI}, \text{AAPL}$\}$ we determine the MCS p-value of each model in $\mathcal{M}_0$. The MCS p-value for the \textit{unpredictive benchmark} model are reported in Table \ref{table:MCS_benchmark}. Lighter green shading corresponds to $\alpha =0.05$ while darker green is $\alpha=0.01$.

\begin{table}[htb]
\centering
\small
\begin{tabular}{|l||c|c|c|c|c|c|c|c|c|}
\hline
   & $h=10$                  & $h=20$                 & $h=30$                & $h=50$                  & $h=100$                & $h=200$              & $h=300$ & $h=500$ & $h=1000$  \\
   \hhline{|=#=|=|=|=|=|=|=|=|=|}
    LILAK & \cellcolor[rgb]{ .663,  .816,  .557}0.00 & \cellcolor[rgb]{ .663,  .816,  .557}0.00 & 0.42  & 1.00  & 1.00  & 1.00  & 1.00  & 1.00  & 1.00 \\
    \hline
    QRTEA & 1.00  & \cellcolor[rgb]{ .776,  .878,  .706}0.01 & \cellcolor[rgb]{ .776,  .878,  .706}0.01 & \cellcolor[rgb]{ .663,  .816,  .557}0.00 & \cellcolor[rgb]{ .663,  .816,  .557}0.00 & \cellcolor[rgb]{ .663,  .816,  .557}0.00 & 0.08 & 0.77  & 0.54 \\
    \hline
    XRAY  & \cellcolor[rgb]{ .663,  .816,  .557}0.00 & \cellcolor[rgb]{ .663,  .816,  .557}0.00 & \cellcolor[rgb]{ .663,  .816,  .557}0.00 & \cellcolor[rgb]{ .776,  .878,  .706}0.01 & 0.10  & 0.23  & 0.48  & 1.00  & 0.93 \\
    \hline
    CHTR  & \cellcolor[rgb]{ .663,  .816,  .557}0.00 & \cellcolor[rgb]{ .663,  .816,  .557}0.00 & \cellcolor[rgb]{ .776,  .878,  .706}0.04 & \cellcolor[rgb]{ .663,  .816,  .557}0.00 & 0.42  & 1.00  & 1.00  & 1.00  & 0.97 \\
    \hline
    PCAR  & \cellcolor[rgb]{ .663,  .816,  .557}0.00 & \cellcolor[rgb]{ .663,  .816,  .557}0.00 & \cellcolor[rgb]{ .663,  .816,  .557}0.00 & \cellcolor[rgb]{ .663,  .816,  .557}0.00 & 1.00  & 1.00  & 0.60  & 1.00  & 1.00 \\
    \hline
    EXC   & 0.26  & \cellcolor[rgb]{ .663,  .816,  .557}0.00 & \cellcolor[rgb]{ .663,  .816,  .557}0.00 & \cellcolor[rgb]{ .663,  .816,  .557}0.00 & \cellcolor[rgb]{ .663,  .816,  .557}0.00 & \cellcolor[rgb]{ .663,  .816,  .557}0.00 & \cellcolor[rgb]{ .663,  .816,  .557}0.00 & 0.51  & 1.00 \\
    \hline
    AAL   & 0.11  & \cellcolor[rgb]{ .663,  .816,  .557}0.00 & \cellcolor[rgb]{ .663,  .816,  .557}0.00 & \cellcolor[rgb]{ .663,  .816,  .557}0.00 & \cellcolor[rgb]{ .663,  .816,  .557}0.00 & 0.71  & 1.00  & 1.00  & 1.00 \\
    \hline
    WBA   & \cellcolor[rgb]{ .663,  .816,  .557}0.00 & \cellcolor[rgb]{ .663,  .816,  .557}0.00 & \cellcolor[rgb]{ .663,  .816,  .557}0.00 & \cellcolor[rgb]{ .663,  .816,  .557}0.00 & \cellcolor[rgb]{ .776,  .878,  .706}0.02 & 0.27  & 0.13  & \cellcolor[rgb]{ .776,  .878,  .706}0.01 & 1.00 \\
    \hline
    ATVI  & \cellcolor[rgb]{ .663,  .816,  .557}0.00 & \cellcolor[rgb]{ .663,  .816,  .557}0.00 & \cellcolor[rgb]{ .663,  .816,  .557}0.00 & \cellcolor[rgb]{ .663,  .816,  .557}0.00 & \cellcolor[rgb]{ .663,  .816,  .557}0.00 & 0.78  & 1.00  & 1.00  & 1.00 \\
    \hline
    AAPL  & \cellcolor[rgb]{ .663,  .816,  .557}0.00 & \cellcolor[rgb]{ .663,  .816,  .557}0.00 & \cellcolor[rgb]{ .663,  .816,  .557}0.00 & 0.23  & 1.00  & 1.00  & 1.00  & 1.00  & 1.00 \\
    \hline
\end{tabular}
\caption{MCS p-values of the \textit{unpredictive benchmark} model for the 10 tickers and 9 horizons under consideration. When the p-value is low at least one of the order book-driven models statistically outperforms the \textit{unpredictive benchmark}, i.e.\ there is order book-driven predictability according to Definition \ref{def:predictability}.} \label{table:MCS_benchmark}
\end{table}

From the results reported in Table \ref{table:MCS_benchmark} we see that at high frequencies predictability is systematically present. For most of the stocks under consideration, we were able to identify predictability up to 50 order book events ahead at the 99\% confidence level.

Except for LILAK, which is the most illiquid stock, we observe a substantial correlation between the persistence in predictability and the average Updates to Price Changes ratio, cf.\ Table \ref{table:stock_characteristics}. Recalling that in our setting the horizon $h$ is measured in order book updates, this can be interpreted as it being easier for the deep learning models to predict returns that are the result of fewer price changes. One might thus expect to obtain a more consistent maximum predictable horizon across stocks when using a price change-driven clock to measure time.

\subsection{Which order book representations perform the best?} \label{sec:best_prediction_experiment}
Having discussed the extent to which the class of models under consideration can identify predictability, it is now natural to ask which of these models performs best. In the MCS framework, this corresponds to determining the specifications which are consistently placed in the set of superior models. We restrict our attention to the horizons and stocks at which predictability is identified and, for each model in consideration, we count the number of times it is identified as a superior model. The results are reported in Table \ref{table:best_MCS}.

\begin{table}[htb]
\centering
\small
\begin{tabular}{lrr}
\cmidrule[\heavyrulewidth]{2-3}
            & $\alpha$ = 0.05 & $\alpha$ = 0.01 \\ 
\toprule
benchmark   & 0\%          & 0\%          \\
\midrule
deepLOB(L1) & 7\%          & 11\%          \\
\midrule
deepOF(L1)  & 23\%         & 22\%         \\
\midrule
deepLOB(L2) & 5\%          & 11\%          \\
\midrule
deepOF(L2)  & 88\%         & 89\%        \\
\midrule
deepVOL(L2) & 65\%         & 84\%         \\
\midrule
deepVOL(L3) & 77\%         & 86\%         \\
\bottomrule
\end{tabular}
\caption{\% of times the model is in the $\alpha$-MCS when predictability is identified at the corresponding level $\alpha$.} \label{table:best_MCS}
\end{table}

From the results in Table \ref{table:best_MCS} (at the 99$\%$ confidence level) we can make the following observations on the way order book representations influence model performance. When considering deep learning models for short-term return prediction having access to L2 data provides a significant advantage over L1 data. We see that models with L1 data are rarely placed in the set of superior models when predictability is identified. When going from L2 to L3 data instead, the increased granularity does not seem to provide a clear advantage: deepVOL(L2) and deepVOL(L3) display similar performance. Our experiment, therefore, suggests that L3 data might be excessively granular when predicting high-frequency returns from order books.

The choice of features used to represent the order book is also crucial when leveraging deep learning methods for return prediction. Using order flow or volume representations provides a significant improvement in performance: the basic deepLOB model ends up being included in the set of best models only 10\% of the time and is outperformed even by the model with only first-level (L1) order flow. Volume- and order flow-based models (with L2/L3 data granularity) display comparable performance, being placed in the set of superior models in 85-90\% of the predictable horizons.

\subsection{Can we use a single model across multiple horizons?} \label{sec:multihorizon_experiment}
In this section, we explore whether using a seq2seq decoder to produce multi-horizon predictions is beneficial. Such models have the clear advantages of having a single set of weights (the network size is only slightly bigger than single horizon networks) and output multiple predictions in very similar run times. This significantly reduces both the memory required to store the models and the time needed to train them. But how do they perform when compared to their single-horizon counterparts?

To answer this question we run the same experiment as in Section \ref{sec:predictability_experiment} and Section \ref{sec:best_prediction_experiment} but enlarge the set of models with the seq2seq specifications. We focus only on prediction horizons $h\in\{10, 20, 30, 50\}$. The results are reported in Table \ref{table:MCS_benchmark_mh} and Table \ref{table:best_MCS_mh}.

From Table \ref{table:best_MCS_mh} we note that, for each input type, the seq2seq specifications outperform their single horizon counterparts. We suggest this behavior might be due to the increased availability of information in a multi-horizon setting. For a given input-target pair, when targets are multi-horizon, more information is available on the ``order book regime'' the inputs should be mapped to. Multi-horizon models can thus learn a more granular map from the input variables to the latent space of ``order book regimes'', which might be beneficial for producing predictions.

\begin{remark} \label{rem:counterintuitive_MCS}
    Note that running the model confidence set procedure again with a larger set of models leads to a counter-intuitive situation where fewer prediction horizons are identified. One would expect that adding models could only increase the number of horizons at which predictability is identified. While this is true in the limit as the number of horizons increases to infinity, it does not hold in the setting where we have access to a finite set of observations. With this observation in mind, we note that results in Table \ref{table:MCS_benchmark_mh} are consistent with those in Table \ref{table:MCS_benchmark}. 
\end{remark}

\begin{table}[htb]
    \centering
    \small
    \begin{tabular}{|l||c|c|c|c|}
    \hline
          & \multicolumn{1}{l|}{$h=10$} & \multicolumn{1}{l|}{$h=20$} & \multicolumn{1}{l|}{$h=30$} & \multicolumn{1}{l|}{$h=50$} \\
    \hhline{|=#=|=|=|=|}
    LILAK & \cellcolor[rgb]{ .663,  .816,  .557}0.00 & \cellcolor[rgb]{ .663,  .816,  .557}0.00 & 0.31  & 0.55 \\
    \hline 
    QRTEA & 1.00  & 0.10  & 0.06  & \cellcolor[rgb]{ .663,  .816,  .557}0.00 \\
    \hline
    XRAY  & \cellcolor[rgb]{ .663,  .816,  .557}0.00 & \cellcolor[rgb]{ .663,  .816,  .557}0.00 & \cellcolor[rgb]{ .663,  .816,  .557}0.00 & \cellcolor[rgb]{ .663,  .816,  .557}0.00 \\
    \hline
    CHTR  & \cellcolor[rgb]{ .663,  .816,  .557}0.00 & \cellcolor[rgb]{ .663,  .816,  .557}0.00 & \cellcolor[rgb]{ .776,  .878,  .706}0.01 & \cellcolor[rgb]{ .663,  .816,  .557}0.01 \\
    \hline
    PCAR  & \cellcolor[rgb]{ .663,  .816,  .557}0.00 & \cellcolor[rgb]{ .663,  .816,  .557}0.00 & \cellcolor[rgb]{ .663,  .816,  .557}0.00 & \cellcolor[rgb]{ .663,  .816,  .557}0.00 \\
    \hline
    EXC   & 0.23  & \cellcolor[rgb]{ .663,  .816,  .557}0.00 & \cellcolor[rgb]{ .663,  .816,  .557}0.00 & \cellcolor[rgb]{ .663,  .816,  .557}0.00 \\
    \hline
    AAL   & 0.08  & \cellcolor[rgb]{ .663,  .816,  .557}0.00 & \cellcolor[rgb]{ .663,  .816,  .557}0.00 & \cellcolor[rgb]{ .663,  .816,  .557}0.00 \\
    \hline
    WBA   & \cellcolor[rgb]{ .663,  .816,  .557}0.00 & \cellcolor[rgb]{ .663,  .816,  .557}0.00 & \cellcolor[rgb]{ .663,  .816,  .557}0.00 & \cellcolor[rgb]{ .663,  .816,  .557}0.00 \\
    \hline
    ATVI  & \cellcolor[rgb]{ .663,  .816,  .557}0.01 & \cellcolor[rgb]{ .663,  .816,  .557}0.00 & \cellcolor[rgb]{ .663,  .816,  .557}0.00 & \cellcolor[rgb]{ .663,  .816,  .557}0.00 \\
    \hline
    AAPL  & \cellcolor[rgb]{ .663,  .816,  .557}0.00 & \cellcolor[rgb]{ .663,  .816,  .557}0.00 & \cellcolor[rgb]{ .663,  .816,  .557}0.00 & 0.07 \\
    \hline
\end{tabular} 
\caption{MCS p-values of the \textit{unpredictive benchmark} model for the 10 tickers and 9 horizons under consideration when seq2seq models are also considered. When the p-value is low at least one of the order book-driven models statistically outperforms the \textit{unpredictive benchmark}, i.e.\ there is order book-driven predictability according to Definition \ref{def:predictability}.} \label{table:MCS_benchmark_mh}
\end{table}

\begin{table}[htb]
    \centering
    \begin{tabular}{lrr}
    \cmidrule[\heavyrulewidth]{2-3}
     & $\alpha = 0.05$ & $\alpha = 0.01$ \\
    \toprule
    benchmark & 0\% & 0\% \\
    \midrule
    deepLOB(L1) & 6\% & 10\% \\
    \midrule
    deepOF(L1) & 13\% & 16\% \\
    \midrule
    deepLOB(L2) & 6\% & 13\% \\
    \midrule
    deepOF(L2) & 63\% & 71\% \\
    \midrule
    deepVOL(L2) & 63\% & 74\% \\
    \midrule
    deepVOL(L3) & 75\% & 84\% \\
    \midrule
    deepLOB(L1, seq2seq) & 13\% & 26\% \\
    \midrule
    deepOF(L1, seq2seq) & 16\% & 16\% \\
    \midrule
    deepLOB(L2, seq2seq) & 16\% & 39\% \\
    \midrule
    deepOF(L2, seq2seq) & 91\% & 97\% \\
    \midrule
    deepVOL(L2, seq2seq) & 78\% & 84\% \\
    \midrule
    deepVOL(L3, seq2seq) & 81\% & 87\% \\
    \bottomrule
    \end{tabular}%
  \caption{\% of times the model is in the $\alpha$-MCS when predictability is identified at the corresponding level $\alpha$.} \label{table:best_MCS_mh}
\end{table}

\subsection{Can we use a single model across multiple stocks?} \label{sec:universality_experiment}
In a similar spirit to \citet{universalLOB}, we investigate questions regarding the universality of order book dynamics. Intuitively, at a microstructural level, securities which are traded by market participants with similar characteristics may be subject to the same trading patterns, independently of the underlying stock's properties. In \citet{universalLOB}, the authors observe that price formation dynamics driven by past order book information, i.e.\ next mid-price moves, display common patterns across different stocks. In this paper, we explore whether similar results hold over longer horizons.

We run the same experiment as in Section \ref{sec:predictability_experiment} but train the models on multiple stocks simultaneously. Specifically, we split the set of 10 stocks under consideration into two; a first ``in-sample'' set of stocks given by $\{$QRTEA, CHTR, EXC, WBA, AAPL$\}$, and a second ``out-of-sample'' set $\{$LILAK, XRAY, PCAR, AAL, ATVI$\}$. For each window $w$, we use all the training-validation data for the ``in-sample'' stocks to train (and validate) the models. We then evaluate the trained models on the test data of both the ``in-sample'' and the ``out-of-sample'' stocks. The results are reported in Table \ref{table:MCS_benchmark_univ}. 

\begin{table}[htb]
    \centering
    \small
	\begin{tabular}{|l||c|c|c|c|c|c|c|c|c|}
		\hline
		& $h = 10$ & $h = 20$ & $h = 30$ & $h = 50$ & $h = 100$ & $h = 200$ & $h = 300$ &  $h = 500$ & $h = 1000$ \\
		\hhline{|=#=|=|=|=|=|=|=|=|=|}
    LILAK & 0.98  & 1.00  & 1.00  & 1.00  & 1.00  & 1.00  & 1.00  & 1.00  & 1.00 \\
    \hline
    QRTEA$(*)$ & \cellcolor[rgb]{ .663,  .816,  .557}0.00 & \cellcolor[rgb]{ .663,  .816,  .557}0.00 & \cellcolor[rgb]{ .776,  .878,  .706}0.01 & 0.06  & \cellcolor[rgb]{ .663,  .816,  .557}0.00 & 0.13  & 0.59  & 0.34  & 1.00 \\
    \hline 
    XRAY  & \cellcolor[rgb]{ .776,  .878,  .706}0.03 & 1.00  & 1.00  & 1.00  & 1.00  & 1.00  & 0.62  & 0.81  & 1.00 \\
    \hline 
    CHTR$(*)$ & 1.00  & 1.00  & 1.00  & 1.00  & 1.00  & 1.00  & 0.94  & 1.00  & 1.00 \\
    \hline
    PCAR  & 0.08  & 1.00  & 1.00  & 1.00  & 1.00  & 1.00  & 1.00  & 1.00  & 1.00 \\
    \hline
    EXC$(*)$ & \cellcolor[rgb]{ .663,  .816,  .557}0.00 & \cellcolor[rgb]{ .663,  .816,  .557}0.00 & \cellcolor[rgb]{ .663,  .816,  .557}0.00 & \cellcolor[rgb]{ .663,  .816,  .557}0.00 & \cellcolor[rgb]{ .663,  .816,  .557}0.00 & 0.39  & 0.69  & 1.00  & 1.00 \\
    \hline
    AAL   & \cellcolor[rgb]{ .663,  .816,  .557}0.00 & \cellcolor[rgb]{ .663,  .816,  .557}0.00 & \cellcolor[rgb]{ .663,  .816,  .557}0.00 & \cellcolor[rgb]{ .663,  .816,  .557}0.00 & \cellcolor[rgb]{ .663,  .816,  .557}0.00 & 0.58  & 0.60  & 1.00  & 1.00 \\
    \hline 
    WBA$(*)$ & \cellcolor[rgb]{ .663,  .816,  .557}0.00 & \cellcolor[rgb]{ .663,  .816,  .557}0.00 & \cellcolor[rgb]{ .663,  .816,  .557}0.00 & \cellcolor[rgb]{ .663,  .816,  .557}0.00 & \cellcolor[rgb]{ .776,  .878,  .706}0.02 & 0.58  & 0.36  & 0.14  & 1.00 \\
    \hline
    ATVI  & \cellcolor[rgb]{ .663,  .816,  .557}0.00 & \cellcolor[rgb]{ .663,  .816,  .557}0.00 & \cellcolor[rgb]{ .663,  .816,  .557}0.00 & \cellcolor[rgb]{ .663,  .816,  .557}0.00 & 0.60  & 0.74  & 0.88  & 1.00  & 1.00 \\
    \hline
    AAPL $(*)$ & 0.07  & 1.00  & 1.00  & 1.00  & 1.00  & 0.59  & 1.00  & 0.65  & 1.00 \\
    
		\hline
	\end{tabular}
	\caption{MCS p-values of the \textit{unpredictive benchmark} model for the 10 tickers and 9 horizons under consideration, universal models only. When the p-value is low at least one of the order book-driven models statistically outperforms the \textit{unpredictive benchmark}, i.e.\ there is order book-driven predictability according to Definition \ref{def:predictability}. $(*)$ are ``in-sample'' stocks.} \label{table:MCS_benchmark_univ}
\end{table}

Focusing on ``in-sample'' stocks, we note that using universal models leads to results that are partially inconsistent with those in Table \ref{table:MCS_benchmark}. A possible interpretation for this is that universal models may be picking up different order book dynamics from those identified by stock-specific models. In this sense, relatively illiquid stocks with overall ``standard'' trading behavior might benefit from universal models thanks to the greater availability of data. We suggest this might be the case for QRTEA and EXC at horizon $h=10$. When, instead, the ticker is mainly subject to stock-specific trading patterns, universal models have a hard time detecting predictability -- we believe this might be the case for CHTR and AAPL. These results are intrinsically tied to the observations of Remark \ref{rem:univ_problems}.

It is rather remarkable that universal models can identify predictability for ``out-of-sample'' stocks. This provides evidence of the presence of common trading patterns in the order books of different tickers. For AAL and ATVI, universal models can consistently outperform the \textit{unpredictive benchmark} predictions without ever learning from the stock's past order book data.

\begin{table}[htb]
    \centering
    \small
	\begin{tabular}{lrr}
		\cmidrule[\heavyrulewidth]{2-3}    \multicolumn{1}{r}{} & \multicolumn{1}{l}{$\alpha = 0.05$} & \multicolumn{1}{l}{$\alpha = 0.01$} \\
		\toprule
		benchmark & 0\% & 0\% \\
		\midrule
		deepLOB(L1, universal) & 0\% & 0\% \\
		\midrule
		deepOF(L1, universal) & 0\% & 0\% \\
		\midrule
		deepLOB(L2, universal) & 8\% & 24\% \\
		\midrule
		deepOF(L2, universal) & 42\% & 62\% \\
		\midrule
		deepVOL(L2, universal) & 71\% & 81\% \\
		\midrule
		deepVOL(L3, universal) & 92\% & 100\% \\
		\bottomrule
	\end{tabular}
\caption{\% of times each model is in the $\alpha$-MCS when predictability is identified at the corresponding level $\alpha$.} \label{table:best_MCS_univ}
\end{table}

Since we believe universal and stock-specific models may be picking up different trading patterns, a natural question to ask is whether the conclusions from Section \ref{sec:best_prediction_experiment} carry over to the universal setting. The results in Table \ref{table:best_MCS_univ} suggest the superior power of L2 over L1 data is retained for universal models. In this case, though, the increased granularity of L3 data appears to be actually beneficial. Moreover, when considering universal models, the volume representation appears to outperform both order book and order flow inputs. These results thus suggest that some predictive universal patterns can be extracted only from the most granular data. This contrasts with stock-specific models which achieve good predictive performance simply based on order flow information (which is, by construction, also contained in volume-based features).

\begin{remark}\label{rem:univ_problems}
	As discussed in Section \ref{sec:responses}, the choice of $\gamma$ used to define the return classes is stock-specific. This means that up/down labels for different securities may correspond to different numbers of mid-price changes. This is in line with our choice of order book-driven clock, which is by definition also stock-specific. It is important to note that the choices of $\gamma$ and clock $t$ might not entirely account for structural trading differences of stocks, making it harder to identify universal trading patterns. This contrasts with \citet{universalLOB}, where the next mid-price move has a straightforward universal structural interpretation. There are a couple more differences with \citet{universalLOB} we want to point out:
	\begin{itemize}
            \item First, in \citet{universalLOB}, the authors consider a much bigger set of stocks, comprised of 500 ``in-sample'' stocks used for training and 500 ``out-of-sample'' stocks, making their model more general and thus more universal. 
		\item Second, in that paper, the authors also investigate questions regarding the stationarity of price formation dynamics. This entails using a single long training window: the greater availability of data leads to more stable results. To employ the model confidence set procedure and allow the models to capture patterns specific to different economic regimes, we did not adopt this approach in our work.
		\item The LSTM model considered in that paper differs from the deepLOB/deepOF/deepVOL architectures. In \citet{universalLOB}, the LSTM specification is an online model, i.e.\ inputs at time $t$ represent the order book state at time $t$ only and are fed through an LSTM-based architecture updating the stored hidden state and producing output predictions. In our models, instead, there is no storage of hidden units between one prediction and the next, i.e.\ at each time step $t$, we input the order book history between $t-T+1$ and $t$ (in the form of raw order books, order flow or volumes) and, after appropriate convolutional feature extraction, apply an LSTM to the processed sequential data to produce the prediction. When studying dependence on order book history $T$, \citet{universalLOB} refer to the cut-off horizon used in the backpropagation through time computation of gradients during the training procedure.
	\end{itemize}
\end{remark}

\section{Conclusions and Outlook} \label{sec:conclusions}
In this paper, we explored empirical questions regarding the predictability of mid-price returns driven by order book data in centralized exchanges. The predictability in price formation dynamics, already considered in the sense of next mid-price change in \citet{universalLOB}, was found to persist up to 50-300 order book updates ahead, horizons over which multiple mid-price changes may occur. Such predictable horizons might vary from a few milliseconds to nearly half a second, depending on the stock under consideration. These results contrast with low-frequency returns, where predictability is much harder to identify but easier to trade once discovered. In fact, in the high-frequency context, predictability is not always exploitable due to technological limitations and market microstructure issues. A related and more challenging question is, therefore, to understand whether the predictability identified in this paper is actually tradable.

The experiments were carried out using specific deep learning architectures. Other than identifying predictable horizons, we aimed to answer questions related to the models under consideration. In particular, we found strong empirical evidence for using L2 data over L1 data. But, from the results presented in this paper, using even more granular order book information (L3) seems to benefit only universal models. The experiments also highlight the importance of carefully choosing a data representation: models based on the basic order book level representation are considerably outperformed by those with order flow or volume inputs. Finally, we found empirical evidence for the presence of universal trading patterns in order book dynamics. Our preliminary results suggest that deep learning architectures may pick up different predictable order book patterns when trained on a pool of stocks instead of a single one. The volume representation has some considerable theoretical advantages, in particular, it is robust to small perturbations, which might explain its superior predictive ability when considering universal models.

Many further theoretical questions regarding predictability which we believe to be of research interest remain unanswered. First, it would be natural to explore whether predictability in returns can be entirely explained by order book structure or if recurring trading patterns play a relevant role. Next, one could compare the predictive performance of the deep learning architectures discussed in this paper to that of the models based on carefully engineered features considered in \citet{AitSahalia2022}. Finally, only Nasdaq-traded stocks were considered in this paper. This is a centralized electronic dealer market on which relatively big companies are listed. It would be interesting to explore whether similar results are obtained for securities traded on exchanges with different market structures. With appropriate experimental setups, we believe the model confidence set procedure used in this paper provides a solid framework to tackle all these questions.

On the practical side, there are some relevant issues one should consider. First, it is essential to note that in this paper we focus on mid-to-mid returns, which are not tradable in practice. When considering specific trading applications, one should thus define the return labels appropriately, for example, working with ask-to-bid returns. Second, in this paper, we use an order book-driven clock. In practice, this means the prediction horizon is intrinsically stochastic, which might be a problem in execution. Another critical aspect that would need to be accounted for in practical high-frequency trading is total execution speed. This consists of order book information latency, model prediction run time, and order submission speed. \citet{deepLOB} and \citet{deepOF} argue that their models (which are also considered in this paper) are sufficiently fast to be used by traders with good connections: see Appendix \ref{app:latencies} for a discussion on the effects of infrastructure latency. 

As with any trading application, one must also account for the impact his own trade will have on the order book. Moreover, if many traders exploit the same inefficiencies, this might erode all predictive power above the lowest tradable latencies. Overall we believe that while the predictability identified in this paper might not be directly tradable through a standalone strategy, it might still help some market players, such as market makers, to gauge the direction of the market and adjust their quotes accordingly.

\clearpage

\begin{appendices}

\section{Appendix} \label{app:appendix}
\subsection{Deep Learning} \label{app:deep_learning}
\subsubsection{Deep Neural Network Architectures}
When approximating a function\footnote{This notation covers both
regression $f(\cdot) = g(\cdot)$ and classification $f(\cdot) = (p_{\downarrow}(\cdot), p_{=}(\cdot), p_{\uparrow}(\cdot))$ tasks.} $f$ from a set of models $\{f_\theta\}_{\theta\in\Theta}$, such as in the return prediction task introduced in Section \ref{sec:predictability_returns}, a quite common approach is to consider a parametric family of deep neural networks. Theoretically motivated by the universality of such models, this approach does not rely on the functional form of $f$ being correctly specified. 

The main idea behind deep neural networks is to lift the input into a higher dimensional space to extract relevant latent features before projecting onto output space. Mathematically a neural network consists of a composition of functions, known as layers, $f_l: \mathbb{R}^{d_{l-1}} \rightarrow \mathbb{R}^{d_l}$ for $l=1,\ldots, L$ where $d_0$ is the dimensionality of input space and $d_L$ is the dimensionality of the output space, for example $d_L=3$ in our three-class classification task. A neural network is considered deep if the number of layers $L$ is large.

In the simplest case the layers $f_l$’s are ``activated'' affine transformations, i.e.\ for input $z_{l-1} \in \mathbb{R}^{d_{l-1}}$
\[f_l(z_{l-1}) = \sigma_l(W_l z_{l-1} + b_l),\] where $W_l \in \mathbb{R}^{d_l \times d_{l-1}}$ and $b_l \in \mathbb{R}^{d_l}$ are learnable parameters, also known as weights, and $\sigma_l$ is a non-linear activation function applied element-wise. Neural networks of the form $f=f_L \circ \cdots \circ f_1$ such that each $f_l$ is of this form have been shown to be universal for the class of continuous functions \citep{approx_Cybenko, approx_Hornik}, i.e.\ any continuous function (on a bounded domain) can be approximated arbitrarily well by a large enough network in the supremum norm\footnotemark \footnotetext{It follows directly that the class of neural networks is dense in $L^p(\mathcal{X})$ space, for bounded $\mathcal{X}\subseteq\mathbb{R}^d$.}. Related works have shown the representational benefits of depth: there are functions that deep networks can construct with polynomially many parameters, which instead require exponentially many parameters when considering shallow networks \citep{approx_Telgarsky}. More complex and domain-specific forms of layers $f_l$ exist; we briefly introduce convolutional and recurrent/LSTM layers.

\begin{figure}[!htb]
    \centering
    \includegraphics[width = 14cm]{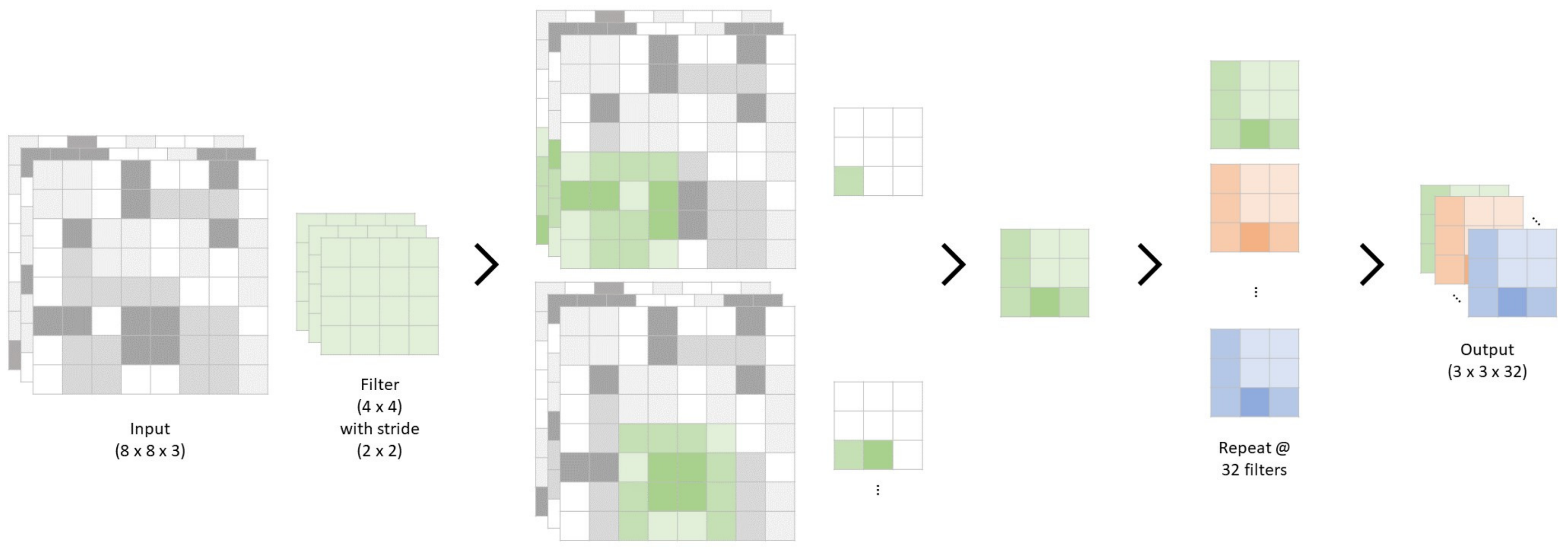}
    \caption{Convolutional layer, input $z_{l-1}\in\mathbb{R}^{8\times8\times3}$, filter size $(n_l\times m_l) = (4 \times 4)$, stride $(s_l \times t_l) = (2 \times 2)$ and number of filters $k_l=32$.} \label{fig:conv}
\end{figure}

\paragraph{Convolutional layers} Convolutional layers are a natural choice when the input is an image, i.e.\ 
\[z_{l-1}\in \mathbb{R}^{d_{l-1}} = \mathbb{R}^{h_{l-1} \times w_{l-1} \times c_{l-1}},\] where $h_{l-1}$ is the height in pixels, $w_{l-1}$ is the width in pixels and $c_{l-1}$ is the number of channels, for example $c_{l-1}=3$ if the image is in RGB. A convolutional layer is a specific case of a generic dense layer with parameter restrictions to account for adjacencies in the input's structure. It consists of $k_l$ weight kernels, also known as filters, which are convoluted with the input image to produce the output. Mathematically a 2-dimensional convolution with $k_l$ $(n_l\times m_l)$ filters and stride $(s_l\times t_l)$ is described by
\begin{align*}
[f_l(z_{l-1})]_{i,j,k} &= \sum_{n=1}^{n_l} \sum_{m=1}^{m_l}  \sum_{c=1}^{c_{l-1}} \Big\{[W_{l}^{(k)}]_{n,m,c} [z_{l-1}]_{s_l (i-1)+n,t_l(j-1)+m,c} + b^{(k)}_{l} \Big\},
\end{align*}
where $W_{l}^{(k)}\in\mathbb{R}^{n_l \times m_l \times c_{l-1}}$, for $k=1, \ldots, k_l$, are the $k_l$ filters, $b_l^{(k)}\in\mathbb{R}$ are bias terms and $[\cdot]_{i,j,k}$ denotes the $i,j,k$-th entry of a three dimensional tensor. The output then lives in $\mathbb{R}^{h_l\times w_l\times k_l}$, where $h_l = (h_{l-1}-n_l)/s_l + 1$ and $w_l = (w_{l-1} - m_l)/t_l + 1$, assuming that $s_l|(h_{l-1}-n_l)$ and $t_l|(w_{l-1}-m_l)$. While this equation might look a bit daunting, convolutional layers are often easily understood via a graphical depiction, as in Figure \ref{fig:conv}. Many empirical studies \citep{AlexNet, ResNet} have investigated the ability of convolutional layers to extract relevant features from input with grid-like topologies. In Section \ref{sec:DL_models}, we discuss how this ability might be leveraged to extract relevant features from order books.

\begin{figure}[!htb]
    \centering
    \includegraphics[width = 7cm]{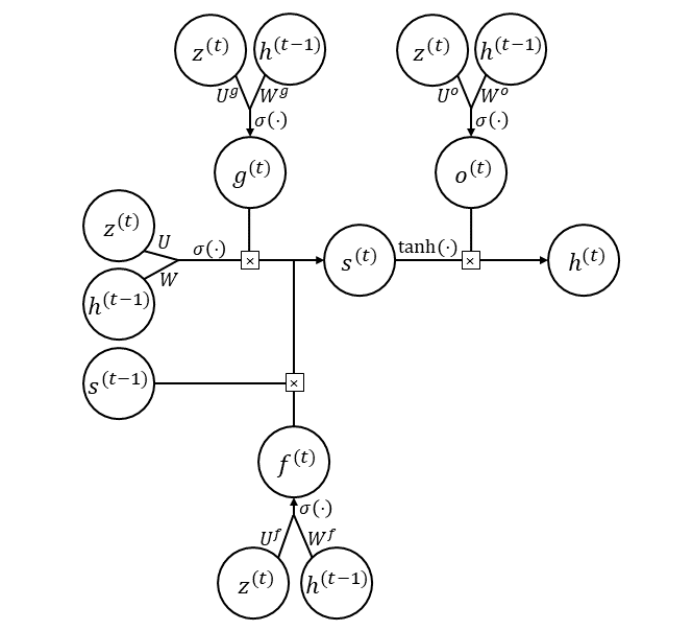}
    \caption{LSTM layer network diagram.} \label{fig:LSTM}
\end{figure}

\paragraph{Recurrent/LSTM layers} When the input has a built-in temporal structure, this can be quite naturally accounted for by recurrent layers. This type of layer retains information over time, discovering temporal dependencies in the data. Recurrent layers may be used with streaming data to obtain online predictions or applied to a whole time series yielding a single output. We focus on the latter case, i.e.\ assuming the input $z_{l-1} \in \mathbb{R}^{d_{l-1}}$ has a temporal structure of the form 
\[z_{l-1} =\Big(z_{l-1}^{(1)}, \ldots, z_{l-1}^{(T_{l-1})}\Big)^\text{T} \in \mathbb{R}^{T_{l-1} \times n_{l-1}},\]
the recurrent layer $z_l = f_l(z_{l-1})$ is given by
\begin{align*}
h_{l-1}^{(0)} \in \mathbb{R}^{m_{l-1}}, \quad
h_{l-1}^{(t)} = \phi_l \Big(h_{l-1}^{(t-1)}, z_{l-1}^{(t)} \Big)\ \text{ for } t = 1, \ldots, T_{l-1},\quad 
z_{l} &= h_{l-1}^{(T_{l-1})},
\end{align*}
where $\phi_l:\mathbb{R}^{m_{l-1}} \times \mathbb{R}^{n_{l-1}} \rightarrow \mathbb{R}^{m_{l-1}}$ is a parameterized recurrent function and the $h_{l-1}^{(t)}$'s are known as hidden states. The most widely used type of recurrent layer is Long Short-Term Memory (LSTM) \citep{lstm}. While other types of recurrent layers may suffer from vanishing or exploding gradients when training, the specific structure of LSTMs largely prevents such problems \citep{vanishing_grad}. In an LSTM layer each hidden state $h$ is augmented with a memory state $s$, and hence the recurrence becomes
\[\Big(h^{(t)}_{l-1}, s^{(t)}_{l-1}\Big) = \phi_l\Big(h_{l-1}^{(t-1)},  s^{(t-1)}_{l-1}, z_{l-1}^{(t)}\Big).\]
The equations governing the LSTM recurrence are given by
\begin{align*}
	h^{(t)}_{l-1} &= o^{(t)}_{l-1} \odot \tanh(s^{(t)}_{l-1}), \\
	s^{(t)}_{l-1} &= g^{(t)}_{l-1} \odot \sigma\Big(b_{l} + U_l z_{l-1}^{(t)} + W_l h_{l-1}^{(t-1)} \Big) + f^{(t)}_{l-1} \odot s^{(t-1)}_{l-1},
\end{align*}
where the input, forget and output gates depend on the context, i.e.\ on $h_{l-1}^{(t-1)}, z_{l-1}^{(t)}$, via:
\begin{align*}
	g^{(t)}_{l-1} &= \sigma\Big(b_l^g + U^g_{l} z_{l-1}^{(t)} + W^g_{l} h_{l-1}^{(t-1)} \Big),\\
	f^{(t)}_{l-1} &= \sigma\Big(b^f_l + U^f_{l} z_{l-1}^{(t)} + W^f_{l} h_{l-1}^{(t-1)} \Big), \\
	  o^{(t)}_{l-1} &= \sigma\Big(b^o_l + U^o_{l} z_{l-1}^{(t)} + W^o_{l}  h_{l-1}^{(t-1)} \Big),
\end{align*}
for sigmoid activation function $\sigma$, element-wise product $\odot$ and parameters $b_l, b^g_l, b^f_l, b^o_l \in\mathbb{R}^{m_{l-1}}$, $U_l, U^g_l, U^f_l, U^o_l \in \mathbb{R}^{m_{l-1}\times n_{l-1}}$ and $W_l, W^g_l, W^f_l, W^o_l \in \mathbb{R}^{m_{l-1}\times m_{l-1}}$. The network diagram of an LSTM layer is given in Figure \ref{fig:LSTM}. LSTMs summarize relevant context information in the memory cell $s_{l-1}^{(t)}$, performing exceptionally well on tasks with a strong temporal dependency \citep{lstm_sota}.

\subsubsection{Training Deep Neural Networks}
As discussed in Section \ref{sec:predictability_returns}, once we have specified a parametric model for approximating the prediction function $f\approx f_\theta$ we wish to learn the best approximation to $f$ from a training set of observed input-output data $\mathcal{D}_{\text{train}} = \{(x_i, y_i)\}_{i \in \mathcal{I}_{\text{train}}}$. To do so, we aim to find the parameter combination $\theta$ which yields predictions closest to the observations:
\begin{equation*}
     \hat \theta = \underset{\theta \in\Theta}{\text{argmin }} L( \theta | \mathcal{D}_{\text{train}}) =  \underset{\theta \in\Theta}{\text{argmin}} \Big\{\frac{1}{|\mathcal{I}_{\text{train}}|} \sum_{i\in\mathcal{I}_{\text{train}}} \mathcal{L}\big(f_{\theta}(x_i), y_i\big)  \Big\},
\end{equation*} 
where $\mathcal{L}(\cdot, \cdot)$ is a loss function quantifying the distance between the prediction and the observed value, and $\theta$ contains all network parameters (weights matrices, bias terms, CNN filters...). The form of $\mathcal{L}$ may be chosen using probabilistic arguments or according to some other task-specific criteria. 

When the model $f_\theta$ is a deep neural network the loss landscape $L(\theta) = L(\theta|\mathcal{D}_{\text{train}})$ may be very complex and finding a ``good'' minimizer is often a difficult task. The most widely used approach is to use (some variant of) gradient descent in which the model parameters are iteratively updated by 
\[\theta^{(n + 1)} = \theta^{(n)} - \eta \nabla_{\theta} L\big(\theta^{(n)}\big), \]
until a convergence criterion is met. Here $\eta>0$ is a fixed learning rate, and the weights $\theta^{(0)}$ are chosen according to an appropriate initialization rule. When considering deep neural network architectures, the number of parameters is often very large, and therefore second-order methods are often intractable or computationally infeasible. Many variants of the first order gradient descent algorithm exist though. For example, the stochastic gradient descent algorithm updates parameter values based on  estimates of the gradient $\nabla_{\theta} L(\theta^{(n)})$ computed from a random subset (known as a batch) of the training data set. More advanced first-order optimizers consider momentum and/or use adaptive learning rates. In our experiments, we use the Adam optimizer \citep{adam} which updates the parameters based on adaptive estimates of first and second-order moments:
\begin{align*}
m^{(n+1)} &= \frac{1}{1-\beta_1^n} \Big\{ \beta_1 m^{(n)} + (1-\beta_1) \nabla_{\theta} L\big(\theta^{(n)}\big) \Big\}, \\
v^{(n+1)} &= \frac{1}{1-\beta_2^n} \Big\{ \beta_2 v^{(n)} + (1-\beta_2) \big[\nabla_{\theta} L \big(\theta^{(n)}\big)\big]^2 \Big\}, \\
\theta^{(n+1)} &= \theta^{(n)} - \eta \frac{m^{(n+1)}}{\sqrt{v^{(n+1)}}+\epsilon},
\end{align*}
for parameters $\eta, \beta_1, \beta_2, \epsilon>0$.

\begin{remark} \label{rem:uncertainty_errors}
Note that, contrary to other model specifications, such as linear models, the fitting procedure of deep neural networks is not closed form. This adds a further layer of uncertainty. Any model is subject to the following two sources of uncertainty: one due to model specification (i.e.\ $f\approx f_{\theta}$) and one due to parameter estimation uncertainty (i.e.\ statistical error arising from estimating $\theta$ from a finite sample). Models that do not have closed forms for $\hat{\theta}$ from observed data are further subject to numerical optimization error in the parameter estimation phase. Sometimes, the optimization procedure may not even converge to a (global) minimum. With deep neural networks, we thus trade off smaller model uncertainty (at least theoretically due to the universality property) for larger parameter estimation error.
\end{remark}

When the task is a classification problem $f_\theta$ is usually chosen so that it maps to $\{p\in [0,1]^C: \|p\|_1= 1\}$, where $C$ is the number of classes in the set $\mathcal{C}$. This allows to give a probabilistic interpretation of the deep neural network outputs, i.e.\
\[f_\theta(x) = \{p_{c,\theta}(x)\}_{c\in\mathcal{C}}, \text{ where }\ p_{c,\theta}(x) = \mathbb{P}(y=c|x, \theta).\] In this case, a natural choice of loss function to use during training is the cross-entropy loss, which can be understood by maximum likelihood arguments: assuming the responses are independent given the features and the distribution of the features is independent of the parameters $\theta$ the negative log-likelihood is
\[L(\theta|\mathcal{D}_{\text{train}}) = - \log \mathbb{P}(\mathcal{D}_{\text{train}}|\theta) \propto - \frac{1}{|\mathcal{I}_{\text{train}}|}\sum_{i\in\mathcal{I}_{\text{train}}} \sum_{c\in\mathcal{C}} \mathds{1}_{\{y_{i} = c\}} \log p_{c,\theta}(x_i). \]

When training samples for the classes are unbalanced, say class $\bar{c}\in\mathcal{C}$ has many more samples than all other classes, the optimization algorithm tends to get stuck in the trivial minimum given by the Dirac measure at $\bar{c}\in\mathcal{C}$. One way to mitigate this effect is to re-weight the categorical cross entropy loss such that the network is less incentivized to move towards the trivial minimum. With appropriate weighting, $\bar{c}\in\mathcal{C}$ observations influence the direction of the gradient less, thus reducing the strength of the attraction towards the trivial minimum. Mathematically one uses the following weighted categorical cross-entropy loss
\[L(\theta|\mathcal{D}_{\text{train}}) = - \frac{1}{|\mathcal{I}_{\text{train}}|}\sum_{i\in\mathcal{I}_{\text{train}}} \sum_{c\in\mathcal{C}} w_c \mathds{1}_{\{y_{i} = c\}} \log p_{c,\theta}(x_i), \]
where the weights $w_c$ are chosen to be inversely proportional to the class's training set density. For example, in our experiments, we set:

\[w_c = w_c(\mathcal{D}_{\text{train}}) = \frac{|\mathcal{I}_{\text{train}}|}{\#\{c_i\in\mathcal{D}_{\text{train}} : c_i =c\}} .\]

\subsection{Model Confidence Sets} \label{app:MCS}
The model confidence set procedure introduced by \citet{mcs} provides a formalized statistical inference framework in which to compare competing models. It does not assume that any of the models is the true data generating process but simply aims to identify a set of models that will contain the best model with a given level of confidence, known as model confidence set, MCS. A model confidence set is, therefore, analogous to a confidence interval for a parameter. To identify the best models, one must compare the model outputs (either as estimated class probabilities or as forecasts) by selecting an appropriate loss/score function.

Let us denote by $\mathcal{M}_0$ the set of models under consideration and by $\{L_{i,w}\}_{w=1}^W$ for $i\in\mathcal{M}_0$ the time series of losses. In the MCS framework for models $i,j\in\mathcal{M}_0$ and window $w\in\{1,\ldots, W\}$ we define:
\[ d_{ij,w} = L_{i,w} - L_{j,w}. \]
Under a stationarity assumption on the $(d_{ij,w})_{w\geq 1}$, we define the set of best models among those in $\mathcal{M}_0$ as
\[\mathcal{M}^* = \{i\in\mathcal{M}_0 : \mathbb{E}[d_{ij}] \leq 0\ \forall j\in\mathcal{M}_0\}. \]
This is the set we want to identify by using the MCS procedure. The MCS algorithm is then defined as follows.
\begin{definition} \label{def:MCS}
    Let $\delta_\mathcal{M}$ and $e_\mathcal{M}$ be an equivalence test and an elimination rule.
    \begin{itemize}[align=right, leftmargin=6.5em]
        \item[{Step 0.}] Initially set $\mathcal{M}=\mathcal{M}_0$.
        \item[{Step 1.}] Test 
        $H_{0,\mathcal{M}}: \mathbb{E}[d_{ij}]=0, \  \forall i,j\in\mathcal{M}$
        using $\delta_\mathcal{M}$ at level $\alpha$.
        \item[{Step 2.}] If $H_{0,\mathcal{M}}$ is accepted, set $\hat{\mathcal{M}}_{1-\alpha}^* = \mathcal{M}$; otherwise, use $e_\mathcal{M}$ to eliminate an object from $\mathcal{M}$ and repeat the procedure from Step 1.
    \end{itemize}
\end{definition}

Under appropriate assumptions on the equivalence test and elimination rule\footnote{We require the equivalence test $\delta_\mathcal{M}$ to have asymptotic level $\alpha$ and asymptotic power 1, while the elimination rule $e_\mathcal{M}$ needs to satisfy $\lim_{W\rightarrow 0} \mathbb{P}(e_\mathcal{M} \in \mathcal{M}^*|H_{1,\mathcal{M}}) =0$.}, the model confidence set $\hat{\mathcal{M}}_{1-\alpha}^*$ has the following asymptotic properties:
\begin{enumerate}
    \item[{(i)}] $\displaystyle \liminf_{W\rightarrow\infty}\mathbb{P}(\mathcal{M}^* \subset \hat{\mathcal{M}}^*_{1-\alpha}) \geq 1-\alpha$,
    \item[{(ii)}] $\displaystyle \lim_{W\rightarrow\infty} \mathbb{P}(i \in\hat{\mathcal{M}}^*_{1-\alpha}) = 0 $ for $i\notin \mathcal{M}^*$,
\end{enumerate}
where $W\rightarrow\infty$ denotes the number of sample periods. If furthermore, the equivalence test and elimination rule satisfy a coherency condition\footnote{The required coherency condition is $\mathbb{P}(\delta_{\mathcal{M}}=1, e_M\in \mathcal{M}^*) \leq \alpha$.}, we have the following finite sample property:
\begin{enumerate}
    \item[{(iii)}] $\mathbb{P}(\mathcal{M}^* \subset \hat{\mathcal{M}}^*_{1-\alpha}) \geq 1-\alpha$.
\end{enumerate}
In \citet{mcs}, the authors give multiple practical examples of equivalence tests and elimination rules. We focus on implementing the MCS procedure with tests constructed from t-statistics and bootstrap estimators. This choice of equivalence test and elimination rule has two practical advantages: it does not require estimating a variance-covariance matrix for the time series $\{ (d_{ij,w})_{i,j\in\mathcal{M}_0}, w\geq 1\}$ which might be difficult when $|\mathcal{M}_0|\approx W$ and it satisfies the coherency condition required for the finite sample property (iii) of the MCS procedure.
Let us assume the following holds:
\begin{assumption}\label{assumption:MCS}
    For some $r > 2$ and $\gamma >0$, it holds that that $\{ (d_{ij,w})_{i,j\in\mathcal{M}_0}, w\geq 1\}$ is strictly stationary, $\alpha$-mixing of order $-r/(r-2)$ and $\text{Var}(d_{ij,w}) > 0$,  $\mathbb{E}|d_{ij, w}|^{r+\gamma} <\infty$ for all $i, j \in\mathcal{M}_0$.
\end{assumption}
For a set of models $\mathcal{M}\subset \mathcal{M}_0$ with time series of relative performances $\{ (d_{ij,w})_{i,j\in\mathcal{M}}, w\geq 1\}$ define the following quantities $\forall i,j\in\mathcal{M}$:
\[\bar{d}_{ij} = \frac{1}{W}\sum_{w=1}^{W} {d}_{ij, w}, \quad \bar{d}_{i\cdot} = \frac{1}{|\mathcal{M}|}\sum_{j\in\mathcal{M}} \bar{d}_{ij},\]
respectively measuring the sample relative performance of the models $i$ and $j$ and the sample relative performance of model $i$ and all other models in $\mathcal{M}$. We can then introduce the following t-statistics $\forall i\in\mathcal{M}$:
\[t_{i\cdot} = \frac{\bar{d}_{i\cdot}}{\sqrt{\widehat{\text{Var}}(\bar{d}_{i\cdot})}},\]
where $\widehat{\text{Var}}(\bar{d}_{i\cdot})$ are appropriate estimators of the variances of the $\bar{d}_{i\cdot}$'s. The associated test statistic is $ T_{\max, \mathcal{M}} = \max_{i\in\mathcal{M}} t_{i\cdot}.$
The equivalence test and elimination rule are given by:
\begin{itemize}[align=left, leftmargin=4em]
\item[{$(\delta_\mathcal{M})$}] Reject $H_{0,\mathcal{M}}$ if $T_{\max, \mathcal{M}} > F^{-1}_\varrho(1-\alpha)$ where $T_{\max}\sim F_\varrho$.
\item[{$(e_\mathcal{M})$}] Eliminate $\text{argmax}_{i\in\mathcal{M}} t_{i\cdot}.$
\end{itemize}
The asymptotic distribution of the test statistic $T_{\max,\mathcal{M}}$ is nonstandard because it depends on the (usually unknown) nuisance parameter $\varrho$ under the null. The value of $\varrho$ is not strictly necessary since we can consistently estimate the distribution of $T_{\max, \mathcal{M}}$ by using a bootstrapping procedure. In this case, the equivalence test compares the observed value $T_{\max, \mathcal{M}}$ with the bootstrap empirical quantile. The resulting procedure preserves the asymptotic\footnote{As both the sample size $W$ and the number of bootstrap samples $B$ increase.} properties (i) and (ii) and the finite sample property (iii) under Assumption \ref{assumption:MCS}. For details of the bootstrapping procedure see the Appendix of \citet{mcs}.

\begin{remark}
    The test based on the t-statistic discussed above relies on the fact that the null hypothesis $H_{0,\mathcal{M}}$ can be equivalently rewritten as 
    \[H_{0,\mathcal{M}}: \mathbb{E}[d_{i\cdot}] = 0, \ \forall i \in \mathcal{M}, \quad \text{where}\quad d_{i\cdot} = \frac{1}{|\mathcal{M}|}\sum_{j\in\mathcal{M}}d_{ij}.\]
\end{remark}

In this paper, we present results using MCS p-values, which are defined as follows.

\begin{definition}
    Let $(\delta_\mathcal{M}, e_{\mathcal{M}})$ be the equivalence test and elimination rule associated with a MCS procedure as defined in Definition \ref{def:MCS}. The elimination rule defines a decreasing sequence of random sets 
    $\mathcal{M}_0 \supset \mathcal{M}_1 \supset \ldots \supset \mathcal{M}_{|\mathcal{M}_0|}$
    by successively eliminating models $e_{\mathcal{M}_0}, \ldots, e_{|\mathcal{M}_0|}$. Let $p_{H_{0,\mathcal{M}}}$ denote the p-value associated with the null hypothesis $H_{0,\mathcal{M}}$ under the equivalence test $\delta_{\mathcal{M}}$ with the convention that $p_{H_{0,\mathcal{M}_{|\mathcal{M}_0|}}} \equiv 1$. Then for model $i = e_{\mathcal{M}_k}\in\mathcal{M}_0$ the MCS p-value is defined as 
    \[p^{\text{MCS}}_i = \max_{k'\leq k}\ p_{H_{0,\mathcal{M}_{k'}}}.\]
\end{definition}
Note that the definition of MCS p-value is such that for $i\in\mathcal{M}_0$
$ p^{\text{MCS}}_i\geq \alpha \iff i \in \hat{\mathcal{M}}^*_{1-\alpha}.$

\begin{remark}
    The MCS procedure is quite flexible. It allows to compare models between each other and to compare them to an \textit{unpredictive benchmark} model (by including the \textit{unpredictive benchmark} in $\mathcal{M}_0$). When only two models are considered, i.e.\ $|\mathcal{M}_0| = 2$, the MCS procedure reduces to testing the null hypothesis $H_0: \mathbb{E}[d_{i_1i_2}] = 0$.
\end{remark}

The model confidence set procedure relies on the assumption that the loss differences time series $\{(d_{ij,w})_{i,j\in\mathcal{M}_0},\, w\geq 1\}$ is stationary and $\alpha$-mixing. This means that even when the loss process $\{(L_{i,w})_{i\in\mathcal{M}_0},\, w\geq 1\}$ is non-stationary, e.g.\ model performance is tied to regime changes, the model confidence set procedure might still be applicable. In Section \ref{sec:experiments}, we assume the time series $\{(d_{ij,w})_{i,j\in\mathcal{M}_0},\, w\geq 1\}$ to be stationary based on intuitive arguments, but we discuss some statistical checks which might be carried out to provide further evidence in this context. First of all, it is worth noting that general tests for stationarity are not available: most tests require assuming a parametric form on the time series. The simplest example would be to impose a linear structure on the time series, i.e.\ assume the difference time series $\{\boldsymbol{d}_w,\, w\geq 1\} = \{(d_{ij,w})_{i,j\in\mathcal{M}_0},\, w\geq 1\}$ follows a Vector Auto-Regressive process of order 1:
\[\boldsymbol{d}_w = \Gamma \boldsymbol{d}_{w-1} + \boldsymbol{\eta}_w, \]
where $\boldsymbol{\eta}_w \sim N(0, \Sigma_\eta)$ are serially uncorrelated. In this setting, one can consider various tests to determine whether the process is stationary. For example, one may apply multivariate homogeneous Dickey-Fuller tests \citep{multivariate_unit_root} to test the null of a unit root against the alternative of stationarity or a multivariate KPSS test to test (short-memory) stationarity as the null and the presence of a unit root as the alternative \citep{multivariate_KPSS}.

\subsection{Data Processing} \label{app:data_proc}
From LOBSTER \citep{lobster} we obtain the order book and message files introduced in Section \ref{sec:dataset_LOBSTER}. These files are jointly processed to obtain the desired order book features, i.e.\ raw order books, order flow and volumes, and responses as described in Section \ref{sec:dataset_processed}. Before extracting the feature-response pairs the data is treated in the following way:
\begin{enumerate}
\item order book states with crossed quotes are removed;
\item states occurring at the same time stamp (at nanosecond precision) are collapsed onto the last state, e.g.\ an aggressive order executing against multiple resting limit orders;
\item the first and last 10 minutes of market activity are dropped;
\item the smoothed returns at the requested horizons --  as defined in Section \ref{sec:responses} -- are computed and matched to the corresponding features.
\end{enumerate}

The data provided by LOBSTER is easily accessible and does not exhibit significant anomalies. The only data inconsistencies we found were on 2019-11-05 for the WBA ticker after the trading halt occurring between 13:32:50.233001415 and 13:37:50.2335639.

\subsubsection{Volume data processing} \label{app:data_proc_vol}
LOBSTER provides level-based data in a form that can be easily manipulated to derive order book and order flow features (we access the first $L=10$ levels). Deriving the volume features from LOBSTER data instead is a non-trivial task. If we set a window $W>L$ for the volume representation, some levels past $L=10$ might have prices that fall in the volume window, i.e.\ $p_{x,t}^{(L+1)} \in \{\pi_{x,t}^{(1)}, \ldots, \pi_{x,t}^{(W)}\}$, and we won't have the required information. To avoid this issue in all our experiments we set $W=L=10$. Reconstruction of L3 queue data has a further complication: every time a quoting price leaves the 10-level range, all orders submitted/canceled at this price are not recorded. Therefore there is no way of reconstructing what happens at such a price level while it is outside of the 10-level range. To minimize the effect of this shortcoming, we adopt the following systematic approach: each time a quoting price (re-)enters the 10-level range, we initialize the corresponding queue with a single order of size given by the total liquidity available at that price level (even if it may be made up of multiple independent orders).

These issues could be mitigated by accessing data deeper in the order book from LOBSTER, say $L=50$ or $100$, while keeping a relatively small window $W << L$.

\subsection{Data set descriptive statistics: ATVI deep dive} \label{app:descriptive_stats}
In this section, we report some descriptive statistics of the processed data used for the experiments. We focus on one representative stock, Activision Blizzard Inc. (ATVI). First, we explore the distribution of the order book, order flow, and volume features. For each level/price tick, we report the aggregated distributional boxplot. Moreover, in order to understand the temporal behavior of the data, we fix one specific level/ price tick and look at how the distribution varies day by day. Finally, we analyze the distribution of the target categorical returns. To do so we fix the first window in the experiment procedure, i.e.\ $w=1$ in Section \ref{sec:experiments}, and plot the unconditional train, validation, and test distributions at each prediction horizon $h\in\{10,20,30,50,100,200,300,500,1000\}$.

\subsubsection{Features} \label{app:descriptive_stats_features}
\begin{figure}[!htb]
    \centering
    \includegraphics[width = 12cm]{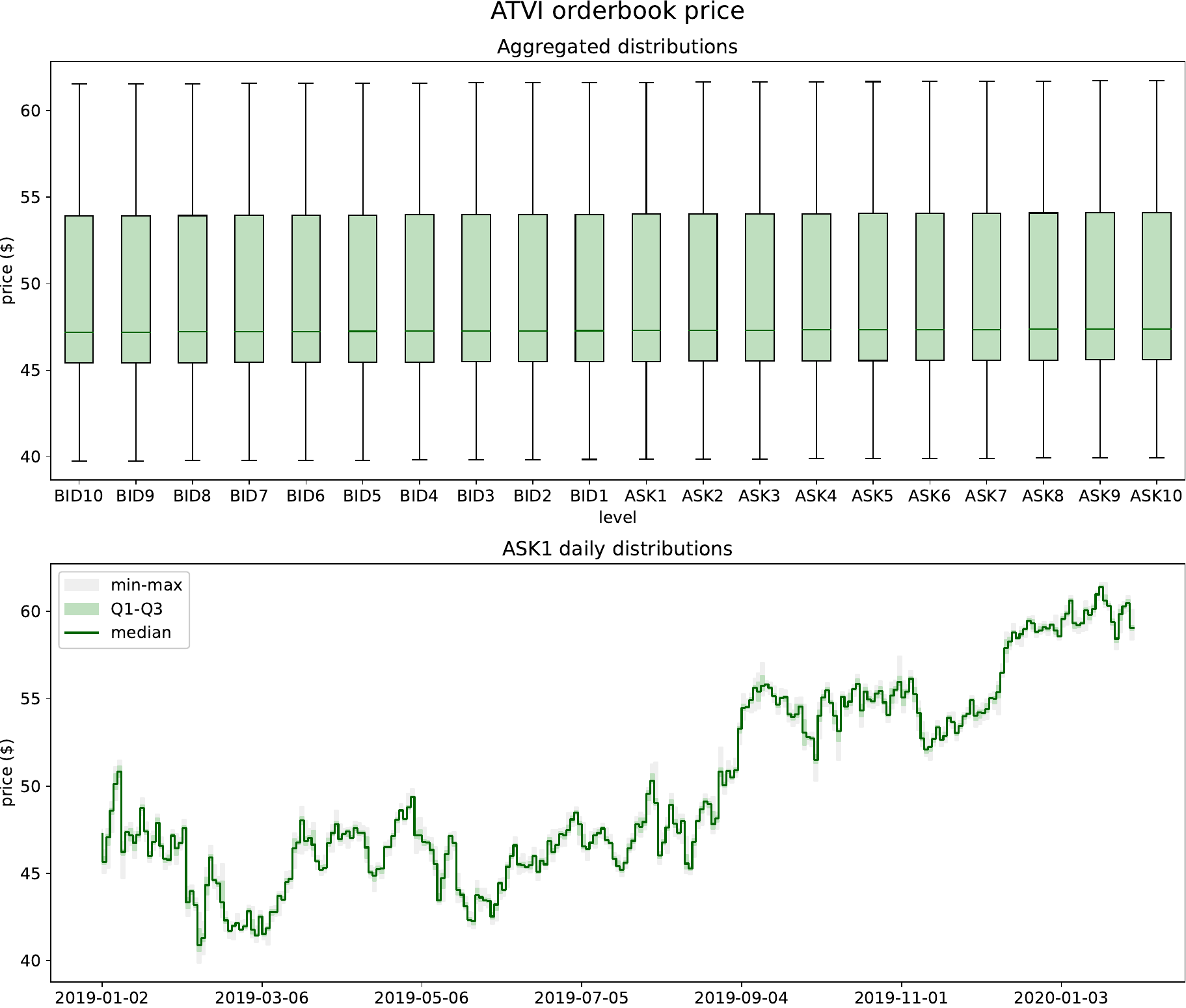}
    \caption{ATVI order book prices during the period from January 2nd, 2019 to Januray 31st, 2020.} \label{fig:ATVI_orderbook_price}
\end{figure}

\begin{figure}[!htb]
    \centering
    \includegraphics[width = 14cm]{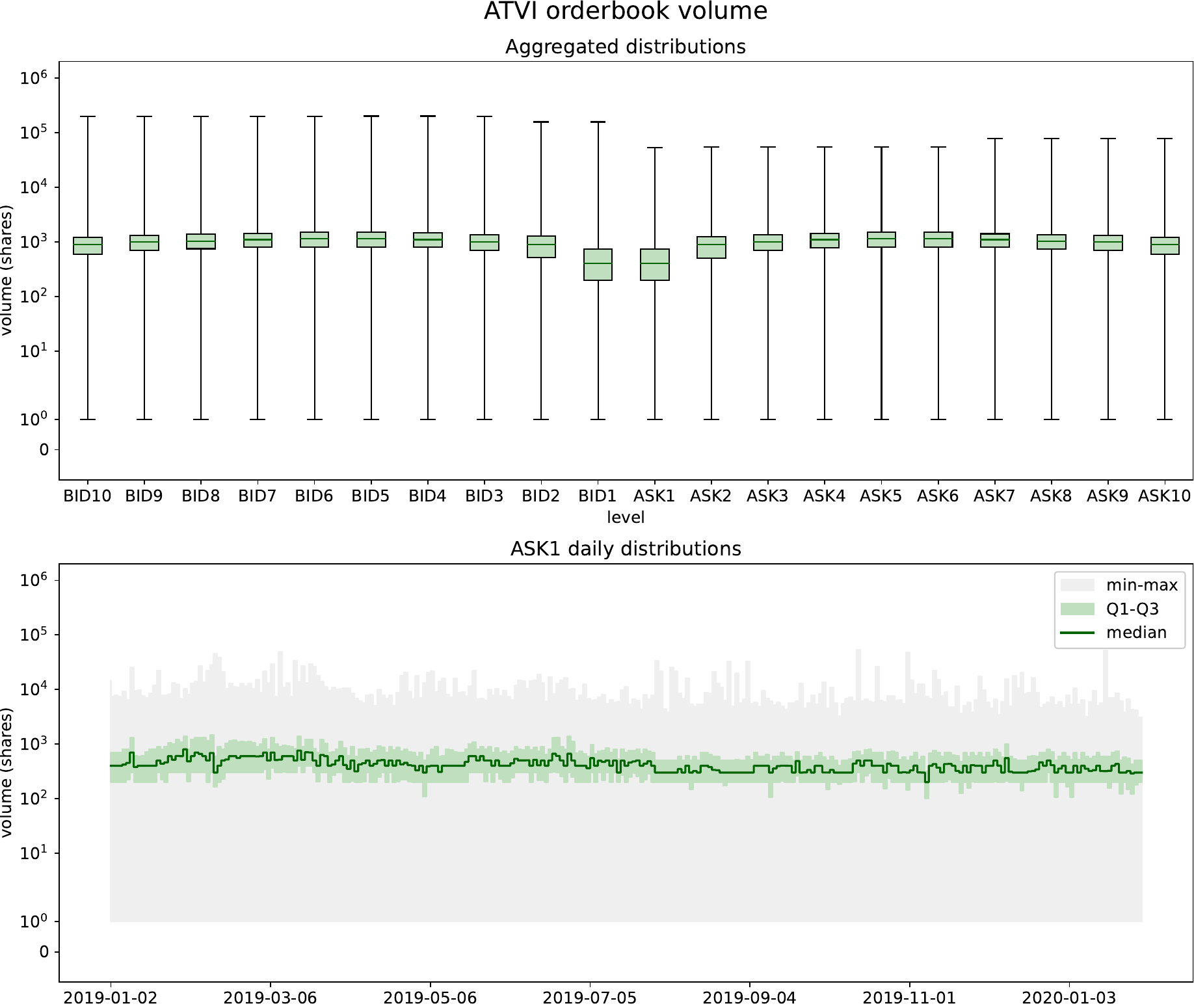}
    \caption{ATVI order book volumes during the period from January 2nd, 2019 to Januray 31st, 2020.} \label{fig:ATVI_orderbook_volume}
\end{figure}

\begin{figure}[!htb]
    \centering
    \includegraphics[width = 14cm]{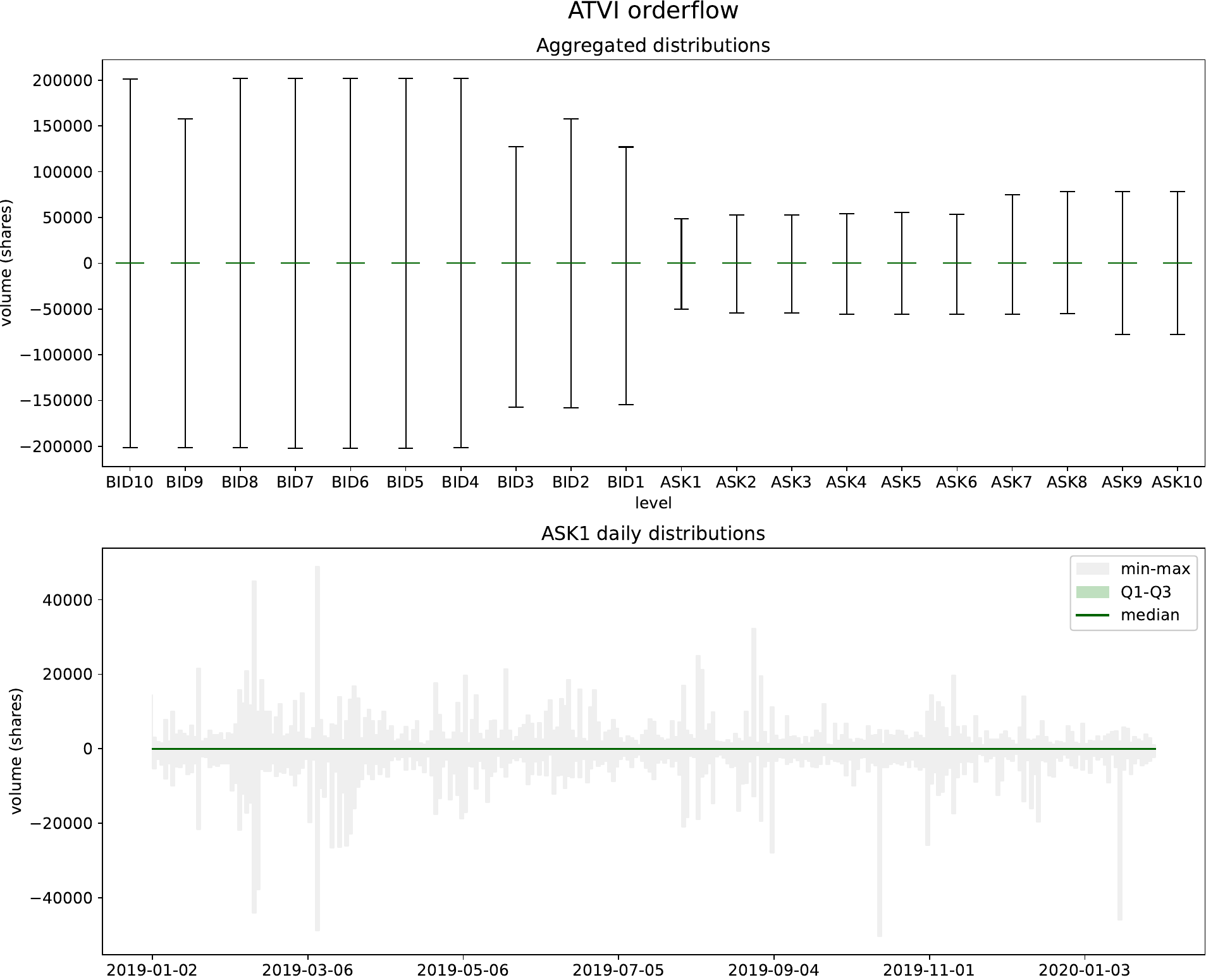}
    \caption{ATVI order flow during the period from January 2nd, 2019 to Januray 31st, 2020.} \label{fig:ATVI_orderflow}
\end{figure}

\begin{figure}[!htb]
    \centering
    \includegraphics[width = 14cm]{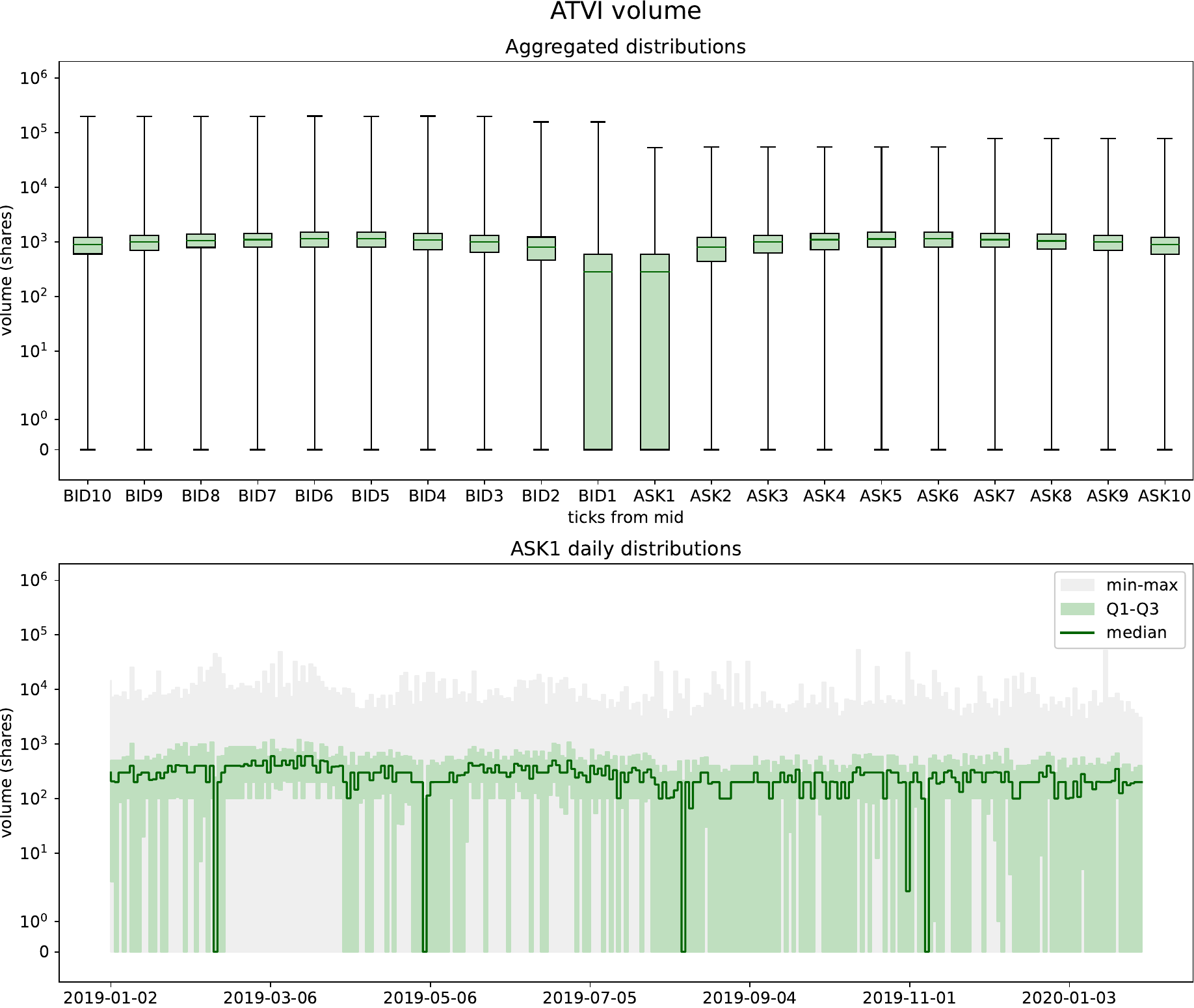}
    \caption{ATVI volumes during the period from January 2nd, 2019 to Januray 31st, 2020.} \label{fig:ATVI_volume}
\end{figure}

\begin{figure}[!htb]
    \centering
    \includegraphics[width = 14cm]{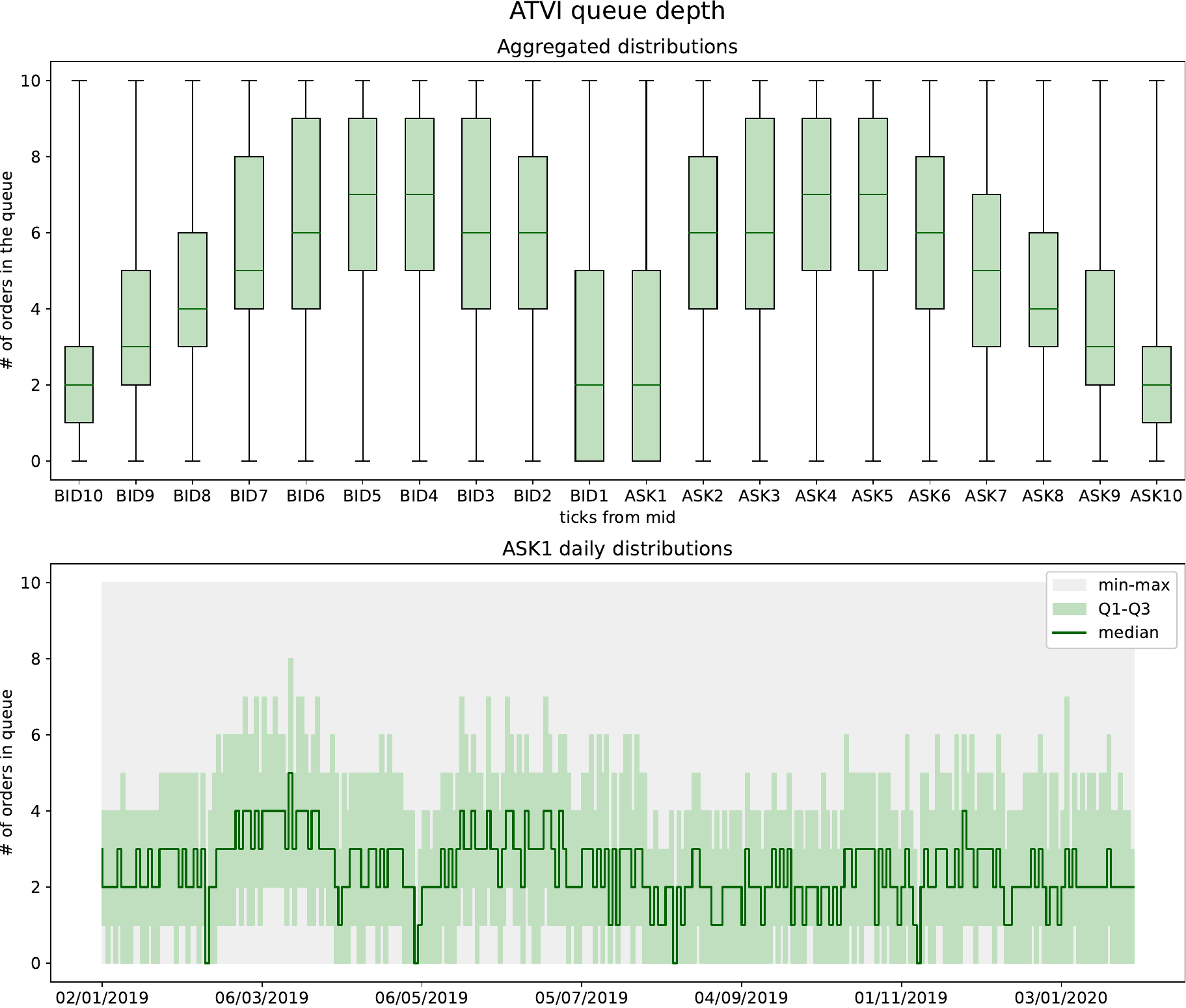}
    \caption{ATVI queue depths during the period from January 2nd, 2019 to Januray 31st, 2020.} \label{fig:ATVI_queue_depth}
\end{figure}

In Figures \ref{fig:ATVI_orderbook_price}--\ref{fig:ATVI_queue_depth} we plot the aggregate and temporal distributions of the features $\mathbf{x}_t$ described in Section \ref{sec:dataset_features}. We note the following:
\begin{itemize}
    \item As expected, order book prices display significant non-stationarity, cf.\ Figure \ref{fig:ATVI_orderbook_price}.
    \item The difference between order book volumes $v_{t,x}^{(l)}$, Figure \ref{fig:ATVI_orderbook_volume}, and volumes $s_{t,x}^{(w)}$, Figure \ref{fig:ATVI_volume}, is that in the former we measure the distance from the mid in levels, i.e.\ the volume is strictly positive, while in the latter we use all, i.e.\ also possibly empty, price ticks.
    \item In Figure \ref{fig:ATVI_orderflow} we note that order flow is zero most of the time. By definition, we have non-zero flow only when an event affects exactly that level or the whole order book shifts.
    \item To provide some insight on the distribution of L3 volume features, we plot queue depth statistics in Figure \ref{fig:ATVI_queue_depth}. We note that most queue depths are less than 9 orders long, this justifies the queue depth cutoff of length 10 chosen in our experiments.
    \item Finally, looking at Figure \ref{fig:ATVI_orderbook_volume}, Figure \ref{fig:ATVI_volume} and Figure \ref{fig:ATVI_queue_depth}, one can clearly note the difference between first level/price tick dynamics and those of the rest of the order book: active price discovery leads to lower volumes and shorter queues.
\end{itemize}

\subsubsection{Responses} \label{app:descriptive_stats_responses}
\begin{figure}[!htb]
    \centering
    \includegraphics[width = 15.5cm]{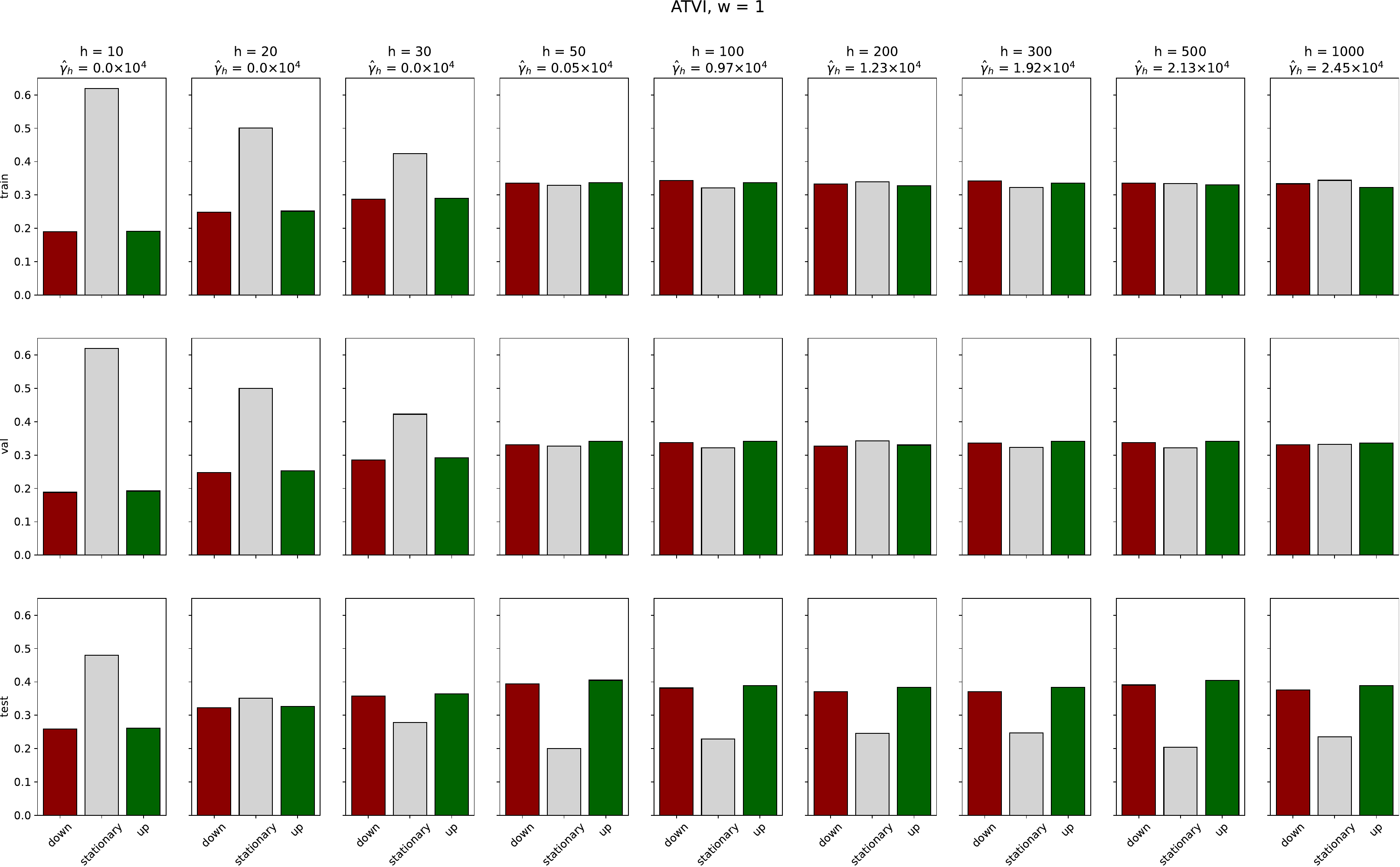}
    \caption{ATVI return labels during the first window of the experiment, i.e.\ from January 14th, 2019 to February 15th, 2019.} \label{fig:ATVI_target_distributions}
\end{figure}

In Figure \ref{fig:ATVI_target_distributions} we plot the unconditional empirical distributions of the target responses defined in Section \ref{sec:responses} for the first window $w=1$. Note that the threshold $\hat\gamma_h$ is based on the training data set only. As detailed in Section \ref{sec:experiments}, the train and validation sets are randomly partitioned from a four week period, i.e.\ $\mathcal{D}_{w,\text{train}}\cup \mathcal{D}_{w,\text{val}}$, and thus are likely to display similar volatility regimes. On the other hand, the test set $\mathcal{D}_{w,\text{test}}$ corresponds to the following week, cf.\ Figure \ref{fig:rolling_window}, and thus may exhibit a different volatility profile. By joining the train and validation distribution we obtain the \textit{unpredictive benchmark} described in Remark \ref{rem:benchmark} and used to define predictability in the experiments in Section \ref{sec:experiments}. 

\begin{figure}[!htb]
    \centering
    \includegraphics[width = 15.5cm]{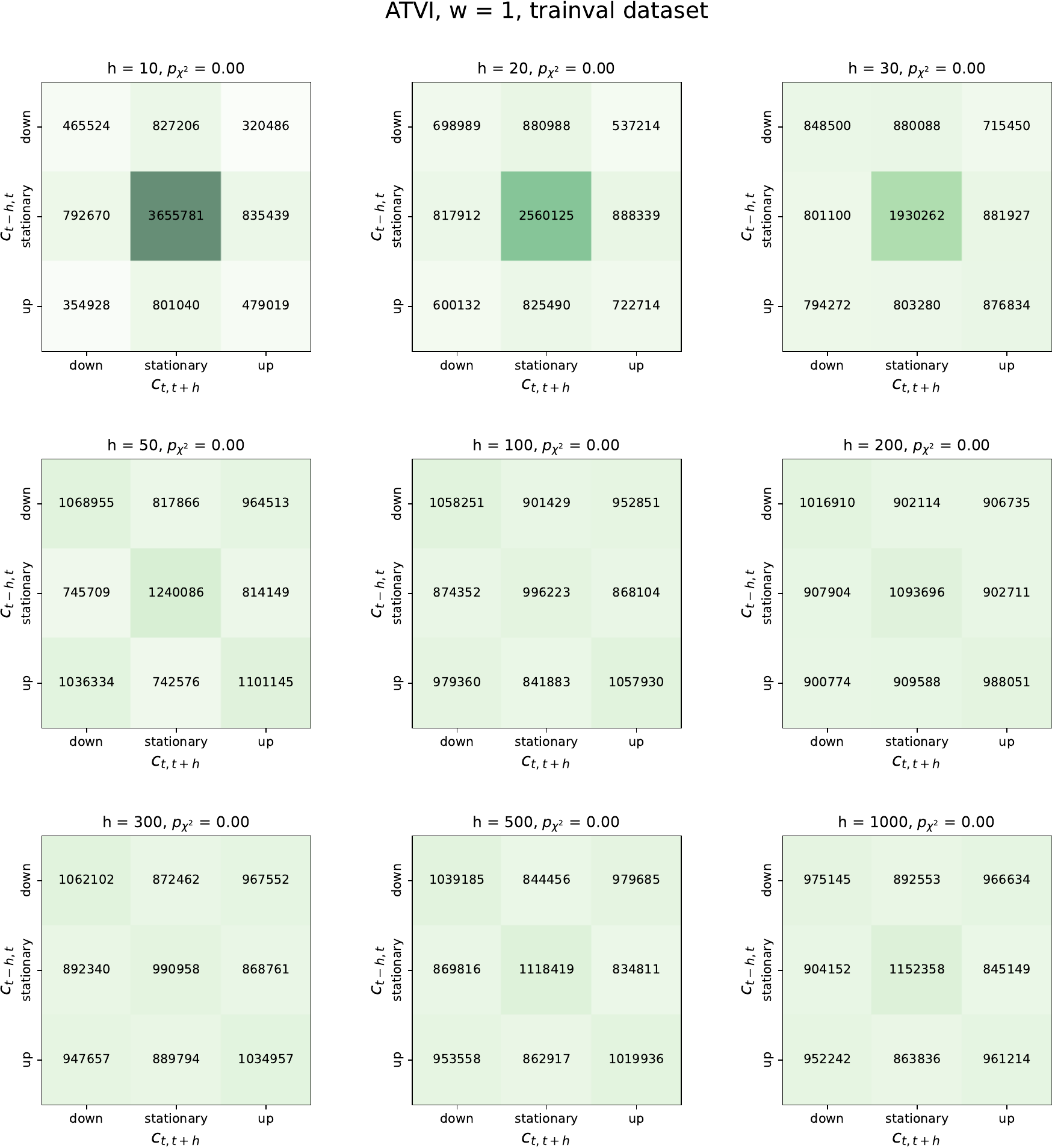}
    \caption{ATVI return labels dependence in the joint training and validation data set during the first window of the experiment, i.e.\ from January 14th, 2019 to February 15th, 2019.} \label{fig:ATVI_dependence_responses}
\end{figure}

In Figure \ref{fig:ATVI_dependence_responses} we explore the dependence structure of the return labels by plotting\footnote{To be more precise, we plot $c_{t-h-k, t}$ against $c_{t, t+h}$ in order to avoid any kind of overlap between the labels, cf.\ Section \ref{sec:responses}. We drop the $k$ for clarity of exposition and notational convenience.} the previous $h$-step return $c_{t-h, t}$ against the next $h$-step return $c_{t, t+h}$. We note that for all horizons $h\in\{10,20,30,50,100,200,300,500,1000\}$ the null hypothesis of independence is rejected by a contingency chi-square test. The fact that return labels display some form of persistency has a two-fold interpretation. On one hand, it provides evidence towards the existence of predictability in the data, i.e.\ against the i.i.d.\ Efficient Market Hypothesis on which the \textit{unpredictive benchmark} is based, cf.\ Remark \ref{rem:benchmark}. On the other, it suggests that a simple one-step ahead Markovian model of returns may display predictive power, i.e.\ a model which predicts the next price move $c_{t, t+h}$ conditional on the last return $c_{t-h,t}$ only. In this respect, we note the return label $c_{t-h,t}$ can be computed from the features $(\mathbf{x}_{t-T+1},\ldots, \mathbf{x}_t)$ for $T$ sufficiently large. In view of the universal approximation property of neural networks, any such simple model can thus be considered as `nested' in the deep learning models treated in this paper. In Appendix \ref{app:empirical_AR_model} we explore how well a specific model based on $c_{t-h,t}$ only performs when compared to the more complex deep learning models considered in this paper.

\subsection{Simpler predictive models: an empirical auto-regressive specification} \label{app:empirical_AR_model}
In this section, we explore how a simple model performs when compared to the deep learning architectures considered in this paper. We base the model on the simple empirical observation that the target return labels display time-dependence, cf.\ Appendix \ref{app:descriptive_stats_responses} and Figure \ref{fig:ATVI_dependence_responses}. We consider a one-step ahead\footnote{To avoid any look-ahead bias we actually condition on $c_{t-h-k, t}$ where $k$ is the smoothing factor introduced in Section \ref{sec:responses}. We write $c_{t-h,t}$ for clarity of exposition.} auto-regressive (AR) model, i.e.\ we aim to estimate $p_*(\star) = \mathbb{P}(c_{t, t+h} = *|c_{t-h,t} = \star)$ for $*,\star\in\{\downarrow,=,\uparrow\}$. We use a non-parametric approach, setting $\hat{p}_*(\star)$ in period $w\in\{1,\ldots, 11\}$ to be the empirical distribution of $c_{t, t+h} =*\in\{\downarrow,=,\uparrow\}$ given the previous return label was $c_{t-h, t} =\star\in\{\downarrow,=,\uparrow\}$ in the training and validation set $\mathcal{D}_{w,\text{train}} \cup \mathcal{D}_{w,\text{val}}$, i.e.\ the row normalised matrices displayed in Figure \ref{fig:ATVI_dependence_responses}. Formally, this is the maximum likelihood estimator of the conditional probabilities $\mathbb{P}(c_{t, t+h} = *|c_{t-h,t} = \star)$ given the data set $\mathcal{D}_{w,\text{train}} \cup \mathcal{D}_{w,\text{val}}$. We repeat the experiments of Sections \ref{sec:predictability_experiment} and \ref{sec:best_prediction_experiment} with this empirical AR model in the set $\mathcal{M}_0$. The updated results are reported in Table \ref{table:MCS_benchmark_with_AR} and Table \ref{table:best_MCS_with_AR} (we only report results at the 99\% confidence level, i.e.\ $\alpha=0.01$).

\begin{table}[htb]
\centering
\small
\begin{tabular}{|l||c|c|c|c|c|c|c|c|c|}
    \hline
   & $h=10$                  & $h=20$                 & $h=30$                & $h=50$                  & $h=100$                & $h=200$              & $h=300$ & $h=500$ & $h=1000$  \\
   \hhline{|=#=|=|=|=|=|=|=|=|=|}
    LILAK & \cellcolor[rgb]{ .663,  .816,  .557}0.00 & \cellcolor[rgb]{ .663,  .816,  .557}0.00 & 0.38  & 0.40  & 0.04  & 0.08  & \cellcolor[rgb]{ .663,  .816,  .557}0.00 & 0.01  & 1.00 \\
    \hline
    QRTEA & \cellcolor[rgb]{ .663,  .816,  .557}0.00 & 0.01  & 0.03  & \cellcolor[rgb]{ .663,  .816,  .557}0.00 & \cellcolor[rgb]{ .663,  .816,  .557}0.00 & \cellcolor[rgb]{ .663,  .816,  .557}0.00 & 0.06  & 0.85  & 0.26 \\
    \hline
    XRAY  & \cellcolor[rgb]{ .663,  .816,  .557}0.00 & \cellcolor[rgb]{ .663,  .816,  .557}0.00 & \cellcolor[rgb]{ .663,  .816,  .557}0.00 & \cellcolor[rgb]{ .663,  .816,  .557}0.01 & 0.09  & 0.35  & 0.35  & 0.73  & 0.85 \\
    \hline
    CHTR  & \cellcolor[rgb]{ .663,  .816,  .557}0.00 & \cellcolor[rgb]{ .663,  .816,  .557}0.00 & 0.03  & \cellcolor[rgb]{ .663,  .816,  .557}0.00 & \cellcolor[rgb]{ .663,  .816,  .557}0.00 & \cellcolor[rgb]{ .663,  .816,  .557}0.00 & 0.06  & 0.14  & 0.56 \\
    \hline
    PCAR  & \cellcolor[rgb]{ .663,  .816,  .557}0.00 & \cellcolor[rgb]{ .663,  .816,  .557}0.00 & \cellcolor[rgb]{ .663,  .816,  .557}0.00 & \cellcolor[rgb]{ .663,  .816,  .557}0.00 & 0.86  & 0.42  & 0.31  & 0.17  & 1.00 \\
    \hline
    EXC   & 0.12  & \cellcolor[rgb]{ .663,  .816,  .557}0.00 & \cellcolor[rgb]{ .663,  .816,  .557}0.00 & \cellcolor[rgb]{ .663,  .816,  .557}0.00 & \cellcolor[rgb]{ .663,  .816,  .557}0.00 & \cellcolor[rgb]{ .663,  .816,  .557}0.00 & \cellcolor[rgb]{ .663,  .816,  .557}0.00 & 0.46  & 0.38 \\
    \hline
    AAL   & 0.04  & \cellcolor[rgb]{ .663,  .816,  .557}0.00 & \cellcolor[rgb]{ .663,  .816,  .557}0.00 & \cellcolor[rgb]{ .663,  .816,  .557}0.00 & \cellcolor[rgb]{ .663,  .816,  .557}0.00 & 0.65  & 0.70  & 0.10  & 0.38 \\
    \hline
    WBA   & \cellcolor[rgb]{ .663,  .816,  .557}0.00 & \cellcolor[rgb]{ .663,  .816,  .557}0.00 & \cellcolor[rgb]{ .663,  .816,  .557}0.00 & \cellcolor[rgb]{ .663,  .816,  .557}0.00 & 0.01  & 0.23  & 0.04  & \cellcolor[rgb]{ .663,  .816,  .557}0.00 & 0.18 \\
    \hline
    ATVI  & \cellcolor[rgb]{ .663,  .816,  .557}0.00 & \cellcolor[rgb]{ .663,  .816,  .557}0.00 & \cellcolor[rgb]{ .663,  .816,  .557}0.00 & \cellcolor[rgb]{ .663,  .816,  .557}0.00 & \cellcolor[rgb]{ .663,  .816,  .557}0.01 & 0.61  & 0.34  & 0.12  & 0.03 \\
    \hline
    AAPL  & \cellcolor[rgb]{ .663,  .816,  .557}0.00 & \cellcolor[rgb]{ .663,  .816,  .557}0.00 & \cellcolor[rgb]{ .663,  .816,  .557}0.00 & 0.16  & 0.08  & 0.03  & 0.05  & 0.06  & 0.09 \\
    \hline
\end{tabular}
\caption{MCS p-values of the \textit{unpredictive benchmark} model for the 10 tickers and 9 horizons under consideration. When the p-value is low at least one of the order book-driven models statistically outperforms the \textit{unpredictive benchmark}, i.e.\ there is order book-driven predictability according to Definition \ref{def:predictability}.} \label{table:MCS_benchmark_with_AR}
\end{table}

\begin{table}[htb]
\centering
\small
\begin{tabular}{lr}
\cmidrule[\heavyrulewidth]{2-2}
            & $\alpha$ = 0.01 \\ 
\toprule
benchmark & 0\% \\
\midrule   
empirical AR model & 28\% \\
\midrule
deepOF(L1) & 12\% \\
\midrule
deepOF(L1) & 26\% \\
\midrule
deepLOB(L2) & 12\% \\
\midrule
deepOF(L2) & 81\% \\
\midrule
deepVOL(L2)  & 77\% \\
\midrule
deepVOL(L3) & 79\% \\
\bottomrule
\end{tabular}
\caption{\% of times the model is in the $\alpha$-MCS when predictability is identified at the level $\alpha=0.01$.} \label{table:best_MCS_with_AR}
\end{table}

The results reported in Table \ref{table:MCS_benchmark_with_AR} are consistent with those in Table \ref{table:MCS_benchmark}. The main differences are due to some additional predictable horizons identified by including the empirical auto-regressive model. 
This mostly occurs for stocks with low liquidity (LILAK, QRTEA, and CHTR) where the low data regime makes it more challenging for deep learning models to train effectively. 

In Table \ref{table:best_MCS_with_AR} we note the greater expressivity of deep learning models provides a significant advantage compared to the simple empirical auto-regressive model. 
While the simple AR model does display predictive power, it is often outperformed by the order book-driven deep learning models deepOF(L2) and deepVOL(L2/L3). In Table \ref{table:comparison_AR_deepVOL} we dive deeper into this analysis by reporting for each ticker-horizon 
combination whether the simple AR model or deepVOL(L3) is in the MCS of best models.

While our results suggest deepOF and deepVOL architectures significantly outperform the simple AR model, a more in-depth study comparing these specifications to the models in \cite{AitSahalia2022} is required to shed some light on the question of whether the expressivity of deep learning techniques is the key to producing good predictive models or if careful feature engineering can be as effective. This analysis is of particular relevance in applications where prediction speed plays a fundamental role, see Appendix \ref{app:latencies}.

\begin{table}[htb]
\centering
\small
\begin{tabular}{|l||c|c|c|c|c|c|c|c|c|}
    \hline
   & $h=10$                  & $h=20$                 & $h=30$                & $h=50$                  & $h=100$                & $h=200$              & $h=300$ & $h=500$ & $h=1000$  \\
   \hhline{|=#=|=|=|=|=|=|=|=|=|}
    LILAK & \cellcolor[rgb]{ .557,  .663,  .859} & \cellcolor[rgb]{ .557,  .663,  .859} & \cellcolor[rgb]{ .682,  .667,  .667} & \cellcolor[rgb]{ .682,  .667,  .667} & \cellcolor[rgb]{ .682,  .667,  .667} & \cellcolor[rgb]{ .682,  .667,  .667} & \cellcolor[rgb]{ .851,  .882,  .949} & \cellcolor[rgb]{ .682,  .667,  .667} & \cellcolor[rgb]{ .682,  .667,  .667} \\
    \hline
    QRTEA & \cellcolor[rgb]{ .851,  .882,  .949} & \cellcolor[rgb]{ .851,  .882,  .949} & \cellcolor[rgb]{ .682,  .667,  .667} & &  &  & \cellcolor[rgb]{ .682,  .667,  .667} & \cellcolor[rgb]{ .682,  .667,  .667} & \cellcolor[rgb]{ .682,  .667,  .667} \\
    \hline
    XRAY  & \cellcolor[rgb]{ .706,  .776,  .906} & \cellcolor[rgb]{ .706,  .776,  .906} & \cellcolor[rgb]{ .706,  .776,  .906} & \cellcolor[rgb]{ .557,  .663,  .859} & \cellcolor[rgb]{ .682,  .667,  .667} & \cellcolor[rgb]{ .682,  .667,  .667} & \cellcolor[rgb]{ .682,  .667,  .667} & \cellcolor[rgb]{ .682,  .667,  .667} & \cellcolor[rgb]{ .682,  .667,  .667} \\
    \hline
    CHTR  &  & \cellcolor[rgb]{ .557,  .663,  .859} & \cellcolor[rgb]{ .682,  .667,  .667} & \cellcolor[rgb]{ .851,  .882,  .949} & \cellcolor[rgb]{ .851,  .882,  .949} & \cellcolor[rgb]{ .851,  .882,  .949} & \cellcolor[rgb]{ .682,  .667,  .667} & \cellcolor[rgb]{ .682,  .667,  .667} & \cellcolor[rgb]{ .682,  .667,  .667} \\
    \hline
    PCAR  & \cellcolor[rgb]{ .706,  .776,  .906} & \cellcolor[rgb]{ .706,  .776,  .906} & \cellcolor[rgb]{ .706,  .776,  .906} & \cellcolor[rgb]{ .706,  .776,  .906} & \cellcolor[rgb]{ .682,  .667,  .667} & \cellcolor[rgb]{ .682,  .667,  .667} & \cellcolor[rgb]{ .682,  .667,  .667} & \cellcolor[rgb]{ .682,  .667,  .667} & \cellcolor[rgb]{ .682,  .667,  .667} \\
    \hline
    EXC   & \cellcolor[rgb]{ .682,  .667,  .667} & \cellcolor[rgb]{ .706,  .776,  .906} & \cellcolor[rgb]{ .706,  .776,  .906} & \cellcolor[rgb]{ .706,  .776,  .906} & \cellcolor[rgb]{ .706,  .776,  .906} & \cellcolor[rgb]{ .706,  .776,  .906} & \cellcolor[rgb]{ .706,  .776,  .906} & \cellcolor[rgb]{ .682,  .667,  .667} & \cellcolor[rgb]{ .682,  .667,  .667} \\
    \hline
    AAL   & \cellcolor[rgb]{ .682,  .667,  .667} & \cellcolor[rgb]{ .706,  .776,  .906} & \cellcolor[rgb]{ .706,  .776,  .906} & \cellcolor[rgb]{ .706,  .776,  .906} & \cellcolor[rgb]{ .706,  .776,  .906} & \cellcolor[rgb]{ .682,  .667,  .667} & \cellcolor[rgb]{ .682,  .667,  .667} & \cellcolor[rgb]{ .682,  .667,  .667} & \cellcolor[rgb]{ .682,  .667,  .667} \\
    \hline
    WBA   & \cellcolor[rgb]{ .557,  .663,  .859} & \cellcolor[rgb]{ .706,  .776,  .906} & \cellcolor[rgb]{ .706,  .776,  .906} & \cellcolor[rgb]{ .706,  .776,  .906} & \cellcolor[rgb]{ .557,  .663,  .859} & \cellcolor[rgb]{ .682,  .667,  .667} & \cellcolor[rgb]{ .682,  .667,  .667} & \cellcolor[rgb]{ .557,  .663,  .859} & \cellcolor[rgb]{ .682,  .667,  .667} \\
    \hline
    ATVI  & \cellcolor[rgb]{ .557,  .663,  .859} & \cellcolor[rgb]{ .706,  .776,  .906} & \cellcolor[rgb]{ .706,  .776,  .906} & \cellcolor[rgb]{ .706,  .776,  .906} & \cellcolor[rgb]{ .706,  .776,  .906} & \cellcolor[rgb]{ .682,  .667,  .667} & \cellcolor[rgb]{ .682,  .667,  .667} & \cellcolor[rgb]{ .682,  .667,  .667} & \cellcolor[rgb]{ .682,  .667,  .667} \\
    \hline
    AAPL  & \cellcolor[rgb]{ .706,  .776,  .906} & \cellcolor[rgb]{ .706,  .776,  .906} & \cellcolor[rgb]{ .706,  .776,  .906} & \cellcolor[rgb]{ .682,  .667,  .667} & \cellcolor[rgb]{ .682,  .667,  .667} & \cellcolor[rgb]{ .682,  .667,  .667} & \cellcolor[rgb]{ .682,  .667,  .667} & \cellcolor[rgb]{ .682,  .667,  .667} & \cellcolor[rgb]{ .682,  .667,  .667} \\
    \hline
\end{tabular}
\caption{Comparison between emprical AR model and deepVOL(L3) MCS results, $\alpha=0.01$. Shading indicates whether the empirical AR model and/or deepVOL(L3) is in the set of best models. White: neither, light blue: empirical AR model only, blue: deepVOL(L3), dark blue: both. Grey shading indicates the ticker-horizon combination was not deemed predictable at the 99\% confidence level.} \label{table:comparison_AR_deepVOL}
\end{table}

\subsection{Return definition} \label{app:return_def}
The way returns are defined is often an overlooked topic which inevitably has a significant effect on model evaluation. Here we state two alternative definitions to that given in Section \ref{sec:responses} taken from the literature and discuss their properties. In all cases, the mid-price is treated as the ``true'' price and returns are defined relative to it. Recall this is not a tradable price.
\begin{itemize}
    \item In \citet{FI2010} the authors define the return at horizon $h$ as
    \begin{align*}
    r_{t,t+h} = \frac{\overline{m}_{t, t+h} - m_t}{m_t},\
    \overline{m}_{t, t+h} = \frac{1}{h}\sum_{i=1}^h m_{t+i},
    \end{align*}
    where $m_t$ denotes the mid-price at time $t$. This definition of return compares the smoothed mid-price over the next $h$ time steps to the current mid-price. In this case, the return can be equivalently understood as the average mid-to-mid return over the next $h$ time steps. As the horizon $h$ becomes longer, so does the window over which we smooth the returns. This has a couple of practical disadvantages in terms of the predictability questions we wish to explore. First, predictability over $h=100$ time steps might be due to changes of the mid-price over the first, say, $h=50$ time steps, and thus it becomes difficult to discuss questions of persistency in predictability. Second, the definition produces correlation between returns at different horizons, $r_{t+h_1}$ and $r_{t+h_2}$, inducing a hidden bias in multi-horizon models.
    \item In \citet{deepLOB} and \citet{deepLOB_multihorizon} the authors define the return at horizon $h$ by
    \begin{align*}
        r_{t,t+h} = \frac{\overline{m}_{t, t+h} - \overline{m}_{t-h, t} }{\overline{m}_{t-h, t}},\
        \overline{m}_{t, t+h} = \frac{1}{h}\sum_{i=1}^h m_{t+i},\
        \overline{m}_{t-h, t} = \frac{1}{h}\sum_{i=0}^{h-1} m_{t-i},
    \end{align*}
    where $m_t$ denotes the mid-price at time $t$. This definition has similar drawbacks to the previous one. But, additionally, we believe it may suffer from look-ahead bias: for example, if the mid-price has been going up in the past $h$ steps (an information which is included in our covariates), then it is more likely the return $r_{t,t+h}$ will be positive. 
\end{itemize}
The definition of returns used in this paper, cf.\ Section \ref{sec:responses}, does not suffer from the issues identified for these two return specifications.

\begin{remark}
    The definition of returns intrinsically depends on the clock used to measure time. Note that in \citet{deepOF} the authors use a physical time clock with stock-specific horizon $h$. In their setting, the authors use simple un-smoothed mid-price returns,
    \begin{align*}
        r_{t,t+h} = \frac{{m}_{t+h} - {m}_{t} }{{m}_{t}}.
    \end{align*}
    With our choice of order book-driven clock, even assuming mid-mid trading, there is no way of placing a trade exactly $h$ order book events ahead, i.e.\ the prediction horizon is random. Smoothing the exit price can thus be understood as averaging out the uncertainty in the execution time. Moreover, in our work, we aim to investigate questions regarding structural market predictability. In our setting, smoothing mid-prices leads to better estimates of the true (latent) prices by weakening idiosyncratic noise effects. As pointed out in the conclusions in Section \ref{sec:conclusions} though, in practical trading applications it may be more appropriate to consider a physical time clock and un-smoothed returns.
\end{remark}

\subsection{Infrastructure latency} \label{app:latencies}
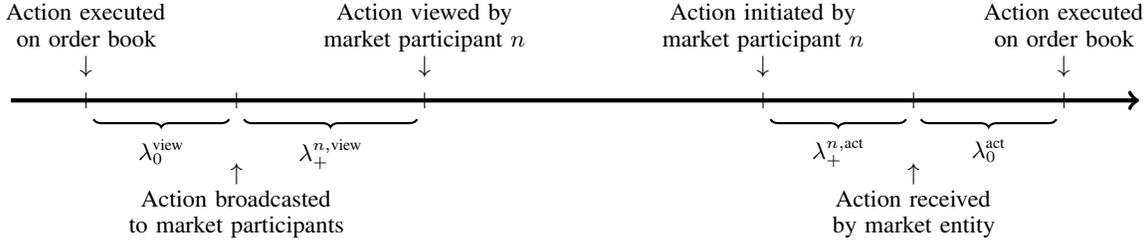
\begin{figure}[ht]
    \centering
    \begin{tikzpicture}
        \small 
        
        \draw[ultra thick, ->] (0,0) -- (15,0);
        
        \foreach \x in {1,3,5.5, 10,12,14}
        \draw (\x cm,3pt) -- (\x cm,-3pt);
        
        \draw[ultra thick] (1,0) node[text width=3cm, above=5pt, align=center] {Action executed on order book \\ $\downarrow$};]
        \draw[black, thick, decorate, decoration={brace, mirror, amplitude=3pt}] (1.1,-0.25) -- (2.9,-0.25) node [black, midway, below=4pt] {$\lambda^{\text{view}}_0$};
        \draw[ultra thick] (3,0) node[text width=3.5cm, below=20pt, align=center] {$\uparrow$ \\ Action broadcasted to market participants};]
            \draw[black, thick, decorate, decoration={brace, mirror, amplitude=3pt}] (3.1,-0.25) -- (5.4,-0.25) node [black, midway, below=4pt] {$\lambda^{n,\text{view}}_+$};
            \draw[ultra thick] (5.5,0) node[text width=3.5cm, above=5pt, align=center] {Action viewed by market participant $n$ \\ $\downarrow$};]

        \draw[ultra thick] (10,0) node[text width=3.5cm, above=5pt, align=center] {Action initiated by market participant $n$ \\ $\downarrow$};]
        \draw[black, thick, decorate, decoration={brace, mirror, amplitude=3pt}] (10.1,-0.25) -- (11.9,-0.25) node [black, midway, below=4pt] {$\lambda^{n,\text{act}}_+$};
        \draw[ultra thick] (12,0) node[text width=3cm, below=20pt, align=center] {$\uparrow$ \\ Action received by market entity};]
            \draw[black, thick, decorate, decoration={brace, mirror, amplitude=3pt}] (12.1,-0.25) -- (13.9,-0.25) node [black, midway, below=4pt] {$\lambda^{\text{act}}_0$};
            \draw[ultra thick] (14,0) node[text width=3cm, above=5pt, align=center] {Action executed on order book \\ $\downarrow$};]
    \end{tikzpicture}
    \caption{Market-wide and idiosyncratic latencies.} \label{fig:latencies}
\end{figure}

In this paper, we explored the predictive value of order book data by assuming instantaneous access to such information. In practice, due to technological constraints, market participants experience varying degrees of delay when engaging with the order book. Let us write $\lambda^{n, \text{view}}$ and $\lambda^{n, \text{act}}$ for the latencies that market participant $n$ experiences when viewing and acting on the order book respectively. For $x\in\{\text{view},\text{act}\}$ we can decompose
\[ \lambda^{n, x} = \lambda^{x}_0 + \lambda^{n,x}_+, \]
where $\lambda^{\text{act}}_0$ is the time it takes the financial entity running the market, e.g.\ the Nasdaq, to execute an action on the order book once received and $\lambda^{\text{view}}_0$ is the time it takes the market entity to send out the information regarding an order book update. Market participants make large efforts to reduce their idiosyncratic latencies $\lambda^{n,x}_+$ for $x\in\{\text{view}, \text{act}\}$ by optimizing software and hardware, one prototypical example being co-location. Note that $\lambda^{n,\text{act}}_+$ also includes the time needed to take a trading decision which, in our setting, requires a forward pass through a pre-trained predictive model, e.g.\ deepLOB/deepOF/deepVOL. Figure \ref{fig:latencies} represents how market-wide and idiosyncratic latencies affect market participants.

When considering practical trading applications one should thus take into account the effect these latencies may have on the definition of returns, cf.\ Section \ref{sec:responses} and Appendix \ref{app:return_def}. For example, denoting by $t=1,2,\ldots$ the order book-driven clock considered in this paper, one may want to predict the return between physical times $\tau_t+\lambda^{n,\text{tot}}$ and $\tau_{t+h}+\lambda^{n,\text{tot}}$ where 
\[ \lambda^{n,\text{tot}} = \lambda^{n,\text{view}} + \lambda^{n,\text{act}} = \lambda_0^{\text{tot}} + \lambda^{n,\text{tot}}_+,\]
is the total round trip latency time to view and act once the order book clock ticks and $\tau:\mathbb{N}\rightarrow[0,\infty)$ is the increasing random function mapping the order book clock to physical time. An interesting analysis in this setting would be to analyze how predictability varies as a function of $\lambda^{n,\text{tot}}_+ = \lambda^{n,\text{view}}_+ + \lambda^{n,\text{act}}_+$.

In order to quantify the magnitude of these time lags and better understand their impact we briefly summarize some infrastructure tests published by market participants.
\begin{itemize}
    \item As of 2023, the Nasdaq reports the door-to-door speed of its matching engine, i.e.\ $\lambda_0^{\text{tot}} = \lambda_0^{\text{act}} + \lambda_0^{\text{view}}$, to be `sub-40 $\mu$s, with the fastest production implementation at 14 $\mu$s' \citep{Nasdaq_matching_engine}. Studies from 2015 report the door-to-door latencies of Brazilian and European futures exchanges in the order of a few hundred microseconds \citep{latency_asset_prices, need_for_speed}.
    \item Leading industry producers of co-located network adapters and feed handlers report the total speed for processing an order book message and sending a trade action, i.e.\ tick-to-trade latency $\lambda^{n,\text{tot}}_+ = \lambda^{n,\text{view}}_+ + \lambda^{n,\text{act}}_+$, to be of the order of a couple of microseconds ($\mu$s) \citep{Enyx_Nasdaq_performance, CSPI_tick_to_trade}. This is achieved via the use of specific hardware known as FPGA cards. These figures assume relatively simple trading decision applications, when working with more sophisticated predictive models, such as the ones considered in this paper, one should account for a couple of hundred microseconds in tick-to-trade latency, see Table III in \cite{deepLOB} for unoptimized forward-pass run times.
\end{itemize}
The total latency $\lambda^{k,\text{tot}}$ is thus less than a fraction of a millisecond (ms) while the average time elapsed between consecutive price changes in our data ranges from 0.18 to 8.38 seconds depending on the liquidity of the stock, cf.\ Table \ref{table:stock_characteristics}. The effect of infrastructural latencies on the definition of returns, cf.\ Section \ref{sec:responses}, is thus negligible.

\subsection{Stock selection} \label{app:stock_selection}
Ideally, we would like to work with the same set of stocks as in \citet{deepOF}. Due to computational limitations, we were able to conduct experiments only on a subset of 10 tickers from the total 115 NASDAQ stocks considered in the original paper. We selected 10 stocks with diverse liquidity characteristics, hopefully providing a sufficiently representative sub-sample of the whole set.

To choose the sub-sample, we use the stock characteristics provided in \citet{deepOF}, Table 6. For each of the liquidity characteristics -- Updates, Trades, Price Changes, Spread -- we compute a sub-score based on the characteristic's rank. For example, the stock with the most updates is assigned an updates score of 1, while the stock with the least number of updates is assigned an updates score of 0. The characteristic-specific scores are then averaged to obtain a general ``liquidity score'' for each stock. The 10 chosen stocks correspond to the 10 evenly spaced quantiles of the ``liquidity score''. The liquidity characteristics of the 10 chosen stocks are reported in Table \ref{table:stock_characteristics}.

\subsection{Model details} \label{app:model_details}
In the experiments we use the model architectures reported in Table \ref{table:deepLOBdeepOF} and Table \ref{table:deepVOL}. To speed up the training procedure, we use batch normalization \citep{batch_norm} (with momentum = 0.6) after every convolutional layer and inception block. To prevent overfitting, we use a dropout layer \citep{dropout} (with noise = 0.2) positioned after the inception module.

All the models are implemented with float32 precision policy: at every layer float32 is used as computation and variable data type. Loss and gradient computations are also carried out with float32 precision.

\begin{landscape}
\vspace*{\fill}
\tiny
\pagestyle{empty}
\setlength\tabcolsep{1.5pt}
\centering
\centerline{
\begin{tabular}{ |p{2.5cm}|p{2cm}||p{3.5cm}|p{5.5cm}|p{1.4cm}||p{3.5cm}|p{5.5cm}|p{1.4cm}|}
 \hline
 \multicolumn{8}{|c|}{Model Architectures} 
 \\
 \hline
 \hline 
 \multicolumn{2}{|c||}{Model}                                                                  & \multicolumn{3}{|c|}{DeepLOB}                      & \multicolumn{3}{|c|}{DeepOF}                                                                                                                                                                                                     \\
 \hline
 \multicolumn{2}{|c||}{}                                                                       & Architecture                                       & Description                                                                                                                                                                            & Output Size 
                                                                                               & Architecture                                       & Description                                                                                                                                                                            & Output Size                             \\
 \hline
 \multicolumn{2}{|l||}{Input Layer}                                                            &                                                    & \shortstack[l]{Raw order book states, \\ $\left\{(p^{(l)}_{a,t-\tau}, v^{(l)}_{a, t-\tau}, p^{(l)}_{b, t-\tau}, v^{(l)}_{b,t-\tau})\right\}_{\substack{\tau=0,\ldots,T-1 \\ l=1,\ldots,L\quad}}$\\ \vspace{-0.05cm}}     & $(T, 4L, 1) $
                                                                                               &                                                    & \shortstack[l]{Multi-level order flow, \\ $\left\{(aOF^{(l)}_{t-\tau}, bOF^{(l)}_{t-\tau}) \right\}_{\substack{\tau=0,\ldots,T-1 \\ l=1,\ldots,L\quad}}$\\ \vspace{-0.05cm}}                                         & $(T, 2L, 1)  $                          \\
 \hline
 \multirow{3}{*}{First Convolutional layer}                             & Spatial convolution  & 32 (1$\times$2) filters, (1$\times$2) stride       & Convolve price and volume information                                                                                                                                                  & $(T, 2L, 32) $
                                                                                               &                                                    &                                                                                                                                                                                        &                                         \\
                                                                        & Temporal convolution & 32 (4$\times$1) filters, padding                   & Convolve through four time-steps                                                                                                                                                       & $(T, 2L, 32) $
                                                                                               &                                                    &                                                                                                                                                                                        &                                         \\
                                                                        & Temporal convolution & 32 (4$\times$1) filters, padding                   & Convolve through four time-steps                                                                                                                                                       & $(T, 2L, 32) $
                                                                                               &                                                    &                                                                                                                                                                                        &                                         \\
 \hline
 \multirow{3}{*}{Second Convolutional layer}                            & Spatial convolution  & 32 (1$\times$2) filters, (1$\times$2) stride       & Convolve bid and ask price-volume information                                                                                                                                          & $(T, L, 32) $
                                                                                               & 32 (1$\times$2) filters, (1$\times$2) stride       & Convolve order flow information                                                                                                                                                        & $(T, L, 32) $                           \\
                                                                        & Temporal convolution & 32 (4$\times$1) filters, padding                   & Convolve through four time-steps                                                                                                                                                       & $(T, L, 32) $
                                                                                               & 32 (4$\times$1) filters, padding                   & Convolve through four time-steps                                                                                                                                                       & $(T, L, 32) $                          \\
                                                                        & Temporal convolution & 32 (4$\times$1) filters, padding                   & Convolve through four time-steps                                                                                                                                                       & $(T, L, 32) $
                                                                                               & 32 (4$\times$1) filters, padding                   & Convolve through four time-steps                                                                                                                                                       & $(T, L, 32) $                          \\
 \hline
 \multirow{3}{*}{Third Convolutional layer}                             & Spatial convolution  & 32 (1$\times$L) filters, (1$\times$1) stride       & Convolve information through the whole order book                                                                                                                                      & $(T, 1, 32) $ 
                                                                                               & 32 (1$\times$L) filters, (1$\times$1) stride       & Convolve information through the whole order book                                                                                                                                      & $(T, 1, 32) $                           \\
                                                                        & Temporal convolution & 32 (4$\times$1) filters, padding                   & Convolve through four time-steps                                                                                                                                                       & $(T, 1, 32) $ 
                                                                                               & 32 (4$\times$1) filters, padding                   & Convolve through four time-steps                                                                                                                                                       & $(T, 1, 32) $                           \\
                                                                        & Temporal convolution & 32 (4$\times$1) filters, padding                   & Convolve through four time-steps                                                                                                                                                       & $(T, 1, 32) $ 
                                                                                               & 32 (4$\times$1) filters, padding                   & Convolve through four time-steps                                                                                                                                                       & $(T, 1, 32) $                           \\
 \hline
 \multirow{6}{*}{\shortstack[l]{Inception module \\ (parallel blocks)}} & Temporal convolution & 64 (1$\times$1) filters                            & Increase dimensionality (Network-in-Network)                                                                                                                                           & $(T, 1, 64) $ 
                                                                                               & 64 (1$\times$1) filters                            & Increase dimensionality (Network-in-Network)                                                                                                                                           & $(T, 1, 64) $                           \\
                                                                        & Temporal convolution & 64 (3$\times$1) filters, padding                   & Convolve through three time-steps, "MA(3)"                                                                                                                                             & $(T, 1, 64) $ 
                                                                                               & 64 (3$\times$1) filters, padding                   & Convolve through three time-steps, "MA(3)"                                                                                                                                             & $(T, 1, 64) $                           \\
                                                                                               \cline{2-8}
                                                                        & Temporal convolution & 64 (1$\times$1) filters                            & Increase dimensionality (Network-in-Network)                                                                                                                                           & $(T, 1, 64) $
                                                                                               & 64 (1$\times$1) filters                            & Increase dimensionality (Network-in-Network)                                                                                                                                           & $(T, 1, 64) $                           \\
                                                                        & Temporal convolution & 64 (5$\times$1) filters, padding                   & Convolve through five time-steps, "MA(5)"                                                                                                                                              & $(T, 1, 64) $ 
                                                                                               & 64 (5$\times$1) filters, padding                   & Convolve through five time-steps, "MA(5)"                                                                                                                                              & $(T, 1, 64) $                           \\
                                                                                               \cline{2-8}
                                                                        & Temporal maxpooling  & (3$\times$1) pool size, (1$\times$1) stride \& padding & Rolling maximum, three time-steps                                                                                                                                                  & $(T, 1, 32) $ 
                                                                                               & (3$\times$1) pool size, (1$\times$1) stride \& padding & Rolling maximum, three time-steps                                                                                                                                                  & $(T, 1, 32) $                           \\
                                                                        & Temporal convolution & 64 (1$\times$1) filters                            & Increase dimensionality (Network-in-Network)                                                                                                                                           & $(T, 1, 64) $ 
                                                                                               & 64 (4$\times$1) filters                            & Increase dimensionality (Network-in-Network)                                                                                                                                           & $(T, 1, 64) $                           \\
 \hline        
 Concatenate and reshape                                                &                      &                                                    & Concatenate and reshape the outputs of the inception module to obtain an $\mathbb{R}^{192}$-valued time series of length $T$                                                           & $(T, 192)  $
                                                                                               &                                                    & Concatenate and reshape the outputs of the inception module to obtain an $\mathbb{R}^{192}$-valued time series of length $T$                                                           & $(T, 192)  $                            \\
 \hline
 LSTM module                                                            & LSTM                 & 64 hidden units                                    & Capture longer term dependencies                                                                                                                                                       & $(64)  $
                                                                                               & 64 hidden units                                    & Capture longer term dependencies                                                                                                                                                       & $(64)  $                                   \\
 \hline
 (a) Output layer                                                       & Dense layer          & Softmax activation                                 & Produce classification output                                                                                                                                                          & $(3)  $
                                                                                               & Softmax activation                                 & Produce classification output                                                                                                                                                          & $(3)  $                                   \\
 \hline
 (b) Decoder module                       & \shortstack[l]{LSTM decoder \\ (seq2seq/attention)}& \shortstack[l]{64 hidden units, \\softmax activation}                & Produce sequential classification output                                                                                                                             & $(5, 3)  $
                                                                                               & \shortstack[l]{64 hidden units, \\softmax activation}                & Produce sequential classification output                                                                                                                             & $(5, 3)  $                                \\
 \hline

\end{tabular}}
\captionof{table}{DeepLOB and DeepOF network description.\\ (a) Single horizon, (b) Multi-horizon.} \label{table:deepLOBdeepOF}

\vspace*{\fill}

\newpage

\vspace*{\fill}

\centerline{
\begin{tabular}{ |p{2.5cm}|p{2cm}||p{3.9cm}|p{5.5cm}|p{1.55cm}||p{3.5cm}|p{5.5cm}|p{1.55cm}|}
 \hline
 \multicolumn{8}{|c|}{Model Architectures} 
 \\
 \hline
 \hline 
 \multicolumn{2}{|c||}{Model}                                                                  & \multicolumn{3}{|c|}{DeepVOL}                      & \multicolumn{3}{|c|}{DeepVOL L3}                                                                                                                                                                                                     \\
 \hline
 \multicolumn{2}{|c||}{}                                                                       & Architecture                                       & Description                                                                                                                                                                            & Output Size 
                                                                                               & Architecture                                       & Description                                                                                                                                                                            & Output Size                             \\
 \hline
 \multicolumn{2}{|l||}{Input Layer}                                                            &                                                    & \shortstack[l]{Volume representation, \\ $\left\{s^{(j)}_{x,t-\tau}\right\}_{\substack{\tau=0,\ldots,T-1,\,j=1,\ldots,W \\ x\in\{a,b\}}}$\\ \vspace{0.1cm}}                                                   & $(T, W, 2, 1) $
                                                                                               &                                                    & \shortstack[l]{\vspace{-0.025cm} \\ L3 volume representation, \\ $\left\{(q^{(j, k)}_{x,t-\tau} \right\}_{\substack{\tau=0,\ldots,T-1,\, j=1,\ldots,W \\ x\in\{a,b\}, k=1,\ldots, D\quad\quad\ }}$\\ \vspace{-0.05cm}}                            & $(T, W, 2, D, 1)$                        \\
 \hline
 First Convolutional layer                                              & Spatial convolution  &                                                    &                                                                                                                                                                                        &  
                                                                                               & 32 (1$\times$1$\times$D) filters, (1$\times$1$\times$1) stride & At each price, convolve the queue volumes                                                                                                                                  & $(T, W, 2, 32)$                         \\
 \hline
 \multirow{3}{*}{Second Convolutional layer}                            & Spatial convolution  & 32 (1$\times$2$\times$2) filters, (1$\times$1$\times$1) stride & Convolve volumes on opposite sides of the mid                                                                                                                              & $(T, W-1, 32)$ 
                                                                                               & 32 (1$\times$2$\times$2) filters, (1$\times$1$\times$1) stride & Convolve volumes on opposite sides of the mid                                                                                                                              & $(T, W-1, 32)$                            \\
                                                                        & Temporal convolution & 32 (4$\times$1) filters, padding                   & Convolve through four time-steps                                                                                                                                                       & $(T, W-1, 32)$  
                                                                                               & 32 (4$\times$1) filters, padding                   & Convolve through four time-steps                                                                                                                                                       & $(T, W-1, 32)$                           \\
                                                                        & Temporal convolution & 32 (4$\times$1) filters, padding                   & Convolve through four time-steps                                                                                                                                                       & $(T, W-1, 32)$ 
                                                                                               & 32 (4$\times$1) filters, padding                   & Convolve through four time-steps                                                                                                                                                       & $(T, W-1, 32)$                            \\
 \hline
 \multirow{3}{*}{Third Convolutional layer}                             & Spatial convolution  & 32 (1$\times$W-1) filters, (1$\times$1) stride     & Convolve information through the whole order book                                                                                                                                      & $(T, 1, 32)$ 
                                                                                               & 32 (1$\times$W-1) filters, (1$\times$1) stride     & Convolve information through the whole order book                                                                                                                                      & $(T, 1, 32)$                            \\
                                                                        & Temporal convolution & 32 (4$\times$1) filters, padding                   & Convolve through four time-steps                                                                                                                                                       & $(T, 1, 32)$  
                                                                                               & 32 (4$\times$1) filters, padding                   & Convolve through four time-steps                                                                                                                                                       & $(T, 1, 32)$                            \\
                                                                        & Temporal convolution & 32 (4$\times$1) filters, padding                   & Convolve through four time-steps                                                                                                                                                       & $(T, 1, 32)$  
                                                                                               & 32 (4$\times$1) filters, padding                   & Convolve through four time-steps                                                                                                                                                       & $(T, 1, 32)$                            \\
 \hline
 \multirow{6}{*}{\shortstack[l]{Inception module \\ (parallel blocks)}} & Temporal convolution & 64 (1$\times$1) filters                            & Increase dimensionality (Network-in-Network)                                                                                                                                           & $(T, 1, 64)$
                                                                                               & 64 (1$\times$1) filters                            & Increase dimensionality (Network-in-Network)                                                                                                                                           & $(T, 1, 64)$                            \\
                                                                        & Temporal convolution & 64 (3$\times$1) filters, padding                   & Convolve through three time-steps, "MA(3)"                                                                                                                                             & $(T, 1, 64)$  
                                                                                               & 64 (3$\times$1) filters, padding                   & Convolve through three time-steps, "MA(3)"                                                                                                                                             & $(T, 1, 64)$                            \\
                                                                                               \cline{2-8}
                                                                        & Temporal convolution & 64 (1$\times$1) filters                            & Increase dimensionality (Network-in-Network)                                                                                                                                           & $(T, 1, 64)$  
                                                                                               & 64 (1$\times$1) filters                            & Increase dimensionality (Network-in-Network)                                                                                                                                           & $(T, 1, 64)$                            \\
                                                                        & Temporal convolution & 64 (5$\times$1) filters, padding                   & Convolve through five time-steps, "MA(5)"                                                                                                                                              & $(T, 1, 64)$  
                                                                                               & 64 (5$\times$1) filters, padding                   & Convolve through five time-steps, "MA(5)"                                                                                                                                              & $(T, 1, 64)$                            \\
                                                                                               \cline{2-8}
                                                                        & Temporal maxpooling  & (3$\times$1) pool size, (1$\times$1) stride \& padding & Rolling maximum, three time-steps                                                                                                                                                  & $(T, 1, 32)$  
                                                                                               & (3$\times$1) pool size, (1$\times$1) stride \& padding & Rolling maximum, three time-steps                                                                                                                                                  & $(T, 1, 32)$                            \\
                                                                        & Temporal convolution & 64 (1$\times$1) filters                            & Increase dimensionality (Network-in-Network)                                                                                                                                           & $(T, 1, 64)$  
                                                                                               & 64 (4$\times$1) filters                            & Increase dimensionality (Network-in-Network)                                                                                                                                           & $(T, 1, 64)$                            \\
 \hline        
 Concatenate and reshape                                                &                      &                                                    & Concatenate and reshape the outputs of the inception module to obtain an $\mathbb{R}^{192}$-valued time series of length $T$                                                           & $(T, 192)$  
                                                                                               &                                                    & Concatenate and reshape the outputs of the inception module to obtain an $\mathbb{R}^{192}$-valued time series of length $T$                                                           & $(T, 192)$                              \\
 \hline
 LSTM module                                                            & LSTM                 & 64 hidden units                                    & Capture longer term dependencies                                                                                                                                                       & $(64)$  
                                                                                               & 64 hidden units                                    & Capture longer term dependencies                                                                                                                                                       & $(64)$                                     \\
 \hline
 (a) Output layer                                                       & Dense layer          & Softmax activation                                 & Produce classification output                                                                                                                                                          & $(3)$  
                                                                                               & Softmax activation                                 & Produce classification output                                                                                                                                                          & $(3)$                                     \\
 \hline
 (b) Decoder module                       & \shortstack[l]{LSTM decoder \\ (seq2seq/attention)}& \shortstack[l]{64 hidden units, \\softmax activation}                & Produce sequential classification output                                                                                                                             & $(5, 3)$  
                                                                                               & \shortstack[l]{64 hidden units, \\softmax activation}                & Produce sequential classification output                                                                                                                             & $(5, 3)$                                  \\
 \hline

\end{tabular}}
\captionof{table}{DeepVOL and DeepVOL L3 network description.\\ (a) Single horizon, (b) Multi-horizon.} \label{table:deepVOL}

\vspace*{\fill}
\end{landscape}

\normalsize
\restoregeometry

\end{appendices}
\pagestyle{plain}

\bibliographystyle{apalike}
\bibliography{references} 

\end{document}